\documentclass[12pt]{article}
\usepackage{epsfig}

\begin{document}



\thispagestyle{empty}

\noindent
\begin{center}
{\large Siberian Branch of Russian Academy of Science \\[2mm]
{\normalsize  BUDKER INSTITUTE OF NUCLEAR PHYSICS}
\\[35mm]
M.N.Achasov, 
V.M.Aulchenko, 
S.E.Baru,
K.I.Beloborodov, 
\\
A.V.Berdyugin, 
A.V.Bozhenok,
A.D.Bukin, 
D.A.Bukin, 
\\
S.V.Burdin,
T.V.Dimova, 
S.I.Dolinsky,
V.P.Druzhinin, 
\\
M.S.Dubrovin, 
D.I.Ganushin, 
I.A.Gaponenko,
V.B.Golubev, 
\\
V.N.Ivanchenko, 
P.M.Ivanov, 
I.A.Koop, 
A.A.Korol, 
\\
M.S.Korostelev, 
S.V.Koshuba, 
A.P.Lysenko,
A.A.Mamutkin, 
\\
I.N.Nesterenko, 
A.V.Otboev,
E.V.Pakhtusova, 
\\
E.A.Perevedentsev, 
A.A.Polunin,
E.E.Pyata, 
\\
A.A.Salnikov,
S.I.Serednyakov, 
V.V.Shary, 
\\
Yu.M.Shatunov,
V.A.Sidorov, 
Z.K.Silagadze, 
\\
A.N.Skrinsky, 
Yu.V.Usov, 
A.A.Valishev, 
\\
A.V.Varganov, 
A.V.Vasiljev,
Yu.S.Velikzhanin
\\[12mm]
Experiments at VEPP-2M with SND detector \\[1.5mm] 
Budker INP 98-65\\
\vfill
NOVOSIBIRSK\\[1mm]
 1998 }
 \end{center}
 

 \thispagestyle{empty}
 \newpage

 \begin{center}
 {\large \bf Experiments at VEPP-2M with SND detector }
  \\[10mm]
{\it 
M.N.Achasov, 
V.M.Aulchenko, 
S.E.Baru,
K.I.Beloborodov, 
\\
A.V.Berdyugin, 
A.V.Bozhenok,
A.D.Bukin, 
D.A.Bukin, 
\\
S.V.Burdin,
T.V.Dimova, 
S.I.Dolinsky,
V.P.Druzhinin, 
\\
M.S.Dubrovin, 
D.I.Ganushin, 
I.A.Gaponenko,
V.B.Golubev, 
\\
V.N.Ivanchenko, 
P.M.Ivanov, 
I.A.Koop, 
A.A.Korol, 
\\
M.S.Korostelev, 
S.V.Koshuba, 
A.P.Lysenko,
A.A.Mamutkin, 
\\
I.N.Nesterenko, 
A.V.Otboev,
E.V.Pakhtusova, 
\\
E.A.Perevedentsev, 
A.A.Polunin,
E.E.Pyata, 
\\
A.A.Salnikov,
S.I.Serednyakov, 
V.V.Shary, 
\\
Yu.M.Shatunov,
V.A.Sidorov, 
Z.K.Silagadze, 
\\
A.N.Skrinsky, 
Yu.V.Usov, 
A.A.Valishev, 
\\
A.V.Varganov, 
A.V.Vasiljev,
Yu.S.Velikzhanin
}
\\[5mm]
  Budker Institute of Nuclear Physics SB RAS \\
  630090 Novosibirsk, Russia
\\[10mm]
\end{center}

\begin{abstract}
Short overview of experiments with SND detector at VEPP-2M $e^+e^-$
collider in the energy range $2E_0 = 400 \div 1400~MeV$ and
preliminary results of data analysis are presented.
\end{abstract}
	
\vfill
\begin{flushright}
\copyright{\small \it
    Budker Institute of Nuclear Physics SB RAS}
\end{flushright}
	
\newpage
	
\vspace{-15mm}

\tableofcontents

\section{Introduction}
The SND detector \cite{SND} was proposed in 1987 \cite{Prep.87}
to continue successful series of experiments with the ND detector 
\cite{ND} at  VEPP-2M $e^+e^-$ collider \cite{VEPP} in Novosibirsk.
The goal of experiments was a detailed investigation 
of $e^+e^-$ annihilation processes
in the same energy range $2E_0 = 400 \div 1400~MeV$,
in particular the final states with photons and other neutral
particles ($\pi^0, \eta, \omega, K_S$), decaying into photons.

The main part of the SND is a spherical electromagnetic calorimeter,
consisted of 1630 $NaI(Tl)$ crystals. The total weight of $NaI(Tl)$
is 3.6~t, the solid angle coverage is $\sim 90\%$ of $4\pi$ steradian.
Energy resolution for photons is $\sigma _E/E = 4.2\% /\sqrt[4\,]{E(GeV)}$
\cite{Calibr.}, angular resolution is about $1.5$ degrees.
The angles of charged particles are measured by two cylindrical
drift chambers covering  95\%
of $4\pi$ steradian solid angle.
Angular accuracy of charged tracks measurements 
($\sigma$) is about $0.4$ and $2.0$ degrees 
in azimuth in polar directions respectively.
From the outside SND detector is covered by muon system,
consisting of streamer tubes \cite{STUB} and plastic scintillation counters.

\section{ PHI-96 Experiment}
Detailed description of PHI-96 experiment
and first physical results, based on about half of the statistics,
were published in \cite{Prep.97}.
The PHI-96 experiment was carried out in the period from February 1996 
until  January 1997.
Seven successive scans PHI\_9601 $\div$ PHI\_9606 and PHI\_9608
were performed in the center of mass energy range $2E_0$ from 980 to 1044~MeV.
Data were recorded at 14 different values of the beam energy.
The total of $4.4~pb^{-1}$ integrated luminosity was collected,
corresponding to 8.3 million $\phi$ mesons produced. 
Most of the results presented in this paper are
based on full recorded statistics.

\section{MHAD-97 Experiment}
The MHAD-97 experiment was described in \cite{Prep.97}.
In the period from January until June 1997 two scans were performed
in the energy range $2E_0$ from 960 to 1380~MeV with a step of 10~MeV
and total integrated luminosity of $6.3~pb^{-1}$.
At present the MHAD9701 scan is accessible for processing completely,
while only part of MHAD9702 scan has been preprocessed.
The results of this paper are based on the
integrated luminosity  $2.8~pb^{-1}$  from both scans, 
corresponding  to production of about 
$1.6\cdot 10^5~\mu^+\mu^-$ pairs.

\section{PHI-97 and OMEGA-98 Experiments }

     The 1997---1998 data taking runs started in October 1997.
The goal was a significant increase of the $\phi$ meson statistics in order
to improve the accuracy of measurements of electrical dipole decays
$\phi \to \pi^0 \pi^0 \gamma$ and
$\phi \to \eta \pi^0 \gamma$, observed with SND for the first time.
Three big scans PHI\_9703, PHI\_9801, PHI\_9802 were carried out, covering
16 energy points in the center of mass energy range $2E_0 = 984 \div 1060~MeV$.
The total recorded integrated luminosity was $10.3~pb^{-1}$,
corresponding to 15 million $\phi$ mesons produced, that is two times
more than in  PHI-96 experiment.

\begin{figure}[htb]
\centerline{\mbox{\epsfig{figure=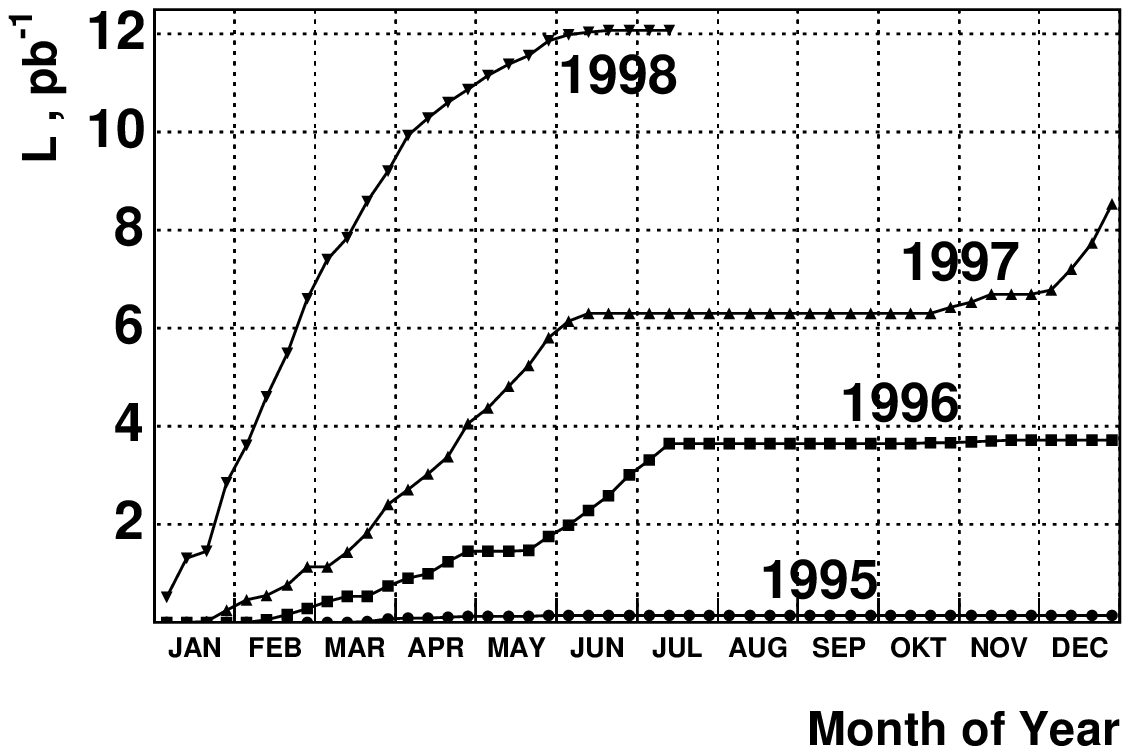,height=65mm}}} 
\vspace*{-5mm}
\caption{Week-by-week graph of integrated luminosity accumulated
during experiments with SND.}
\label{LumWeek}

\centerline{\mbox{\epsfig{figure=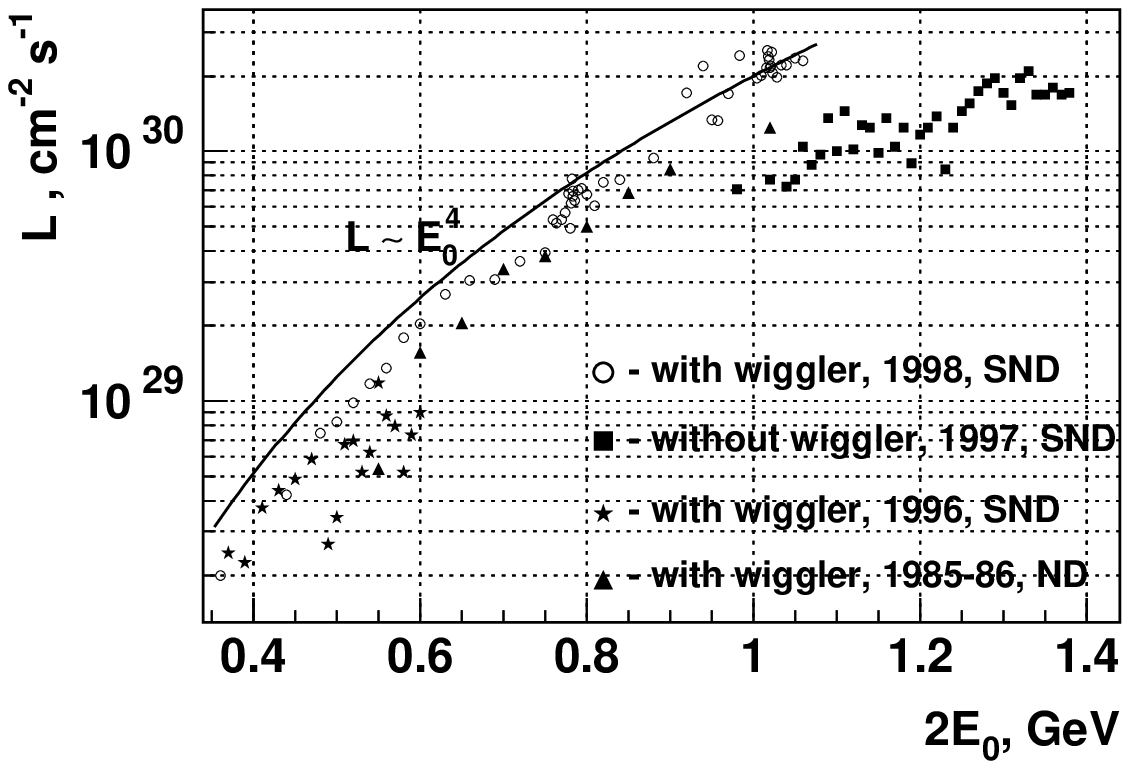,height=70mm}}} 
\vspace*{-5mm}
\caption{Averaged over live time luminosity of the VEPP-2M collider
in ND experiments in
1983$\div$1987 and SND experiments in 1995$\div$1998.
Solid line corresponds to the $L \sim E_0^4$ law and normalized
to $L_0=2\cdot 10^{30}~cm^{-2}s^{-1}$ at $E_0 = 500~MeV$.}
\label{LumvsE}
\end{figure}

From March to June 1998 both SND and CMD-2 detectors
took data in the energy range $2E_0$ from
$970~MeV$ down to $360~MeV$ (OME\_9803 scan).
The goal of this run was measurement of hadronic production cross sections
and search for rare decays of $\rho$ and $\omega$ mesons.
Integrated luminosity of $3.4~pb^{-1}$ was collected,
corresponding to $\sim 4.4\cdot 10^5$ $\mu^+\mu^-$ pairs 
and $\sim 1.9 \cdot 10^6$ $\rho$ mesons produced.
Significant part of data --- $1.2~pb^{-1}$
were accumulated in the energy range $2E_0 = 760 \div 800~MeV$,
where $\sim 1.2 \cdot 10^6$ $\omega$ mesons were produced.
In Fig.~\ref{LumWeek} a week-by-week graph of statistics, recorded by
SND is shown. Fig.~\ref{LumvsE} shows  the VEPP-2M luminosity
averaged over experimental time as a function of energy.
Solid curve in Fig.~\ref{LumvsE} corresponds to
$E_0^4$ dependence, normalized to the experimental value
$L_0=2\cdot 10^{30}~cm^{-2}s^{-1}$ at $E_0=500~MeV$.
All listed above experiments were carried out with a
superconducting wiggler \cite{Wiggler}. 

A distribution of full integrated luminosity accumulated with SND
up to now is shown in Fig.~\ref{ILDE}.
The data processing of last runs PHI-97 and OMEGA-98 
will be started in the end of 1998. 
We believe, that preprocessing 
of this data  will be finished in half a year.
After that, new data will be available for analysis.

\section{General remarks on data processing}

\subsection{The analysis of ``neutral'' modes}
\label{Neut.An}

SND has an advantage in detection  of pure ``neutral''
final states. Up to now the PHI-96 and MHAD-97
experimental data were analyzed in order to study or search for 
the following $\phi$-meson decays:

\begin{equation}
e^+e^- \to \phi \to \pi^0 \pi^0 \gamma \to 5\gamma,
\label{p0p0g}
\end{equation}

\vspace*{-4mm}
\begin{equation}
e^+e^- \to \phi \to \eta \pi^0 \gamma \to 5\gamma,
\label{etap0g}
\end{equation}

\vspace*{-4mm}
\begin{equation}
e^+e^- \to \phi \to \eta \pi^0 \pi^0 \gamma \to 7\gamma,
\label{etap0p0g}
\end{equation}

\vspace*{-4mm}
\begin{equation}
e^+e^- \to\phi\to\eta\prime (958)\gamma \to\eta\pi^0\pi^0\gamma
\label{etapgn}
\end{equation}

\vspace*{-4mm}
\begin{equation}
e^+e^- \to \phi \to \eta \gamma \to 3\pi^0 \gamma \to 7\gamma,
\label{etagn}
\end{equation}

\vspace*{-4mm}
\begin{equation}
e^+e^- \to \phi \to K_S K_L \to neutrals,
\label{kskln}
\end{equation}

\vspace*{-4mm}
\begin{equation}
e^+e^- \to \phi \to \eta\gamma \to 3\gamma, 
\label{etag3g}
\end{equation}

\vspace*{-4mm}
\begin{equation}
e^+e^- \to \phi \to \pi^0\gamma \to 3\gamma,
\label{p0g3g}
\end{equation}

as well as non-resonant or non $\phi$-meson processes:

\begin{equation}
e^+e^- \to \omega \pi^0 \to \pi^0 \pi^0 \gamma,
\label{omp0n}
\end{equation}

\vspace*{-4mm}
\begin{equation}
e^+e^- \to \gamma \gamma,
\label{2g}
\end{equation}

\vspace*{-4mm}
\begin{equation}
e^+e^- \to 3 \gamma.
\label{QEDggg}
\end{equation}

\vspace*{-4mm}
\begin{equation}
e^+e^- \to 4 \gamma, 5\gamma.
\label{QED45}
\end{equation}

\vspace*{-4mm}
\begin{equation}
e^+e^- \to K_S K_L,
\label{ksklmhad}
\end{equation}

\vspace*{-4mm}
\begin{equation}
e^+e^- \to K_S K_L \gamma,
\label{ksklg}
\end{equation}

\vspace*{-4mm}
\begin{equation}
e^+e^- \to K_SK_L \pi^0,
\label{ksklp0}
\end{equation}

\vspace*{-4mm}
\begin{equation}
e^+e^-  \to \eta \gamma \to 3\pi^0 \gamma,
\label{etagnmhad}
\end{equation}

\vspace*{-4mm}
\begin{equation}
e^+e^- \to K_S K_L,\,\, K_S \to \pi^0 \pi^0, 
\label{kspp}
\end{equation}

\vspace*{-4mm}
\begin{equation}
e^+e^- \to \omega \pi^0 \pi^0 \to 3 \pi^0 \gamma,
\label{op0p0}
\end{equation}

\vspace*{-4mm}
\begin{equation}
e^+e^- \to f_2(1270) \to \pi^0 \pi^0.
\label{vslequ02}
\end{equation}

In analysis of any processes listed  above, 
all remaining processes  should be considered as a background.
Processes with any number of photons in the final state
could mimic any certain process, particularly in case, when
the cross-section of the background process is much larger.
The number of found photons in an event could be less than 
the number of produced photons,
due to the merging of close showers in the calorimeter,
or when some of particles escape detection. 
On the other hand, additional photons could be found in events
due to shower splitting or beam background photons hitting the detector.

Parameters, which are widely used in different analyses are listed below:
\\
$N_{\gamma}$ --- number of found photons.
\\
$E_{tot}$ --- total energy deposition.
\\
$P_{tot}$ --- absolute value of total momentum.
\\
$E_{tot}/2E_0$ --- total energy deposition in the calorimeter,
normalized by the center of mass energy.
\\

$P_{tot}/E_{tot}$ --- absolute value of total momentum of all detected 
particles in assumption, that all particles are electrons and photons,
normalized by the total energy deposition.
\\
$E_{np}/2E_0$ --- total energy deposited by neutral particles,
normalized by the center of mass energy $2E_0$.
\\
$E_i$ --- energy deposition of  $i$-th particle.
\\
$E_{\gamma ~max}$ --- maximum energy of neutral particle in an event.
\\
$\theta_i$ --- polar angle of $i$-th particle (particles are enumerated
in the following way: charged particles first, then neutral particles
in descending order in energy within each group).
\\
$\theta_{min}$ --- minimum polar angle between particle and beam direction
in an event.
\\
$\phi_i$ --- azimuth angle of $i$-th particle.
\\
$\zeta$ --- the likelihood
of a hypothesis, that given transverse energy profile of a cluster of
hit crystals in the calorimeter can be attributed to a single photon
\cite{XINM},\cite{Ivanchenko/H-97}.
This parameter allows one to separate events with isolated photon showers,
from events, which have
overlapping showers or group of hit crystals from $K_L$-meson
decay or nuclear interaction.
\\
$\chi_E^2$ --- parameter, describing the degree of energy-momentum balance
in an event under assumption that all particles are photons
and electrons.
\\
$\chi_f^2$ --- parameter, similar to
$\chi_E^2$, describing likelihood of assumption, 
that there is an intermediate state $f$ in event
(for instance $f$ could be $\pi^0 \pi^0 \gamma$, $\eta \pi^0 \gamma$,
etc.).
\\
$N_{\pi ^0}$ --- number of found $\pi ^0$-s.

For example for primary selection of  5-$\gamma$ events of the processes
 (\ref{p0p0g}), (\ref{etap0g}), and (\ref{omp0n}),
the following cuts were imposed:

\begin{eqnarray}
\label{5gamma} 
N_{\gamma}=5;             & N_{cp}=0;                \nonumber \\
0.8 < E_{tot}/2E_0 < 1.1; & P_{tot}/E_{tot} < 0.15;  \\
\theta_{min} > 27^0.      &                 \nonumber 
\end{eqnarray}

\subsection{Analysis of the processes with charged particles}
\label{Char.An}

Up to now the following processes with charged particles
in the final state were analyzed:


\vspace*{-4mm}
\begin{equation}
e^+e^- \to \phi \to \eta e^+e^- \to e^+e^- \gamma \gamma,
\label{etaee}
\end{equation}

\vspace*{-4mm}
\begin{equation}
e^+e^- \to \phi \to \eta \gamma,~ \eta \to e^+e^- \gamma,
\label{TDeq3}
\end{equation}

\vspace*{-4mm}
\begin{equation}
e^+e^- \to \phi \to \eta' \gamma,~ \eta' \to \pi^+ \pi^- \eta,
\eta \to \gamma \gamma, 
\label{etapg}
\end{equation}

\vspace*{-4mm}
\begin{equation}
e^+e^- \to \phi \to \pi^+ \pi^- \pi^+ \pi^- ,
\label{pppp}
\end{equation}

\vspace*{-4mm}
\begin{equation}
e^+e^- \to \omega \pi^0 \to \pi^+ \pi^- \pi^0 \pi^0,
\label{omp0c}
\end{equation}

\vspace*{-4mm}
\begin{equation}
e^+e^- \to e^+ e^- \gamma; 
\label{eeg}
\end{equation}

\vspace*{-4mm}
\begin{equation}
e^+e^- \to e^+e^- \gamma \gamma;
\label{eegg}
\end{equation}

\vspace*{-4mm}
\begin{equation}
 e^+e^- \to\phi\to K^+K^-;
\label{kpkn}
\end{equation}

\vspace*{-4mm}
\begin{equation}
 e^+e^- \to\phi\to K_SK_L, \,\,\, K_S \to\pi^+\pi^-;
\label{ksklc}
\end{equation}

\vspace*{-4mm}
\begin{equation}
 e^+e^- \to \pi^+ \pi^- \pi^0; 
\label{ppp}
\end{equation}

\vspace*{-4mm}
\begin{equation}
 e^+e^- \to \phi \to \pi^+\pi^-\pi^0;
\label{phi3pi}
\end{equation}

\vspace*{-4mm}
\begin{equation}
 e^+e^- \to\phi\to\eta\gamma, \,\,\, \eta\to\pi^+\pi^-\pi^0,
\label{etagc}
\end{equation}

\vspace*{-4mm}
\begin{equation}
 e^+e^- \to \pi^+\pi^-\pi^0\pi^0.
\label{ppp0p0}
\end{equation}

\vspace*{-4mm}
\begin{equation}
\label{vseq2}
e^+e^-\to \pi ^+\pi ^-\pi ^+\pi ^-
\end{equation}

Let us list typical parameters, used in analyses of events with charged 
particles:

$N_{cp}$ --- number of detected charged particles;

$R_i$ --- distance between $i$-th particle track and beam axis
in $R-\phi$ plane;

$Z_i$ --- $Z$ coordinate of the point on the
track of $i$-th particle, closest to the beam axis;

$N_{wire}$ --- number of fired wires in the drift chamber.

$\alpha_{ij}$ --- spatial angle between $i$-th and $j$-th particles.

$dE/dx_i$ --- energy deposition in drift chamber for $i$-th particle.

Parameters of kinematic fit like $\chi_E^2$ are also used
with additional assumptions about masses of charged
particles in a similar way as $\chi_f^2$.

\section{Physical results from the PHI-96 experiment}

\subsection{The $\phi\to\eta\gamma$ decay}

The $\phi\to\eta\gamma$ decay is a radiative  magnetic
dipole transition of $\phi$ into $\eta$ meson,
studied previously in many experiments \cite{PDG}. In this
work we measured the $\phi\to\eta\gamma$ decay rate into
multi-photon final state:
$e^+e^-\to\phi\to\eta\gamma\to 3\pi^0\gamma\to7 \gamma$
(\ref{etagn}). To suppress the background
the events were selected satisfying the following criteria:
\\$\bullet$ 
$N_{\gamma}=6, 7, 8$; $N_{cp}=0$;
$\theta_{min}>27^\circ$;
\\$\bullet$ 
$P/2E_0 < 0.15$;
$0.8 < E_{tot}/2E_0 < 1.1$;
$\chi^2_E < 30$.  
\\The presence of  $\pi^0$ mesons in event
was not required in the kinematic fit.
As a result of such selection cosmic background was completely rejected and
main background process
(\ref{kskln}) was suppressed significantly, as one can see
in the distribution in
$m_{rec.\gamma}$, recoil mass of the most energetic photon in the
event 
  (Fig.\ref{etg98}), where peak at $\eta$ mass dominates.
For final selection of $\eta\gamma$ events  the soft
cut on $m_{rec.\gamma}$ was used:
$400~MeV < m_{rec.\gamma} < 620~MeV$.
The estimated detection efficiency is close to $10\%$ and
background contribution,
obtained using the number of events from the
$m_{rec.\gamma}$ interval $620 \div 840~MeV$, is about $2\%$. 
The ratio of the background events numbers in these  two regions was 
taken from simulation of the process (\ref{kskln}). 
The fit of visible cross section was done for each of the 6 scans
separately. Free parameters of the fit were the shift of the
energy scale of VEPP-2M, $\phi$ meson width, and the branching ratio
$B(\phi\to\eta\gamma)$. 
Also taken into account were collider beam energy spread,
instability of average beam energy, the radiative corrections,
the VDM contribution of $\rho$ and $\omega$ resonances,
and presence of dead calorimeter channels. 
The fit results are listed in Table~\ref{IVN:ETG}.
\begin{table}[htb]
\caption{The fit results for the $\phi\to\eta\gamma$
decay in 6 independent scans.}
\label{IVN:ETG}
\begin{tabular}{llll}
\hline
Experiment & $N_{events}$ & $\Gamma_{\phi}$, MeV  & 
$B(\phi\to\eta\gamma), \%$ \\
\hline
PHI\_9601 & 1045 & $4.34\pm0.43$ & $1.141\pm0.066$ \\
PHI\_9602 & 1436 & $4.00\pm0.34$ & $1.188\pm0.044$ \\
PHI\_9603 & 2163 & $4.29\pm0.33$ & $1.192\pm0.036$ \\
PHI\_9604 & 1241 & $4.53\pm0.39$ & $1.127\pm0.059$ \\
PHI\_9605 & 2222 & $4.38\pm0.24$ & $1.271\pm0.049$ \\
PHI\_9606 & 1709 & $3.97\pm0.27$ & $1.339\pm0.059$ \\
\hline
\end{tabular}
\end{table}
Averaging the data from the Table~\ref{IVN:ETG} one can
obtain the branching ratio
\[B(\phi\to\eta\gamma)=(1.209\pm 0.028\pm0.050)\%.\]
Here the first error is statistical and the second is a systematic one,
estimated to be  $4.2~\%$. The systematic error
is mainly determined by the following contributions:
\\$\bullet$ the systematic uncertainty of normalisation ($3\%$);
\\$\bullet$ the background subtraction error ($1\%$);
\\$\bullet$ the error in the detection efficiency  ($1\%$);
\\$\bullet$ the error in $B(\phi\to e^+e^-)$ ($2\%$ \cite{PDG});
\\$\bullet$ the error in $B(\eta\to3\pi^0)$ ($1.2\%$ \cite{PDG});
\\$\bullet$ the error in value of $\rho-\phi$ interference term  ($1\%)$.

The results in Table~\ref{IVN:ETG} show some
difference between separate PHI-96 scans.
The scale factor for the branching ratio,
calculated according to PDG recommendations, is equal to 1.4. It 
was taken into account in the presented statistical error.
 
The result obtained in the present work
is in agreement with PDG value $(1.26\pm0.06)\%$ \cite{PDG}.
At the moment it is the most accurate single
measurement of $B(\phi\to\eta\gamma)$.

\subsection{Analysis of the
 $\phi \to \eta e^+e^- \to \gamma\gamma e^+e^-$ decay} 

The decay  $\phi \to \eta e^+e^-$ is closely related to the
radiative decay $\phi \to \eta \gamma$,
where, instead of a real photon, virtual one is produced,
decaying via the channel $\gamma^* \to e^+e^-$. This type 
of decays is called Dalitz or conversion one.
Experimental study of these processes is a test of
quantum electrodynamics. Besides this,
existence of transition form factor 
\cite{Landsberg} affects invariant mass spectrum $m_{e^+e^-}$, 
which was evaluated in different models. Theoretical value of
$B(\phi \to \eta e^+e^-)$
for a unit form factor is $1.1\cdot10^{-4}$

This decay was observed for the first time
in our experiment with the detector ND~\cite{NDETAEE} in 1985, 
with a branching ratio of
 $(1.3^{+0.8}_{-0.6})\cdot 10^{-4}$.
Later similar result was obtained with the
CMD-2~\cite{CMD96} detector:
$B(\phi \to \eta e^+e^-)=(1.10\pm 0.49\pm 0.19)\cdot 10^{-4}$.

In present analysis
the $\phi \to \eta e^+e^-\to \gamma\gamma e^+e^-$ decay
was studied in the reaction
 $e^+e^- \to \phi \to \eta e^+e^- \to e^+e^- \gamma \gamma$
(\ref{etaee}). The following event selection criteria were used:
$$
\begin{array}{ll}
N_\gamma=2, & N_{cp}=2, \\
E_{tot}/2E_0 > 0.8, & R_1,R_2<0.2\mbox{ cm}, \\
10^\circ<\alpha_{1,2}<110^\circ , & L_{\gamma\gamma ee}<6, \\
\end{array}
$$
where $L_{\gamma\gamma ee}$ is a logarithmic likelihood function,
obtained as a result of kinematic fit,
using the energies and angles of all particles
(invariant mass of two photons
$m_{\gamma\gamma}$  being evaluated as well).
All 7 $ \phi$-meson  scans were processed with
these selection criteria.

\begin{figure}[htb] 

\centerline{\epsfig{figure=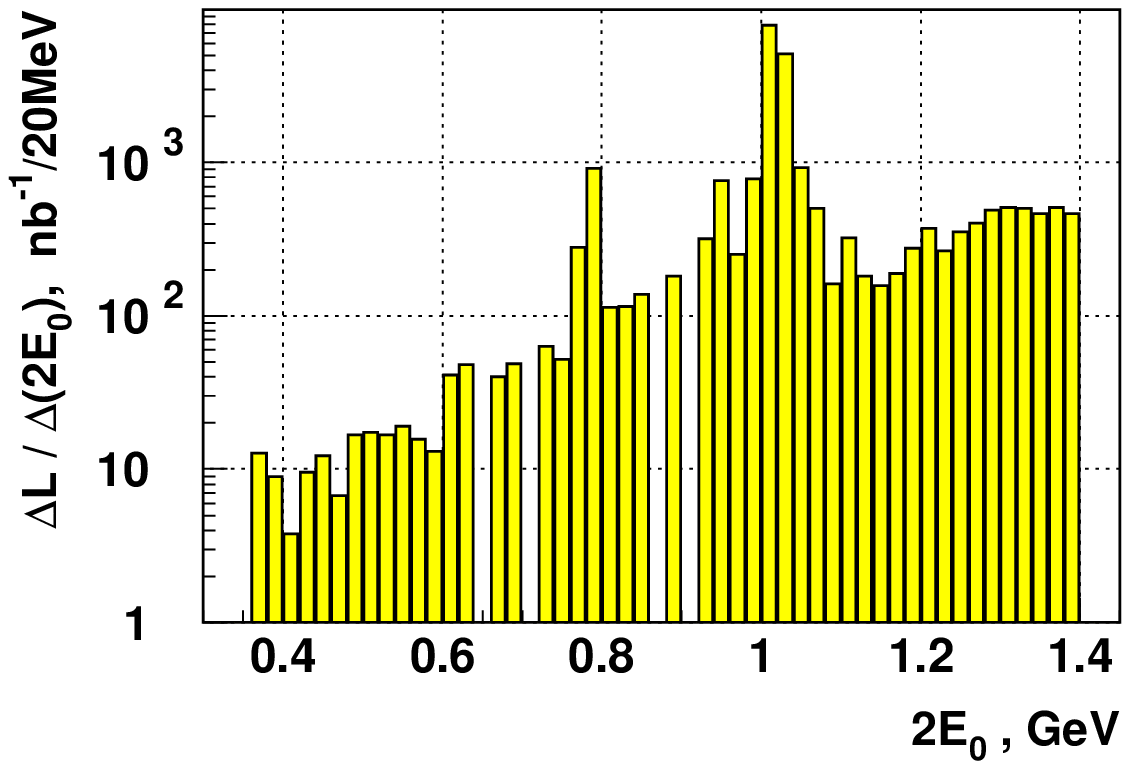,width=0.95\textwidth}} 
\vspace*{-5mm}
\caption{
Integrated luminosity,
accumulated with SND up to July 1998, as a function of energy.
The histogram bin width is equal to $20~MeV$.}
\label{ILDE}

\epsfxsize=0.49\textwidth
  \begin{minipage}[t]{0.47\textwidth}
   \centerline{\epsfbox{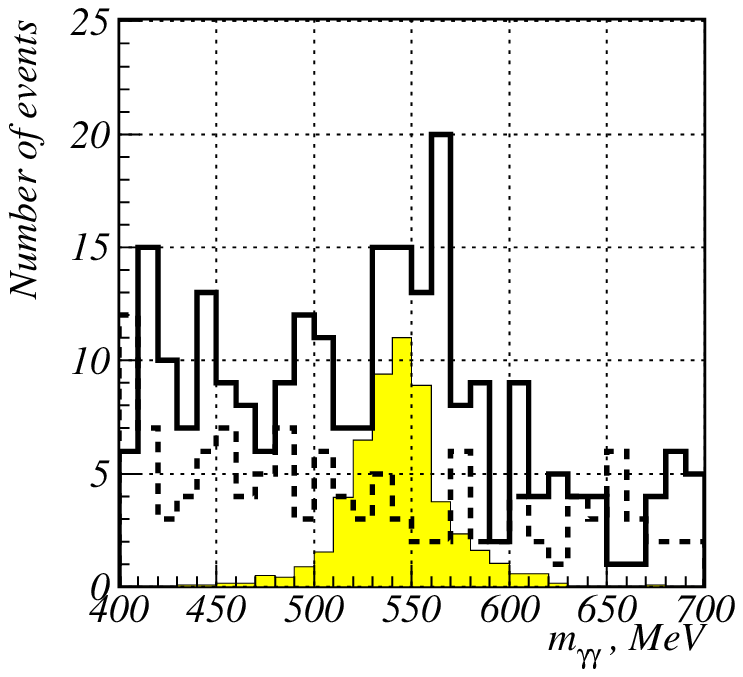}} 
\vspace*{-5mm}
\caption{
Two-photon mass distribution in search for the process
$e^+e^- \to \phi \to \eta e^+e^-$, $\eta \to \gamma \gamma$.
Histogram -- experimental data;
hatched histogram -- simulation of the decay
$\phi\to\eta e^+e^-$;
dashed line -- simulation of the QED process
$e^+e^-\to e^+e^-\gamma\gamma$.}
    \label{Etaee_inv_mass}
  \end{minipage}
\hfill
  \begin{minipage}[t]{0.47\textwidth}
   \centerline{\epsfbox{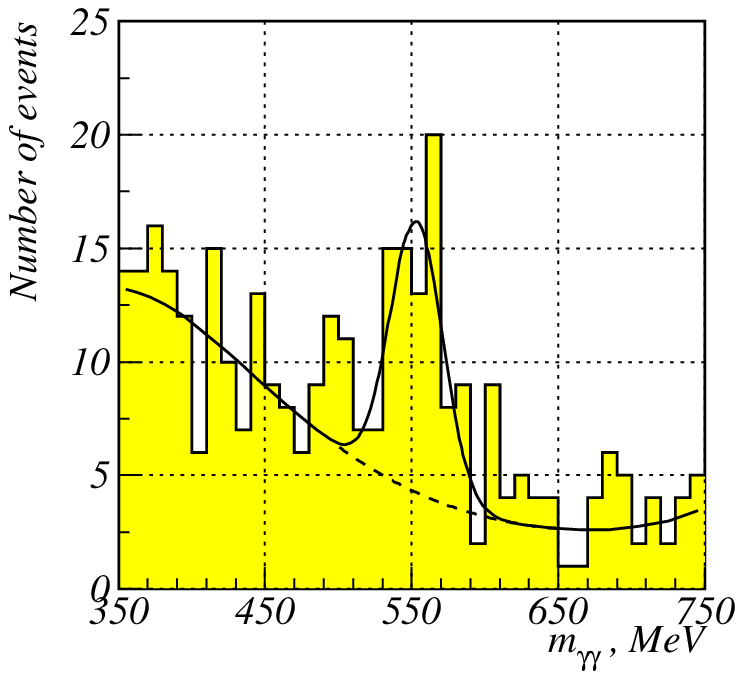}} 
\vspace*{-6mm}
\caption{
Optimal fit of the measured two-photon invariant mass distribution
in search for the process
$e^+e^- \to \phi \to \eta e^+e^-$, $\eta \to \gamma \gamma$.}
\label{Fit_inv_mass_2gamma}
  \end{minipage}
\end{figure}

The distributions of experimental events over invariant
mass of two photons together with corresponding distributions
of simulated $\phi\to\eta e^+e^-$
and QED $e^+e^-\to e^+e^-\gamma\gamma$ events
are shown in Fig.\ref{Etaee_inv_mass}. 
The simulated distributions are normalized to the  
branching ratio $B(\phi\to\eta e^+e^-)=1.3\cdot 10^{-4}$
from \cite{PDG} for $\phi\to\eta e^+e^-$, and 
to the total integrated luminosity for QED 
$e^+e^-\to e^+e^-\gamma\gamma$ reaction.
One can see that the experimental background 
is almost entirely provided by QED process, which has
identical final state as the one, searched for.
Now let us evaluate the result in two ways.

First, the experimental distribution over $m_{\gamma\gamma}$
was fitted with a background, approximated by a
third order polynomial
and Gaussian, representing the $\phi\to\eta e^+e^-$ process.
As a result of the fit  (Fig.\ref{Fit_inv_mass_2gamma})
we obtain the number of events in the peak $N_0 = 54\pm 15$.
Branching ratio reads:
\begin{equation}
B(\phi\to\eta e^+e^-)=\frac{N_0}{N_{\phi} \epsilon} =
(1.42\pm 0.39)\cdot 10^{-4},
\end{equation}
where $N_{\phi} = 8.3\cdot 10^{6}$ is the number of
produced $\phi$-mesons,
$\epsilon=4.45\pm 0.18\%$ is a detection efficiency, obtained by simulation.
Systematic error in $\epsilon$ here was neglected.

Second, let us derive branching ratio
$\phi\to \eta e^+e^-$, using the energy dependence of visible cross section
of the events, satisfying the additional condition
 $500~MeV < m_{\gamma\gamma} < 600~MeV$.
The fitted resonance curve is shown in Fig.\ref{Etaee_res_fit}.
 $\phi$-meson mass and width were fixed
 at the values $m_\phi=1019.41~MeV$
and $\Gamma_\phi=4.43~MeV$ \cite{PDG}.
As a result of the fit the following parameters values were
obtained:
$B(\phi\to\eta e^+e^-)=(1.71\pm 0.42)\cdot 10^{-4}$,
background level $\sigma_{bg}=8.7^{+3.3}_{-2.7}~pb$.
Apparently, the value of
 $B(\phi\to\eta e^+e^-)$ is excessive due to
 resonant background admixture
(supposedly a greater part is due to $\phi\to\pi^+\pi^-\pi^0$).
Correction for this resonant background we estimate
in the following way.
Let
$\xi$ be the ratio of the resonant and
non-resonant background level.
The values of $\xi$, derived by fitting
a resonance curve to energy distributions of background events
of the intervals
 $m_{\gamma\gamma}=
350\pm 50$, $450\pm 50$, $650\pm 50$ and $750\pm 50~MeV$,
are well approximated with a straight line.
Thus at the value of
 $m_{\gamma\gamma}\sim 550~MeV$ we obtain
$$\left. \xi \right| _{m_{\gamma\gamma}=550}=
\left(5.8\pm 2.7\right)\cdot 10^{-2}.$$
After subtraction of the resonant background one gets:
\begin{equation}
B(\phi\to\eta e^+e^-)=(1.21\pm 0.51) \cdot 10^{-4}
\end{equation}

Obtained in essentially different ways the two values of
 $B(\phi\to\eta e^+e^-)$: (35) and (36)
are in a good agreement.
As a final result one can choose (35) as the most precise of them
and consider their difference as a measure of
systematic error of $\pm 0.21$ or $\pm 15\%$.
After taking into account other known sources
of systematics the  error estimate increases up to $16\%$ and the final result
reads as
\begin{equation}
B(\phi\to\eta e^+e^-)=(1.42\pm 0.39\pm 0.23) \cdot 10^{-4}.
\end{equation}
It is close to the theoretical value and
does not contradict previous measurements.

\subsection{ Study of the process $ \phi \to  \eta \gamma$ ,
            $ \eta \to  e^+ e^- \gamma$ } 

The Dalitz decay
\(\eta \to e^+ e^- \gamma \) was studied in the reaction
\(e^+ e^- \phi \to \eta \gamma, ~ \eta \to e^+ e^- \gamma \)
(\ref{TDeq3}). The ratio of probabilities of the
Dalitz decay and the two photon decay of $\eta$ meson is given
by the following expression\cite{Landsberg}:
$$                                                      
\frac{d}{d m_{ee}^2}                             
\frac{ B(\eta \to e^+ e^- \gamma )}{B(\eta \to \gamma \gamma)} =
\frac{2 \alpha }{3 \pi} \frac{|F_{\eta}(m_{ee}^2)|^2}{m_{ee}^2}
\left (1+ \frac{2 m_e^2}{m_{ee}^2}\right )                                  
\left (1-\frac{4 m_e^2}{m_{ee}^2}\right )^{1/2}
\left (1-\frac{m_{ee}^2}{m_\eta^2}\right )^3       
$$
where $m_{ee}$ is a  $e^+e^-$ invariant mass, $ F_{\eta}(m_{ee}^2)$ -
$\eta$-meson transition form factor.
The main contribution is given by small  $m_{ee}$,
where $|F_{\eta}| \approx 1$ . The expected branching ratio is:
$ B(\eta \to e^+ e^- \gamma) = 6.3 \cdot 10^{-3} $.

This decay was measured in only one experiment \cite{TDNEEG},
where in two successive papers the results
differ by a factor of 3.
 We believe, that our measurement could clarify the situation with
this decay.

{\bf Data analysis}

The following event selection criteria were used :
\\ $\bullet$ positive charged event trigger;
\\ $\bullet$ $N_{cp}=2$; $2 \leq N_{\gamma} \leq 3$;
$ R_1, R_2 < 0.5~cm $; $ |Z_1|, |Z_2| < 10~cm $;
\\ $\bullet$ $\theta _{min} > 36^\circ$; $\Delta \phi_{ee} > 5^\circ$;
$E_{tot}/2E_0 > 0.8 $,  $ P/E_{tot} < 0.15$;
$\chi^2_E < 15 $;
\\ $\bullet$ the events with invariant mass of photon pairs
close to masses of $\pi^0$ и $\eta$ mesons
$110~MeV < m_{\gamma \gamma} < 170~MeV$;
$500~MeV < m_{\gamma \gamma} < 600~MeV$;
are excluded.
\\ $\bullet$ $E_{\gamma~min} > 50~MeV$;  
recoil mass of one photon is close to
$\eta$ meson mass in the decay $ \phi \to  \eta \gamma$ :
$0.6<E_{\gamma}/E_0<0.7$ ;
$0.1 < m_{e^+e^-}/m_{\eta} < 0.7$.

After application of these selection criteria 60 events survived.
We estimated their origin in the following way:
\begin{center}
\begin{tabular}{||l|r||}
\hline
Experiment (total) & 60 \\
Contribution from the process $e^+e^-\to e^+e^-\gamma\gamma$ & 35 \\
Contribution from the process $\phi \to \eta e^+ e^-$ & 0.5 \\
Contribution of the process under study& 24.5 \\
\hline
\end{tabular}
\end{center}

Taking the efficiency $4.15\%$ from simulation \cite{Landsberg},
we have

$$ B(\eta \to e^+ e^- \gamma) = \frac{24.5}
{0.0415 \cdot N_\phi \cdot B(\phi\rightarrow \eta\gamma)} 
= 6.1 \cdot 10^{-3}. $$

  The energy dependence was fitted by the sum of QED non-resonant
background and resonant contribution from the decay under study 
(fig.\ref{tdpic3}). The following fit  result was obtained
$ B(\eta \to e^+ e^- \gamma) = (6.8 \pm 1.1 )\cdot 10^{-3} $.
The  systematic error was estimated to be about $10\%$.
The final result is:
$$ B(\eta \to e^+ e^- \gamma) = 
(6.8 \pm 1.1 \pm 0.7)\cdot 10^{-3}. $$
The result obtained in this work is in good agreement 
with the PDG value $(4.9 \pm 1.1)\cdot 10^{-3}$ 
\cite{PDG} and with the prediction  $6.3 \cdot 10^{-3}$.
The accuracy of our measurement is close to the world average 
value\cite{PDG}.

\subsection{Search for the $\phi \to \eta \pi^0 \pi^0 \gamma $ decay}

Radiative decays of vector mesons $V\to M\gamma$, where $M$
is a scalar or pseudo-scalar state, is an important source of information on
the structure of these states. For $V=\phi(1020)$ only main decays
$\phi\to\eta\gamma ,\pi^0\gamma$ \cite{Ph.Rep} were studied by now
at a relatively 
high level of accuracy of $\leq 10\%$. Regarding rarer decays like
$\phi\to\eta' (958)\gamma,~f_0(980)\gamma,~a_0(980)\gamma$, 
even the latest data \cite{Serednyakov/H-97}, \cite{Ivanchenko/H-97}, 
\cite{CMD-2} are still preliminary and have rather low accuracy.

    The $\phi\to\eta'\gamma$ decay was observed in the only experiment
with CMD-2 detector at VEPP-2M collider \cite{CMD-2} with a
branching ratio of $(1.2^{+0.7}_{-0.5})\cdot 10^{-4}$. The quark model
prediction for this value \cite{Q-model}
is $(0.7 \div 1.0)\cdot 10^{-4}$ under assumption that there is
no gluonium component in $\eta'$. In case of 
$\eta'$ consisting of pure gluonium, the expected
branching ratio of the decay
$\phi\to\eta'\gamma$ would be 4$\cdot 10^{-6}$ \cite{DES82}.   

    In this work the decay
$\phi\to\eta\pi^0\pi^0\gamma\to7\gamma$ was studied.
The pseudo-scalar $M=\eta\pi^0\pi^0$ final state could appear from the
decay of $\eta'$(958) or from states with higher masses and large
widths like $\eta$(1295) or $\eta$(1440). The expectation for
the  $\phi\to\eta'\gamma\to\eta\pi^0\pi^0\gamma$
branching ratio, based on quark model predictions \cite{Q-model},
is $B(\phi\to\eta'\gamma)\cdot 
B(\eta'\to\eta\pi^0\pi^0)=(1.5\div2.0)\cdot 10^{-5}$.

{\bf The selection criteria} 
In this work the reactions
(\ref{etap0p0g}) and
(\ref{etapgn}) 
were studied, both producing a 7-photon final states. 
The main background comes from the $\phi$-meson decays 
(\ref{etagn}) and
(\ref{kskln}).

     The events of the processes under study 
(\ref{etap0p0g}) and (\ref{etapgn}) were selected
using the following criteria:

Sel.1 -- event configuration does not contradict energy and momentum
conservation: $\chi^2_E<15$. 
This requirement suppresses the background from the process 
(\ref{kskln}).
 
\begin{figure}[htb]

\epsfxsize=0.49\textwidth
  \begin{minipage}[t]{0.47\textwidth}
   \centerline{\epsfbox{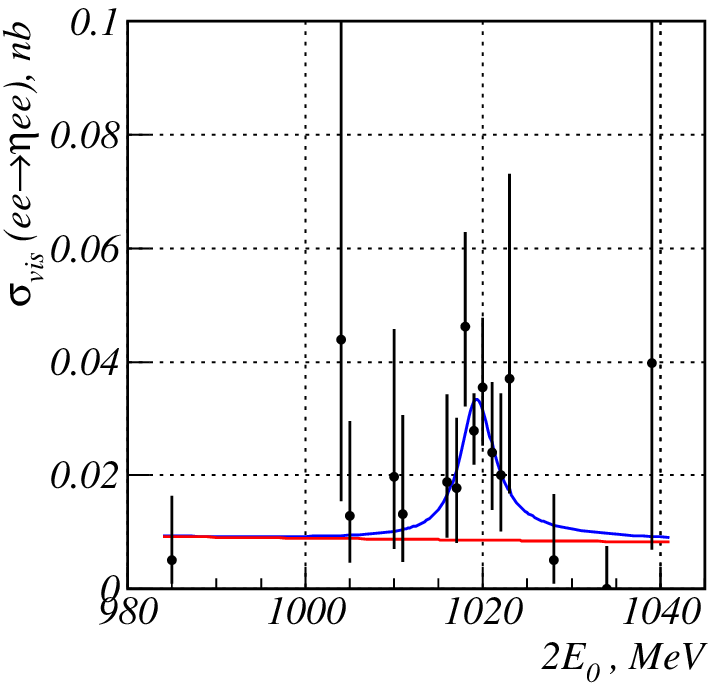}} 
\vspace*{-5mm}
\caption{
Optimal fit of the energy dependence of
$e^+e^- \to \phi \to \eta e^+e^-$,
$\eta \to \gamma \gamma$ visible cross section
(for $500~MeV < m_{\gamma\gamma} < 600~MeV$).}
\label{Etaee_res_fit}
  \end{minipage}
\hfill
  \begin{minipage}[t]{0.47\textwidth}
   \centerline{\epsfbox{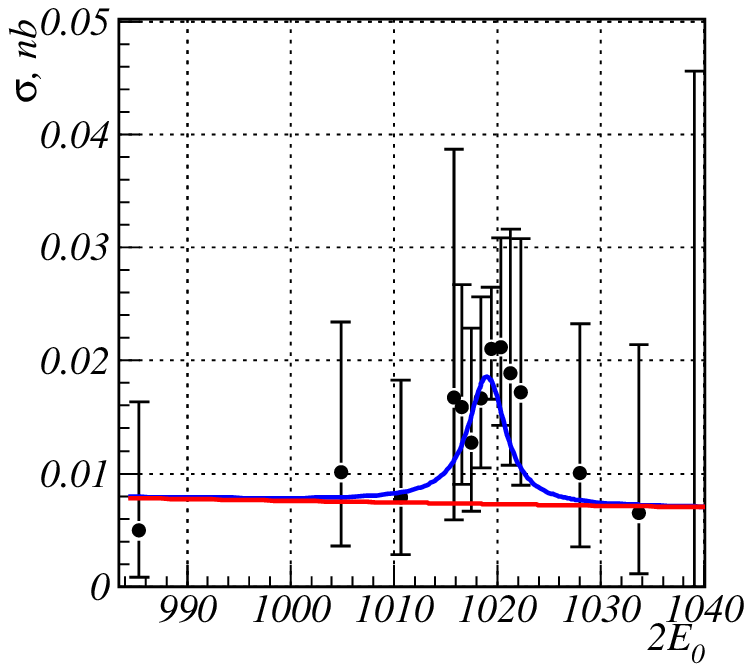}} 
\vspace*{-5mm}
\caption{
Optimal fit of energy dependence of
$\phi \to \eta \gamma$ , $\eta \to e^+e^- \gamma$ 
visible cross section.}
\label{tdpic3}
  \end{minipage}
\vfill
  \begin{minipage}[t]{0.47\textwidth}
\centerline{\epsfbox{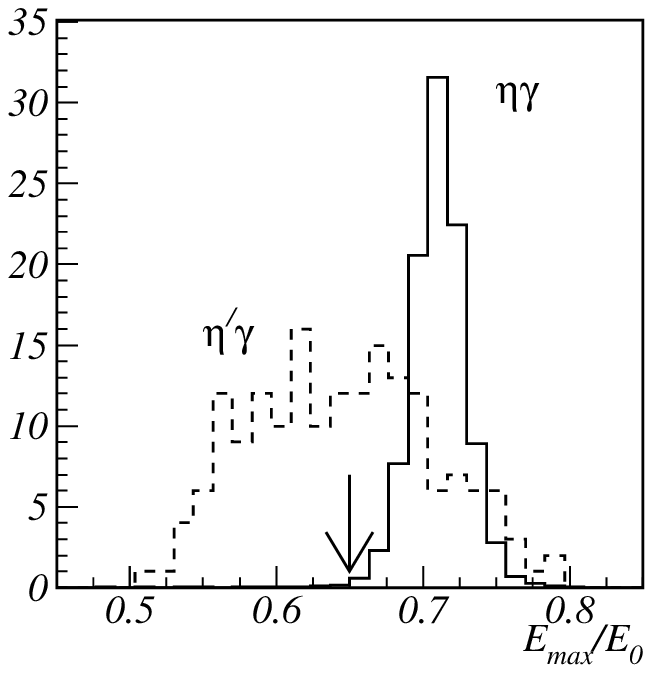}} 
\vspace*{-5mm}
\caption{
The normalized energy spectra of the most energetic photon in
experimental events of the reaction  $e^+e^- \to\phi\to\eta\gamma\to
7\gamma$ and simulated
events of the reaction $e^+e^- \to\phi\to\eta\prime\gamma\to 7\gamma$.
The cut at $E_{max}=0.65E_0$ is shown by arrow.}
\label{ssf1}
  \end{minipage}
\hfill
  \begin{minipage}[t]{0.47\textwidth}
\centerline{\epsfbox{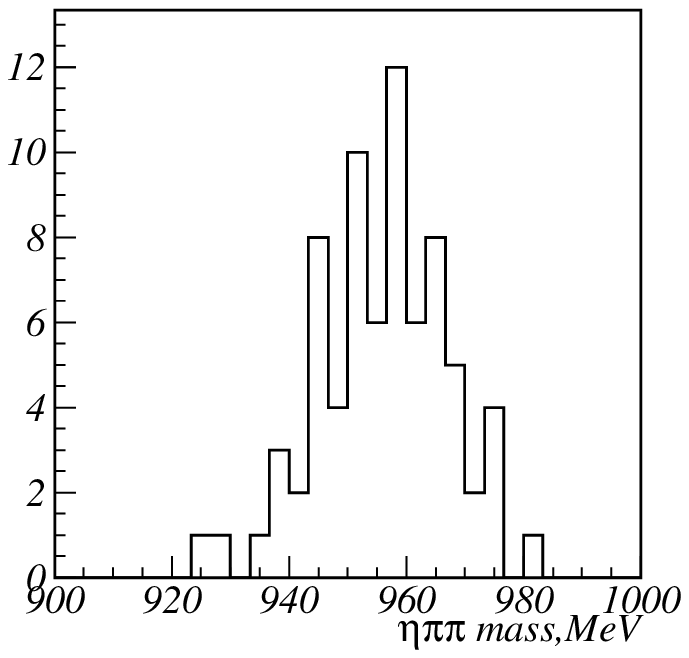}} 
\vspace*{-5mm}
\caption{
The reconstructed mass spectrum of  $\eta\pi^0\pi^0$ system
for simulation of the process
$e^+e^- \to \phi \to \eta \gamma \to 3\pi^0 \gamma \to 7\gamma$.}
\label{ssf2}
  \end{minipage}
\end{figure}

Sel.2 -- the energy of the most energetic photon
in the event $E_{max}/E_0 < 0.65 $. Fig.\ref{ssf1} shows,
that such a threshold almost completely rejects 
the background from process (\ref{etagn}), although the detection 
efficiency for the process (\ref{etapgn}) becomes twice smaller. 

Sel.3 -- the transverse spread of each shower in the calorimeter does not
exceed certain standard value, characteristic for individual showers
$\zeta <0$. 
This cut suppresses the background events from the process 
(\ref{kskln}), where some close photon showers merge together.  

Sel.4 -- among all photon pairs in the event, three meson candidates
must be found:
two $\pi^0$-mesons with the effective mass
$110<m_{\gamma\gamma}<160~MeV$ and one $\eta$-meson with
$520<m_{\gamma\gamma}<580~MeV$. Analysis of simulated events of process
(\ref{etapgn})
showed, that the pair of most energetic photons in the event is
produced by $\eta(550)\to2\gamma$ decay. This fact significantly reduces
combinatorial background in the search of  $\pi^0$
and  $\eta$ candidates in the event.

Sel.5 -- the effective mass of $\eta\pi^0\pi^0$-system  is close to 
$\eta'$ mass. The obtained mass spectra  of $\eta'$  for 
simulated events of process (\ref{etapgn}) is shown in Fig.\ref{ssf2}.

\begin{figure}[htb]
\epsfxsize=0.49\textwidth

  \begin{minipage}[t]{0.47\textwidth}
  \centerline{\epsfbox{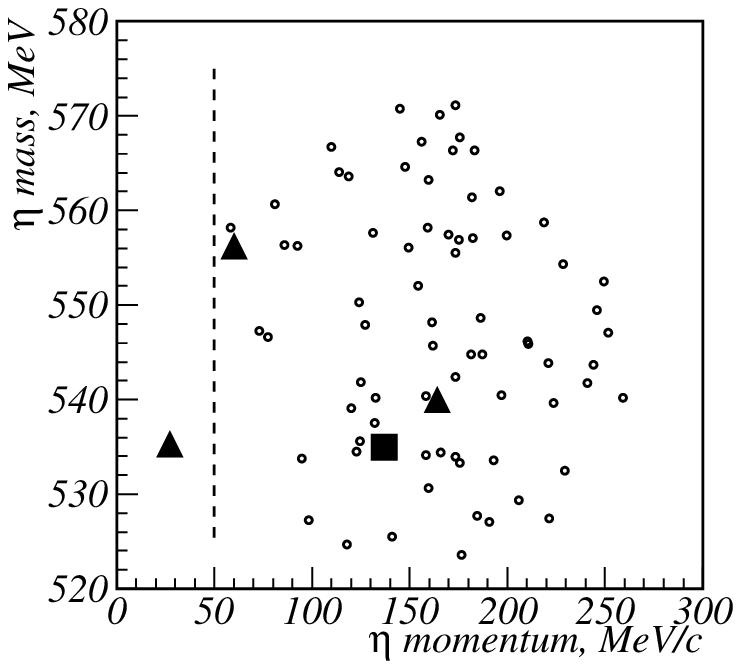}} 
\caption{
Two-dimensional plot of the measured mass of $\eta(550)$
meson versus its momentum in search for $\eta\pi^0\pi^0\gamma$
events.
Circles are simulated events of the reaction
\(e^+e^- \to\phi\to\eta' (958)\gamma \to\eta\pi^0\pi^0\gamma\),
square represents one event from simulation of the process
\(e^+e^- \to \phi \to \eta \gamma \to 3\pi^0 \gamma \to 7\gamma\),
triangles are experimental events.
The vertical
dashed line shows the momentum cut at $P_{\eta}>$50~MeV.}
\label{ssf3}
  \end{minipage}
\hfill
  \begin{minipage}[t]{0.47\textwidth}
  \centerline{\epsfbox{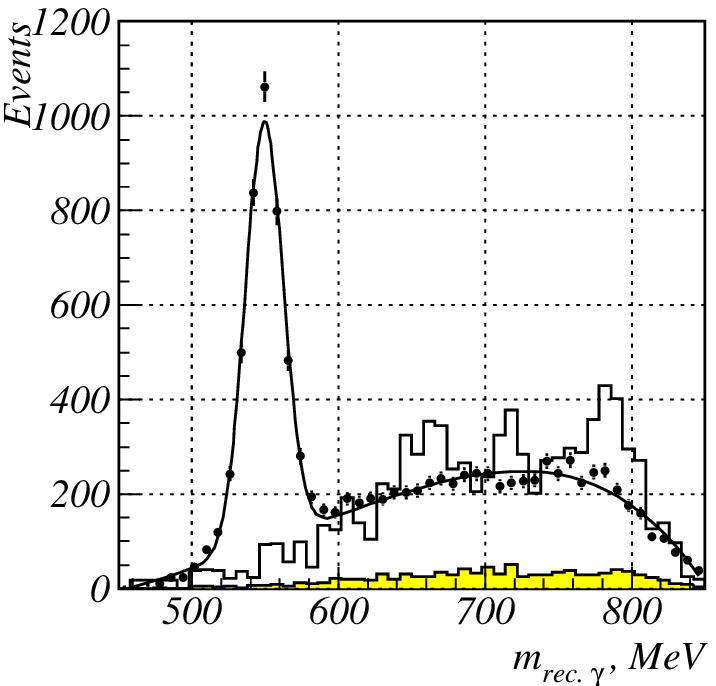}} 
\caption{
Recoil mass distribution of the most energetic photon
in search for the $\phi \to \eta' \gamma$ decay.
Circles with error bars -- experimental data (11239 events);
curve -- optimal fit of the experimental data;
histogram -- estimated sum of contributions from background processes
$e^+e^- \to \phi \to \pi^+ \pi^- \pi^0$
(\ref{phi3pi}),
$e^+e^- \to \phi \to \omega \pi^0 \to \pi^+ \pi^- \pi^0 \pi^0 $
(\ref{omp0c});
hatched histogram -- contribution from the process
$e^+e^- \to \phi \to \omega \pi^0 \to \pi^+ \pi^- \pi^0 \pi^0 $
(\ref{omp0c}).}
\label{dm_erp3}
  \end{minipage}
\end{figure}

{\bf The data analysis.}
The $\phi\to\eta'\gamma$ decay  detection efficiency
with respect to all events of this decay
$\epsilon_{1}=(0.73\pm 0.09)\%$ was obtained using simulation of the process 
(\ref{etapgn})
with the selection criteria Sel.1 --- Sel.5. Corresponding
efficiency with
respect to events of the process (\ref{etapgn})
with exactly 7-photons in the final state
is much higher $\sim$ 9\%.
Among all experimental events $N_x$=3
were found, satisfying selection criteria. They are shown as black
triangles in the scatter plot Fig.\ref{ssf3}. The vertical axis of
the plot depicts the effective mass of the found $\eta$(550) meson,
while the horizontal axis shows its momentum.  After the additional cut
on $\eta$(550) momentum $P_{\eta}>50~MeV/c$ two experimental events survive. 
Simulated events of the process (\ref{etapgn}) are shown in the 
Fig.\ref{ssf3} as circles. 
One event, shown as a square,  was found in the analysis of the simulated
$\phi\to\eta\gamma$ background sample. The estimated detection 
efficiency for this background is close to $10^{-5}$. 

In the analysis of 4.7$\cdot 10^5$ simulated events of the reaction 
(\ref{kskln}) no
events of $\phi\to K_SK_L$ decays, satisfying selection criteria
Sel.1---Sel.5 were found, but one should take into account,
that $\phi\to K_SK_L$ simulation statistics is
5 times smaller than the experimental one. 
Because the simulation is not precise, certain
contribution of the process (\ref{kskln}) into the sample of two
selected experimental events cannot be completely excluded.  

Both found experimental events correspond to a collider beam energy
near $\phi$(1020) maximum, which confirms their resonant origin. If 
both events  originate from the process (\ref{etapgn}), 
the branching ratio would be equal to
($0.4\pm 0.3)\cdot 10^{-4}$. But as it was pointed out above, the background
origin of these events from processes 
(\ref{etagn}) and (\ref{kskln}) could not be excluded.
Thus only upper limit on the branching ratio of the decay 
(\ref{etapgn}) could be reliably established.
At a confidence level of 90\% it is equal to
\begin{equation}
Br(\phi\to\eta'\gamma)<\frac{N_{x}k}{N_{\phi}\epsilon_1}=1.1\cdot 
10^{-4},  
\label{sseq1}
\end{equation}
where coefficient $k=2.7$ comes from 
Poisson distribution of observed decays \cite{PDG}.

   To check the whole procedure on possible systematic errors we studied
another decay chain
(\ref{etagn}) with a similar 7-$\gamma$ final state.
Its probability is known and could be compared with an obtained value.
The selection criteria Sel.1$\otimes$Sel.3 were imposed.
In addition  it was required for each selected event
to contain three $\pi^0$ candidates.
We found \(B(\phi\to\eta\gamma)= 1.05\cdot 10^{-2}\), that is 18\%
lower than the table value. Such large discrepancy could be
attributed to excessively stringent
selection criteria Sel.1$\otimes$Sel.3, used in the search of process 
(\ref{etapgn}). Now we
can correct the detection efficiency $\epsilon_1$ obtained using simulated
events by 18\% down to a value of
$\epsilon_1=0.6\%$, and use it within the expression (\ref{sseq1}) as an
experimentally obtained detection efficiency. 

   In search for the process (\ref{etap0p0g})
   without any assumptions about its dynamics
the cuts Sel.1---Sel.4 were applied. The detection efficiency  $\epsilon_2$
was determined from  a sample of simulated events of the reaction 
(\ref{etap0p0g})with the
matrix element similar to that of the process (\ref{etapgn}) 
and effective masses of
$\eta\pi^0\pi^0$ system ranging from 850 to 1000 MeV. The obtained value of
$\epsilon_2$ is nearly constant in this effective mass range and equals
to $\sim$3\% with the correction described above imposed.
In the experimental sample $N_x$=3 events were found,
one event was found in simulated sample of background process (\ref{etagn})
and none were found in the simulated sample of (\ref{kskln}).
After formal background subtraction
the estimated the number of events from the process (\ref{etap0p0g}) is
$N_x=(3-1)\pm \sqrt{3+1}=2\pm 2$. Given such a small number of events
and not reliable estimation of background, only upper
limit of the branching ratio can be placed using the expression 
(\ref{sseq1}):

$B(\phi\to\eta\pi^0\pi^0\gamma)<2 \cdot 10^{-5}$.

{\bf Conclusion.}
In the analysis of 7-$\gamma$ events the following upper limits at $90\%$
confidence level were placed:

\(B(\phi\to\eta'\gamma)<1.1\cdot 10^{-4},\)

\(B(\phi\to\eta\pi^0\pi^0\gamma) < 2\cdot 10^{-5}\)
 (incl.$\eta'\gamma$).

\subsection{Observation of the $\phi \to \eta' \gamma $ decay}

The result of the first measurement of
$\phi \to \eta' \gamma $ branching ratio
in $\eta' \to \eta \pi^+ \pi^-$, $\eta \to \gamma \gamma $
decay mode
with the CMD-2  detector at VEPP-2M \cite{CMD-2} is
$ B(\phi \to \eta' \gamma) = (1.2^{+0.7}_{-0.5}) \cdot 10^{-4}$.
At the same time
in SND experiment only upper limit was obtained \cite{Prep.97}
in another decay mode:
$\eta' \to \eta \pi^+ \pi^- $, $\eta \to \pi^+ \pi^- \pi^0$:
$ B(\phi \to \eta' \gamma) < 1.7 \cdot 10^{-4}$ at 90\% confidence
level.

This work is devoted to the search for the
$\phi \to \eta' \gamma $ decay in the decay mode 
$e^+e^- \to \eta' \gamma$, $\eta' \to \eta \pi^+ \pi^- $,
$\eta \to \gamma \gamma$ (\ref{etapg}).
The background processes are the following:
$e^+e^- \to \eta \gamma$, $\eta \to \pi^+ \pi^- \pi^0$ (\ref{etagc});
$e^+e^- \to \phi \to \pi^+ \pi^- \pi^0$ (\ref{phi3pi});
\(e^+e^- \to \phi \to \omega \pi^0 \to \pi^+ \pi^- \pi^0 \pi^0 \)
(\ref{omp0c}). The main problem in search  for (\ref{etapg}) is,
that the process (\ref{etagc}) has the same final state,
while its cross section is two orders of magnitude higher.
Processes (\ref{phi3pi}, \ref{omp0c}) give similar signatures
and could contribute to the process under study.
\\
{\bf Event selection.}
Full statistics of 7 scans was analyzed, corresponding to
the production of $8.3\cdot 10^6$ $\phi$ mesons.
The expected numbers of experimental events as well as numbers of
simulated events are listed in the Table~\ref{DMtab1}.
\begin{table}[htb]
\small
\caption{
The expected numbers of experimental events and numbers of
simulated events for investigated and background processes.}
\label{DMtab1}
\begin{tabular}{||c|c|c|c||}
\hline
Process &
\parbox[t]{30mm}{Expected number of experimental events --- $N_{exp.}$} &
\parbox[t]{20mm}{Number of simulated events --- $N_{mod.}$} &
\parbox[t]{20mm}{$ N_{exp.} / N_{mod.}$} \\
\hline
$\phi \to \eta' \gamma $ & 830, at $B(\phi \to \eta' \gamma) = 10^{-4}$ &
9996 & 0.083 \\
$\phi \to \eta \gamma $       & 104580           & 368176 & 0.284 \\
$\phi \to \pi^+ \pi^- \pi^0 $ & $1.29\cdot 10^6$ &  75000 & 17.3  \\
$e^+ e^- \to \omega \pi^0 $   & $38\cdot 10^3$   &  33784 & 1.12  \\
$\phi \to K^+K^- $            & $4.1\cdot 10^6$  &  24991 & 163   \\
$\phi \to K_SK_L $            & $2.8\cdot 10^6$  & 194551 & 14.5  \\
\hline
\end{tabular}
\end{table}
The following cuts  were applied for the preliminary event selection:
\\ $\bullet$ charged particles trigger;
\\ $\bullet$ $N_{cp}=2$; $N_{\gamma}=3$; 
   $R_1, R_2 < 0.3~cm $; $ |Z_1|, |Z_2| < 6~cm $; 
   $8 \leq N_{wires} \leq 12$;
\\ $\bullet$ $\chi^2_R < 50 $; $\chi^2_E < 50 $;
\\ $\bullet$ $\theta _{min} > 27^\circ$; $\alpha _{\pi \pi} < 145^\circ$; 
   $E_{tot}/2E_0 < 0.8 $;
   $E_{NP~tot}/2E_0 > 0.4 $.
\\The cut on $\alpha _{\pi \pi}$ allows to exclude the background process
$e^+e^- \to \phi \to K_SK_L$, $K_S \to \pi^+ \pi^-$,
$K_L \to neutral~particles$ and possible background from
Bhabha scattering and higher order QED processes.
21103 experimental events passed these cuts:

The  analysis of the kinematics of the reaction (\ref{etapg}) shows, 
that the photon from the $\eta'$ radiative decay
always has the smallest energy $E_{\gamma 3}\simeq 60~MeV$,
while the  energy of  photons from the decay
$\eta \to \gamma \gamma$ lies in the interval
$170~MeV < E_{\gamma 1}, E_{\gamma 2} < 440~MeV$.
In the process (\ref{etagc}) the radiative photon  has a maximum
energy $E_{\gamma 1} \simeq 362~MeV$, while the energy  of
photons from the decay $\pi^0\to \gamma \gamma$  
lies in the interval
$12~MeV < E_{\gamma 2}, E_{\gamma 3} < 364~MeV$.
It should be mentioned, that a probability for photons 
to reach the boarders of mentioned intervals is small due to the phase space
suppression. Hence, there is almost no combinatorial background for 
these processes, even taking into account the  finite resolution 
of the detector.

In Fig.~\ref{dm_erp3} the distribution of 11239 experimental events
over recoil mass of the most energetic photon is shown. It was obtained
with the additional cut $|m_{23}-m_{\pi^0}|<35~MeV$
($\pi^0$ meson mass resolution is equal to
$\sigma _{m_{\pi^0}}\simeq 12~MeV$).
The clear peak on $\eta$ mass
from the process (\ref{etagc}) is seen together with a
broader distribution from the processes (\ref{phi3pi}, \ref{omp0c}).
Fitting the data by the sum of Gaussian distribution and
7-th order polynomial one could estimate the number of events in the peak
$N_{\eta \gamma}=3623 \pm 60$. The order of polynomial is not
important, if the mass range is limited to $m_{rec.\gamma}<700~MeV$.
The detection efficiency is equal to
$\varepsilon _{\eta \gamma}
B(\eta \to \pi^+ \pi^- \pi^0)=3.63 \pm 0.04\%$.
It leads to the branching ratio
$B(\phi \to \eta \gamma) = 1.20 \pm 0.03 \%$, 
where the error is a pure statistical one.
The obtained result is only 5\% smaller than the table value 
\cite{PDG},  confirming validity of our analysis and absence 
of significant systematic errors.
With additional cut, excluding
$\pi^0$ meson $|m_{23}-m_{\pi^0}|>35~MeV$, the  peak at
$\eta$ mass vanishes.  So to exclude the events of the 
process (\ref{etagc}) the cut

$( \frac{ m_{rec.1}-m_{\eta} }{50} )^2 +
( \frac{ m_{23}   -m_{\pi^0}}{35} )^2 > 1 \)$ $
\\ was imposed, where  each denominator is equal to three times the
resolution ($3\sigma$) for corresponding meson mass.

\begin{figure}[htb]
\epsfxsize=0.49\textwidth
  \begin{minipage}[t]{0.47\textwidth}
  \centerline{\epsfbox{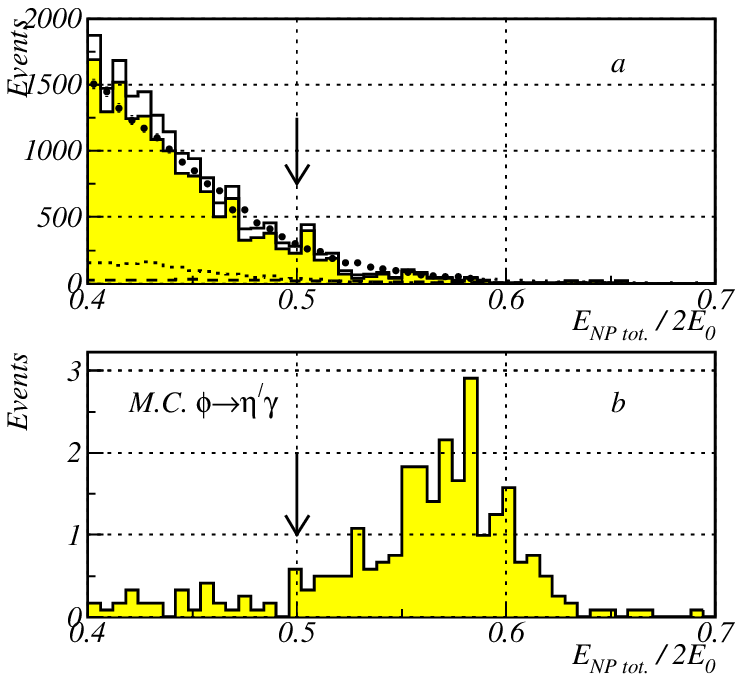}} 
\caption{
Distributions of normalized energy deposition of neutral particles,
in search for the decay $\phi \to \eta' \gamma$.
a) Circles with error bars -- experimental data;
histograms -- estimated contributions from background processes:
hatched histogram --
$e^+e^- \to \phi \to \pi^+ \pi^- \pi^0$
(\ref{phi3pi});
dotted line  --
$e^+e^- \to \phi \to \omega \pi^0 \to \pi^+ \pi^- \pi^0 \pi^0$
(\ref{omp0c});
dashed line --
$e^+e^- \to \eta \gamma$, $\eta \to \pi^+ \pi^- \pi^0$
(\ref{etagc}).
b) Expected contribution from the process under study
at $Br(\phi \to \eta' \gamma)=10^{-4}$.}
\label{dm_eneu}
\end{minipage}
\hfill
\begin{minipage}[t]{0.47\textwidth}
  \centerline{\epsfbox{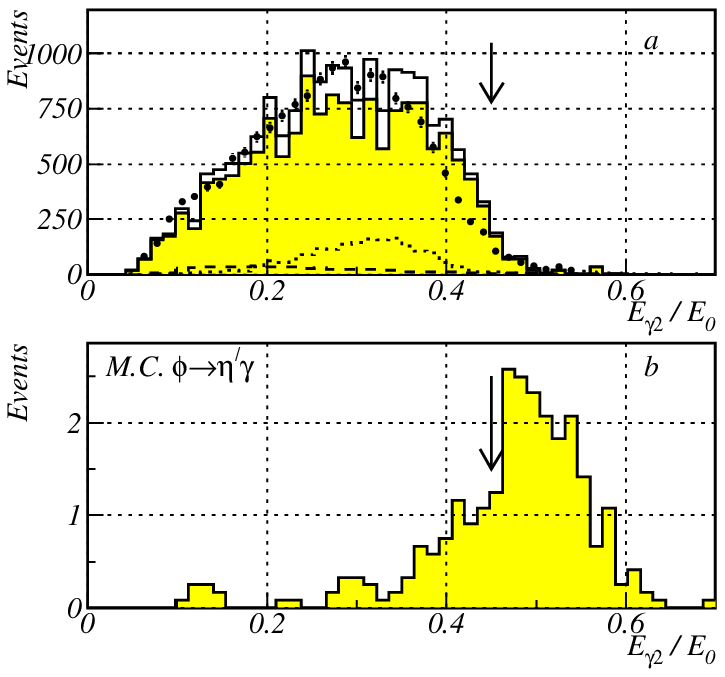}} 
\caption{
Distributions of normalized energy of the 2-nd photon
in search for the decay $\phi \to \eta' \gamma$.
a) Circles with error bars -- experimental data;
histograms -- estimated contributions from background processes:
hatched histogram --
$e^+e^- \to \phi \to \pi^+ \pi^- \pi^0$
(\ref{phi3pi});
dotted line --
$e^+e^- \to \phi \to \omega \pi^0 \to \pi^+ \pi^- \pi^0 \pi^0$
(\ref{omp0c});
dashed line --
$e^+e^- \to \eta \gamma$, $\eta \to \pi^+ \pi^- \pi^0$
(\ref{etagc}).
b) Expected contribution from the process under study
at $Br(\phi \to \eta' \gamma)=10^{-4}$.}
\label{dm_el4n}
\end{minipage}
\end{figure}

In order to reach  more effective suppression of background events,
the characteristic kinematic features of the investigated process were used:
the sum of three photons energy exceeds $607~MeV$;
the photon energy from the decay $\eta \to \gamma \gamma$ 
exceeds $170~MeV$.  For 16512 experimental events the distribution
over the energy deposition of neutral particles and over the 
reconstructed energy of the 2-nd photon is  shown in Fig.~\ref{dm_eneu} 
and Fig.~\ref{dm_el4n} as well as estimated contribution of background 
processes. These figures show that the processes (\ref{phi3pi}) and 
(\ref{omp0c}) are dominant.  To reduce their contribution the following cuts 
\\ $\bullet$ $E_{NP~tot}/2E_0 > 0.5 $, $E_{\gamma2}/E_0 > 0.45 $,\\
were applied, shown by arrows 
in Fig.~\ref{dm_eneu} and Fig.~\ref{dm_el4n}.
A cut ensuring the quality of photon showers in calorimeter was also 
used \\ $\bullet$ $\zeta<0$. \\367 events passed the cuts listed above.

\begin{figure}[htb]
\centerline{\mbox{\epsfig{figure=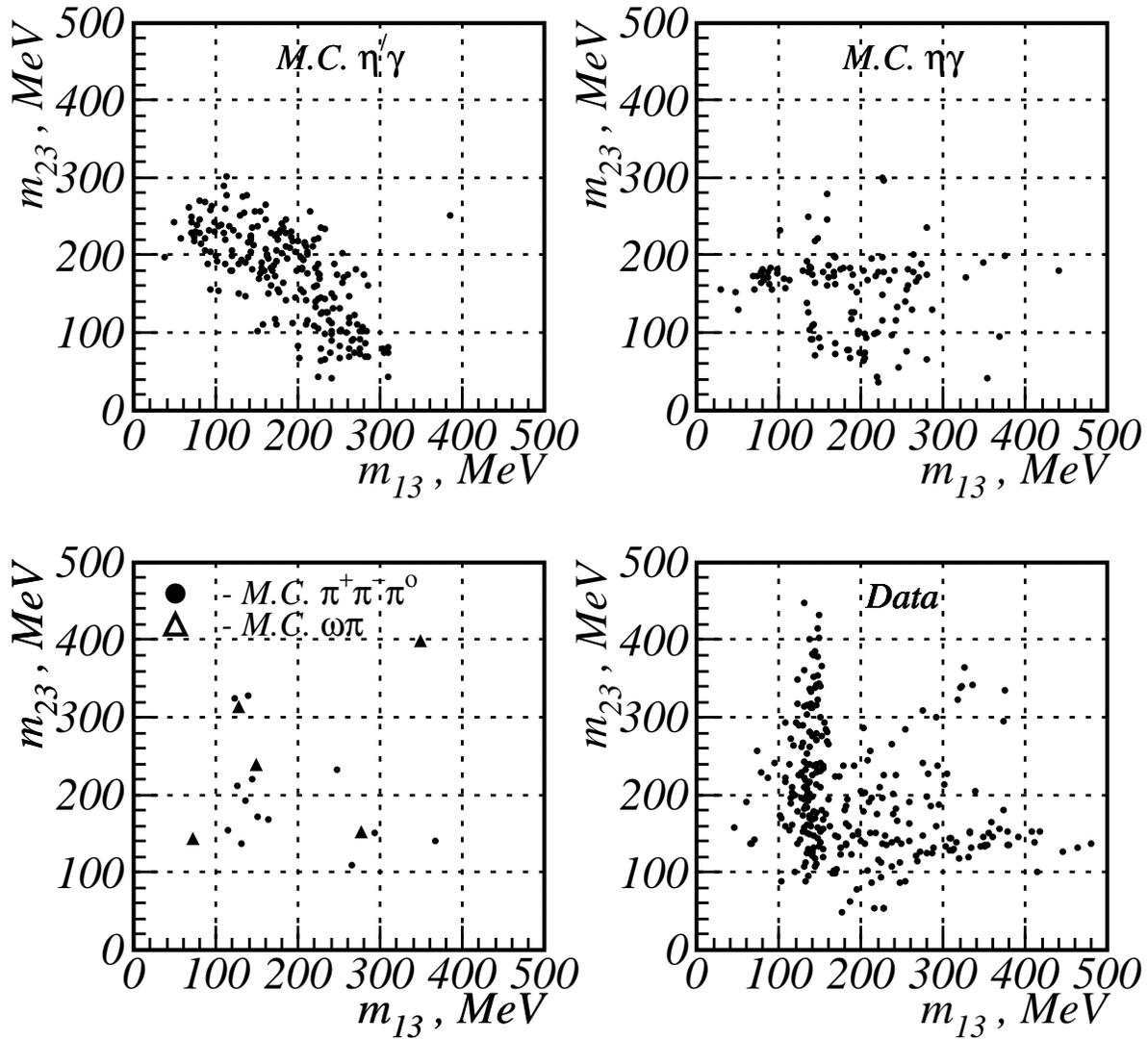,width=\textwidth}}} 
\caption{
Distributions of experimental and simulated events over
invariant masses of photon pairs in search for the
$\phi \to \eta' \gamma$ decay.
Number of simulated events is not normalized.}
\label{dm_ms45vs35}
\end{figure}

A scatter plot of invariant masses of photon pairs
$m_{13}$, $m_{23}$ and $m_{12}$
(for example see Fig.~\ref{dm_ms45vs35}) 
shows a concentration of events along
the lines corresponding to $\pi^0$ masses, which also confirms their 
background nature.
To exclude events of this kind the cuts
\\ $\bullet$ $|m_{13}-m_{\pi^0}|>35~MeV$, $|m_{23}-m_{\pi^0}|>35~MeV$
\\were added. Second of these cuts simply tightens
the condition, used for
event selection of the decay $\phi \to \eta \gamma$.

\begin{figure}[htb]
\centerline{\epsfig{figure=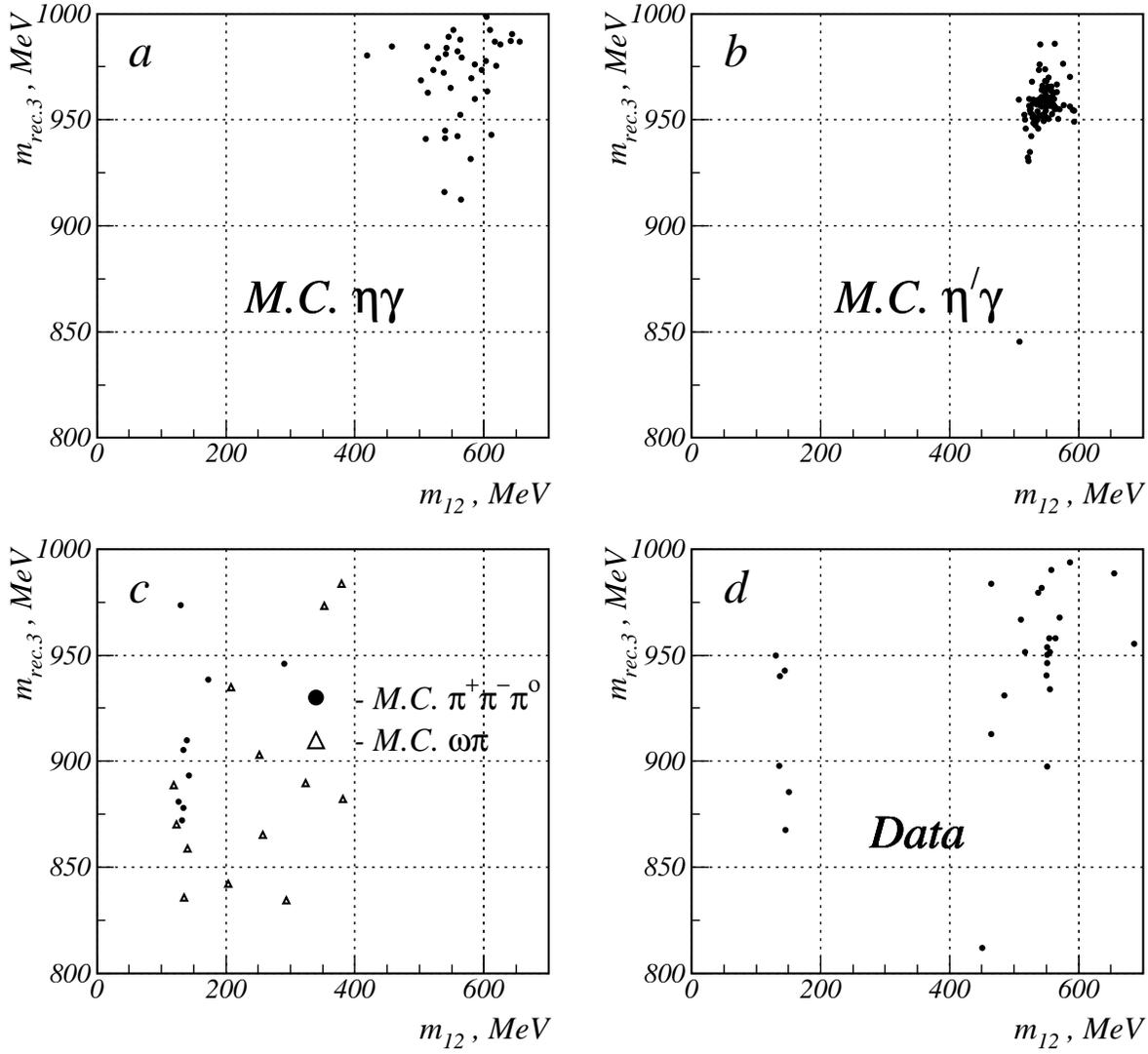,width=\textwidth}} 
\caption{
Distributions of experimental (d) and simulated (a -- c)
events over  recoil mass of the photon with minimal energy
versus invariant mass of the pair of photons with maximal energy
in search for the decay $\phi \to \eta' \gamma$.
Numbers of simulated events are not normalized.}
\label{dm_erp5vsmp34}
\end{figure}

Scatter plot of 28 remaining experimental events in recoil mass
of the photon with minimal energy --- $m_{rec.3}$ versus
invariant mass of pair of most energetic photons --- $m_{12}$
is shown in Fig.~\ref{dm_erp5vsmp34}d.
The same distribution of simulated events for searched and 
main background processes is shown in
Fig.~\ref{dm_erp5vsmp34}a,b,c.
Numbers of simulated events are not normalized.
In order to display contributions from the processes
(\ref{phi3pi}) and (\ref{omp0c}), events in  
Fig.~\ref{dm_erp5vsmp34}c were selected without the cut on
$E_{\gamma 2} / E_0$ parameter.

\begin{figure}[htb]
\epsfxsize=0.49\textwidth

\begin{minipage}[t]{0.47\textwidth}
  \centerline{\epsfbox{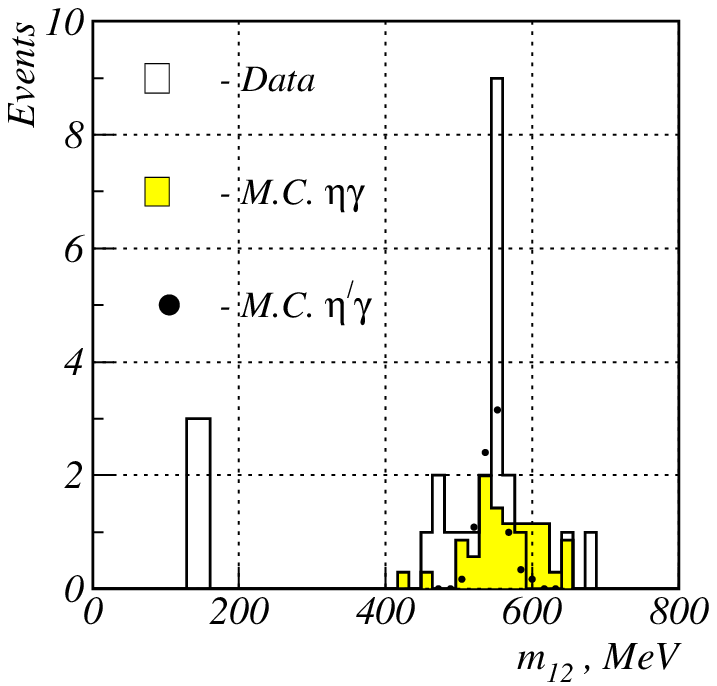}} 
\vspace*{-5mm}
\caption{
Distribution of experimental and simulated events over
invariant mass of two photons with maximal energy
in search for the decay $\phi \to \eta' \gamma$.}
\label{dm_mp34}
\end{minipage}
\hfill
\begin{minipage}[t]{0.47\textwidth}
  \centerline{\epsfbox{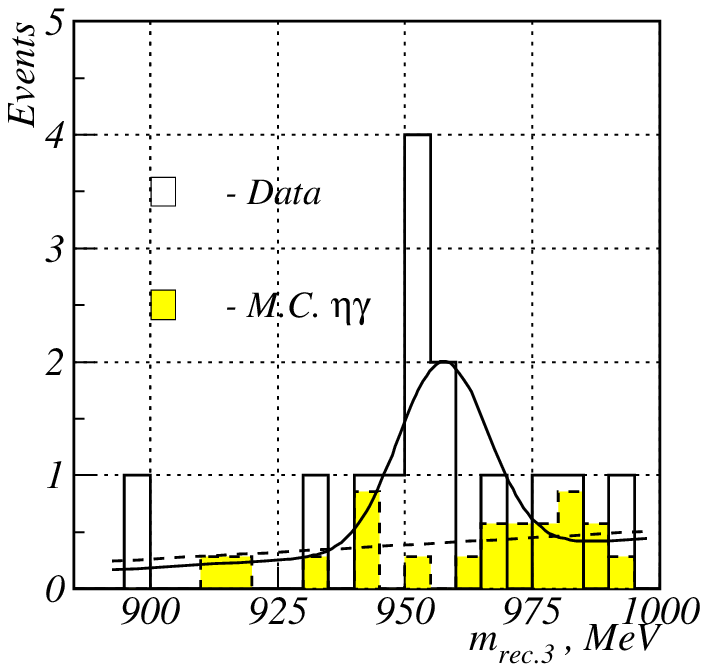}} 
\vspace*{-5mm}
\caption{
Distribution of experimental and simulated events over
recoil mass of the photon with minimal energy
in search for the $\phi \to \eta' \gamma$ decay.
Hatched histogram and dashed line show the estimated contribution
of the background process
$e^+e^- \to \eta \gamma$, $\eta \to \pi^+ \pi^- \pi^0$
(\ref{etagc}) and its linear approximation.
Histogram and smooth curve show the distribution of experimental
events and its optimal approximation.}
\label{dm_erp5}
\end{minipage}
\vfill
\vspace*{-30mm}
\begin{minipage}[t]{0.47\textwidth}
  \centerline{\epsfbox{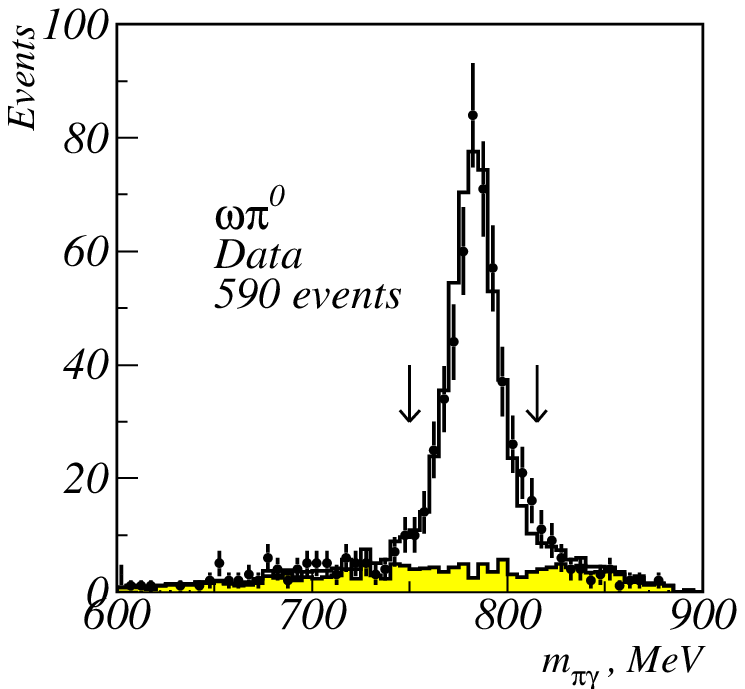}} 
\vspace*{-5mm}
\caption{
Distribution of $\pi^0\gamma$ mass for
$m_{\pi\pi}<700~MeV$ in search for $\phi\to\pi^0\pi^0\gamma$ decay.
Points -- data, histogram -- simulation, shaded histogram --
sum of simulated contributions from $\phi\to\eta\gamma$ and
$\phi\to\pi^0\pi^0\gamma$ decays,
arrows -- selection cuts for  $\omega\pi^0$ class.}
\label{fig4}
\end{minipage}
\end{figure}

Projection of two dimensional plot on the axis
$m_{34}$ is plotted in Fig.~\ref{dm_mp34}. Calculated contribution
from the process (\ref{etapg})
at $B(\phi \to \eta' \gamma) = 10^{-4}$, as well as the main background
process (\ref{etagc}) are plotted in Fig.~\ref{dm_mp34}.
Besides the peak at $\eta$ mass, background events of 
the processes (\ref{phi3pi}) и (\ref{omp0c}) form the peak at 
$\pi^0$ mass.

In Fig.~\ref{dm_erp5} the second projection of two-dimensional
scatter plot on the axis $m_{rec.5}$ is shown with additional
cut $|m_{12}-m_{\eta}|<35~MeV$. The peak at $\eta'$ mass
is observed, confirming the existence of the $\phi\to\eta'\gamma$ decay.
Histogram and smooth curve show distribution of 14
experimental events and its optimal fit by
the sum of linear function and Gaussian with fixed
parameters:
$m_{\eta'} = 957.5~MeV$ and $\sigma_{m_{\eta'}}=6.2~MeV$,
obtained from simulation of the process (\ref{etapg}).
Hatched histogram and dashed line show the estimated contribution
from the background process (\ref{etagc}) and its approximation
by the linear dependence. One can see, that the difference
in the background estimation from experimental data and 
simulation does not exceed 0.6 events in the interval
$m_\eta \pm 20~MeV$ (2.6 and 3.2 events respectively).
The number of useful events in the peak equals to
$N_{\eta ` \gamma} = 5.2^{+2.6}_{-2.2}$.
The detection efficiency of the final state is equal to
$\varepsilon _{\eta' \gamma}=5.5\pm 0.6\%$. Taking into account
table values of 
$B(\eta' \to \eta \pi^+ \pi^-)$ and $B(\eta \to \gamma \gamma)$
\cite{PDG}, we obtain
\begin{equation}
 B(\phi \to \eta' \gamma) = (6.7^{+3.4}_{-2.9}) \cdot 10^{-5},
\label{phiepg}
\end{equation}
which is nearly twice lower than CMD-2 result 
\cite{CMD-2}, but does not contradict it
because of large error in both measurements.

\subsection{ Study of the $\phi \to \pi^0 \pi^0 \gamma $ decay}

Search for $\phi\to\pi^0\pi^0\gamma$ decay was first carried out
in the ND experiment at the VEPP-2M $e^+e^-$
 collider in 1987 and the upper limit
$B(\phi\to\pi^0\pi^0\gamma) < 10^{-3}$ \cite{IVN1,Ph.Rep} was placed.
As it was shown later by Achasov \cite{IVN2}, study of this decay can provide a
unique information on the structure of the light scalar $f_0(980)$ meson.
Subsequent studies \cite{IVN3,IVN4,IVN5,IVN6,IVN7,IVN8,IVN9}
proved this idea. In these works different
models of the  $f_0(980)$-meson structure were considered. The most
popular were two-quark model \cite{IVN12}, four-quark model \cite{IVN13}, and a
molecular model \cite{IVN6}.

  In 1997 the first indications of the process
\({e^+e^-\to\phi\to\pi^0\pi^0\gamma}\) (\ref{p0p0g})
were reported by SND \cite{Ivanchenko/H-97}. 
In this work the results of analysis of 
full statistic of this experiment are presented.
Main resonant background to the decay  (\ref{p0p0g})
comes from the neutral decay
\({e^+e^-\to\phi\to\eta\gamma\to3\pi^0\gamma}\)
(\ref{etagn})
due to the merging of
photons and/or loss of photons through the openings in the
calorimeter.
The main source of non-resonant background is a process
\({e^+e^-\to\omega\pi^0\to\pi^0\pi^0\gamma}\)
(\ref{omp0n}).
The background from the $\phi\to\rho\pi^0\to\pi^0\pi^0\gamma$
decay turned out to be small \cite{IVN5,IVN11}, nevertheless, its amplitude
was taken into account in the simulation of the process (\ref{omp0n}).
The background from the QED 5-$\gamma$ annihilation process was
found to be negligible.

In order to suppress the contribution of background
the events were selected satisfying the following
criteria:
\\$\bullet$ $N_{\gamma}=5$; $N_{cp}=0$; $\theta_{min} > 27^\circ$;
\\$\bullet$ $0.8 < E_{tot}/2E_0 < 1.1$; $P/2E_0 < 0.15$;
$\zeta < 0$;
$\chi^2_{\pi^0\pi^0\gamma} < 20$.
\\Parameter $\zeta$ facilitates efficient
separation of the events with well isolated photons from the
events with merged photons or those produced by  $K_L$ mesons.
The  $\phi\to K_S K_L \to\pi^0\pi^0K_L$ decay
can contribute due to nuclear interactions
of $K_L$ mesons in the material of the calorimeter but after
described cuts it is not seen at present level of statistics.
The $\chi^2_{\pi^0\pi^0\gamma}$,
kinematic fit parameter, describes the degree of
energy-momentum conservation in the event
with additional
requirement of presence of two $\pi^0$ mesons. During this fit all
possible combinations of photons 
in the event were checked in a search for
invariant masses $m_1$ and  $m_2$, satisfying the condition \\
$\sqrt{(m_1-m_{\pi})^2+(m_2-m_{\pi})^2}<25~MeV/c^2$.
In the energy region of this experiment the invariant mass of the
pion pair in the process (\ref{omp0n}) is less than $700~MeV$. In the events,
selected under this condition the clear $\omega$(782) peak is seen
in $m_{\pi\gamma}$ distribution (Fig.~\ref{fig4}), proving
the domination of the process (\ref{omp0n}) in this kinematic region.
The $m_{\pi\gamma}$ parameter was defined as an invariant mass of the
recoil photon and one of $\pi^0$ mesons, closest to the mass of $\omega$ meson.
The 499 events found in the invariant mass region
$750~MeV < m_{\pi\gamma} < 815~MeV$ were assigned to an $\omega\pi^0$
class. 189 events with $m_{\pi\gamma}$ outside this interval and 
$m_{\pi\pi}>700~MeV$ were assigned to the $\pi^0\pi^0\gamma$ class.
Subtracting calculated contribution of the process (\ref{p0p0g}) and 
using estimated probabilities of  events misidentification for
the processes (\ref{p0p0g}) and (\ref{omp0n}) the number
of the events of the process  (\ref{omp0n}) in the $\omega\pi^0$
class were estimated to be equal 449.
The corresponding number of events of the decay (\ref{p0p0g})
in the   $\pi^0\pi^0\gamma$ class is 164.
The background from the process (\ref{omp0n}) was estimated using events
of the $\omega\pi^0$ class, no additional knowledge of
the actual production cross section of this process was necessary.

\begin{figure}[htb]
\epsfxsize=\textwidth
  \centerline{\epsfbox{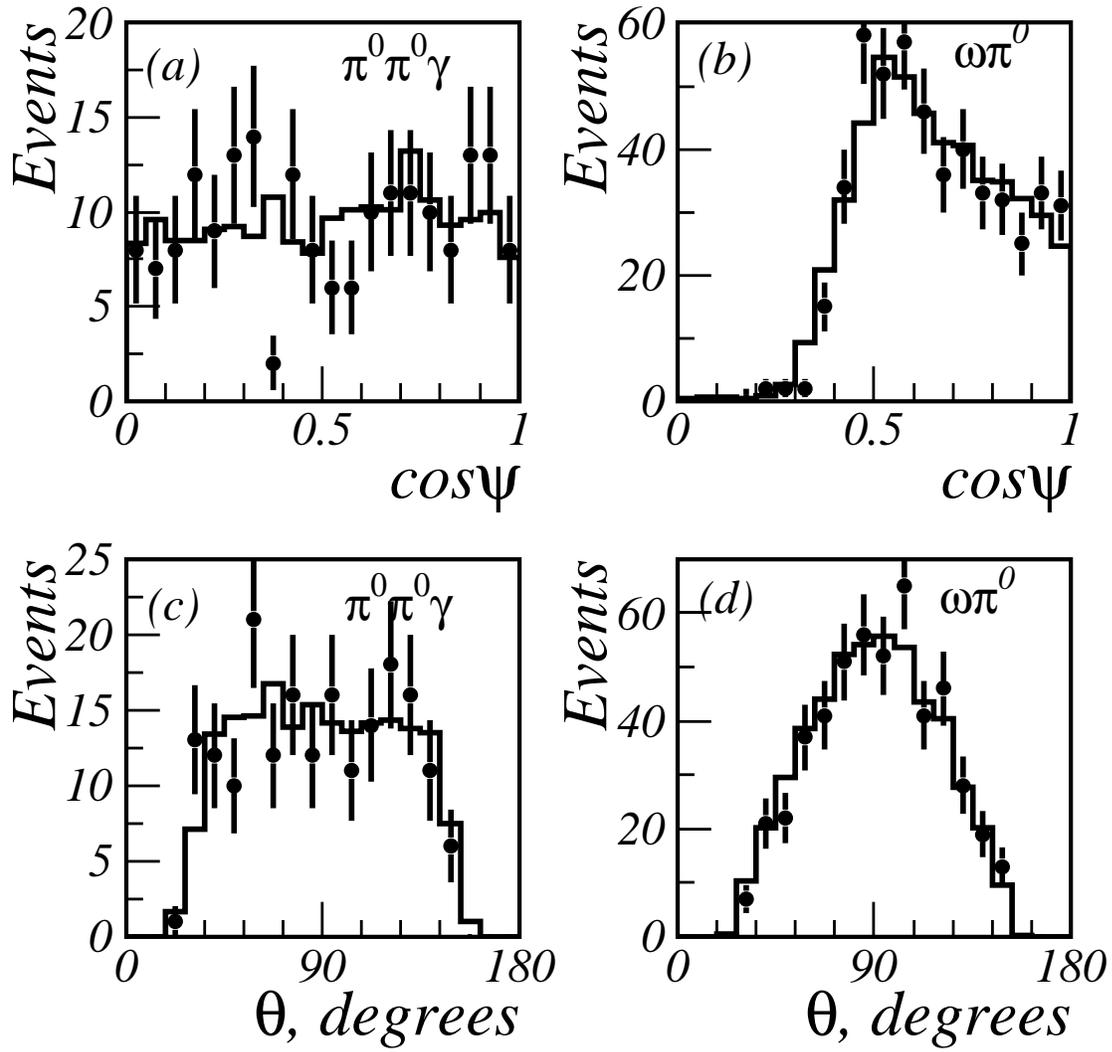}} 
\caption{a, b -- distributions of cosine of $\psi$ --
the angle between directions of $\pi^0$ and recoil $\gamma$
in the rest frame of $\pi^0\pi^0$ system;
c, d --  distributions of  $\theta$ -- angle of the
recoil $\gamma$ with respect to the beam.
Points -- data, histogram -- simulation.}
\label{fig5}
\end{figure}

Then, for the events of the $\pi^0\pi^0\gamma$ and $\omega\pi^0$
classes the comparison of experimental and simulated
distributions in  $\psi$ and $\theta$ angles was made.
Here $\psi$ is an angle between the recoil photon and the direction of
pion in the two-pion center of mass reference frame, $\theta$ is an
angle of the recoil photon with respect to the beam.
As it is known, the distribution in  $\theta$ for  $\pi^0\pi^0\gamma$
class is proportional to $1+cos^2\theta$ and is uniform in $cos\psi$.
The comparison (Fig.~\ref{fig5}a,c) shows, 
that in this class of events 
pions are actually produced in scalar state. 
On the contrary, the experimental
events of the $\omega\pi^0$ class (Fig.~\ref{fig5}b,d)
also well match the hypothesis of the intermediate $\omega\pi^0$ state
with quite different $\psi$ distribution.

\begin{figure}[htb]
\epsfxsize=0.49\textwidth

\begin{minipage}[t]{0.47\textwidth}
  \centerline{\epsfbox{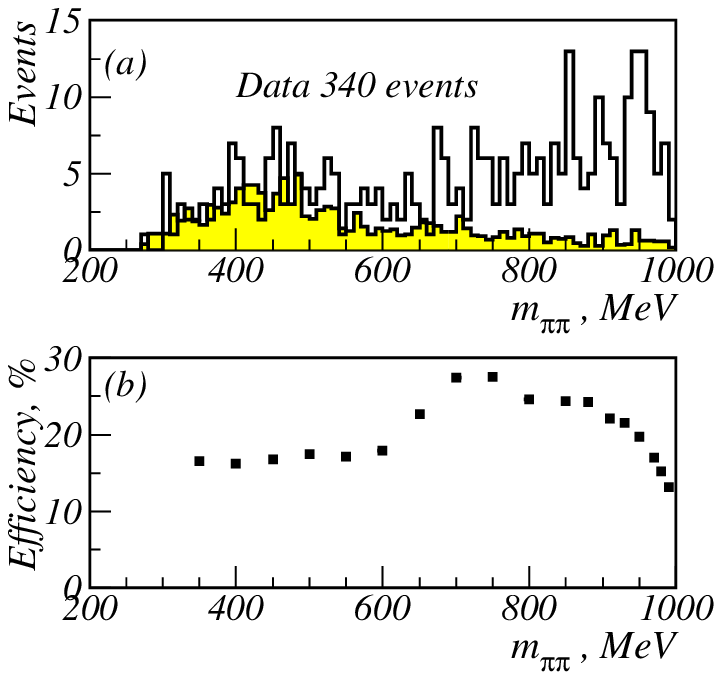}} 
\vspace*{-5mm}
\caption{a -- invariant mass distribution
of $\pi^0\pi^0$ pairs for selected $\pi^0\pi^0\gamma$ events. 
Histogram -- data, shaded histogram -- estimated background
contribution from $e^+e^-\to\omega\pi^0$ and
$\phi\to\eta\gamma$;
b -- detection efficiency for  $\pi^0\pi^0\gamma$ events.}
\label{fig6}
\end{minipage}
\hfill
\begin{minipage}[t]{0.47\textwidth}
  \centerline{\epsfbox{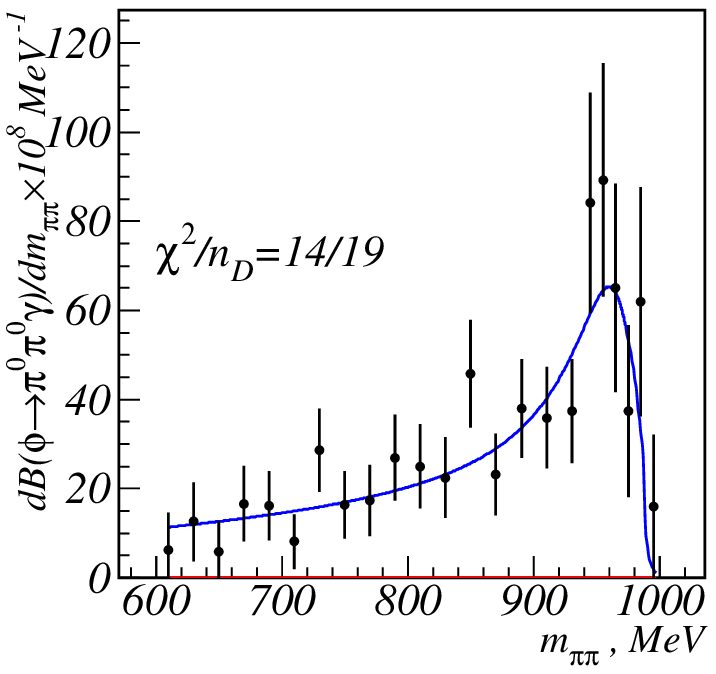}} 
\vspace*{-5mm}
\caption{Measured $\pi^0\pi^0$ invariant mass spectrum
with efficiency corrections.
Points -- data, solid line -- fit.}
\label{fig8}
\end{minipage}
\vfill
\vspace*{-20mm}
\begin{minipage}[t]{0.47\textwidth}
 \centerline{\epsfig{figure=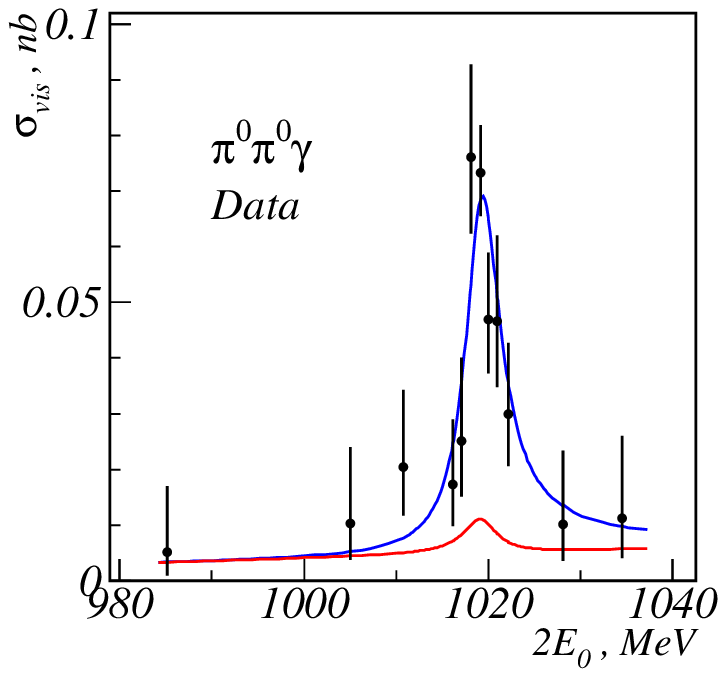,width=\textwidth}} 
\vspace*{-5mm}
\caption{ Energy dependence of the visible
$e^+e^-\to\pi^0\pi^0\gamma$ cross section.
Points -- data, solid line -- fit, 
dotted line --  estimated background from
$e^+e^-\to\omega\pi^0$ and
$\phi\to\eta\gamma$.}
\label{fig7}
\end{minipage}
\hfill
\begin{minipage}[t]{0.47\textwidth}
 \centerline{\epsfig{figure=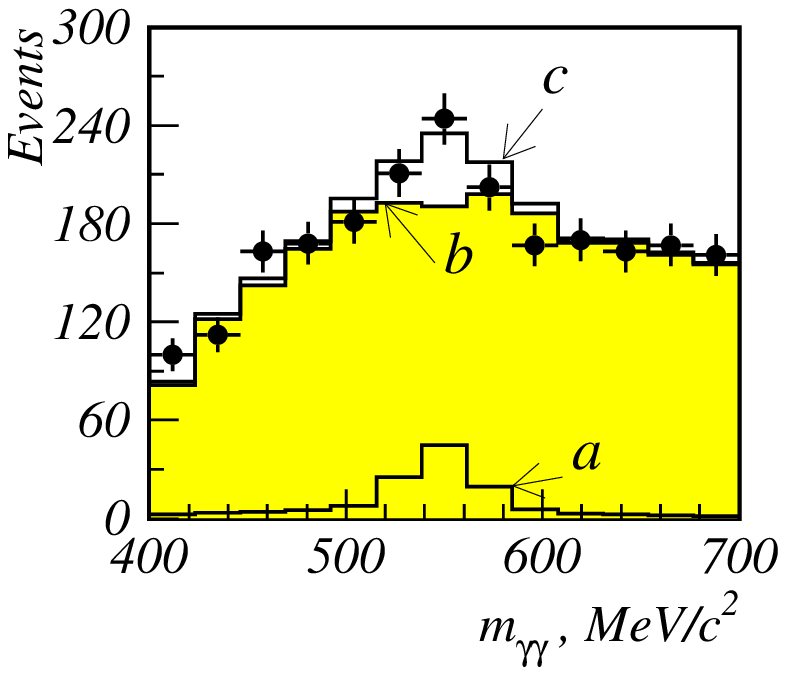,width=1.3\textwidth}} 
\vspace*{-5mm}
\caption{ Invariant masses of pairs of most energetic photons
in search for $\phi\to\eta\pi^0\gamma$ decay:
circles with error bars -- experimental data,
(a) -- simulated signal from $\phi\to\eta\pi^0\gamma$ decay corresponding to a
branching ratio of $0.7\cdot10^{-4}$, (b) -- estimated background
from the $e^+e^-\to\omega\pi^0$ and $\phi\to\eta\gamma,
f_0 \gamma$ events, (c) -- sum of (a) and (b).}
\label{INCLUSIVE}
\end{minipage}
\end{figure}

The $\pi^0\pi^0$ invariant mass distribution of the events with 
$m_{\pi\gamma}$ outside the $750~MeV < m_{\pi\gamma} < 815~MeV$ interval
(Fig.~\ref{fig6}a) 
shows significant excess over background
at large $m_{\pi\pi}$. At $m_{\pi\pi}<600~MeV$
the sum of
background contributions dominates.
Detection efficiency (Fig.~\ref{fig6}b)
for the process (\ref{p0p0g}) was determined using MC simulation 
of the process   $\phi\to S \gamma \to \pi^0 \pi^0\gamma$ with a scalar states
$S$, with different masses ranging from 300 to $1000~MeV$ and zero width. 
In addition, this simulation provided information on $\pi^0\pi^0$
invariant mass resolution and event misidentification probability
as a function of $m_{\pi\pi}$.
After background subtraction  and correction for detection
efficiency the mass spectrum was obtained (Fig.~\ref{fig7}). 
For masses in the  $600\div 850~MeV$ interval the invariant mass resolution
is equal to $12~MeV$, so the $20~MeV$ bin size was chosen. At higher masses
the resolution improves, reaching $7.5~MeV$ at $950~MeV$, thus the bin size
of $10~MeV$ was used for higher masses.
 
For the spectrum normalization the 
events of the process (\ref{etagn}) with 7 photons
in the final state and reconstructed $3 \pi^0$ mesons were analyzed. All other
selection criteria were the same as in the present analysis. The number of
observed events of the process (\ref{etagn}) 
together with the PDG Table value for
$B(\phi\to\eta\gamma)=(1.26\pm0.06)\%$ \cite{PDG} 
provided independent estimation
of the total number of $\phi$ mesons produced.
Such an approach minimizes systematic errors corresponding
to inaccurate simulation of tails of distributions over parameters, used for
event selection.
The measured branching ratio
of the decay (\ref{p0p0g}) for the restricted
mass range $m_{\pi\pi}>700~MeV$ is equal
\begin{equation}
{B(\phi\to\pi^0\pi^0\gamma)=(1.00\pm 0.07\pm0.12)\cdot10^{-4},}
\label{EQ5}
\end{equation}
and for $m_{\pi\pi}>900~MeV$
\begin{equation}
{B(\phi\to\pi^0\pi^0\gamma)=(0.50\pm 0.06\pm0.06)\cdot10^{-4}.}
\label{EQ6}
\end{equation}
Here the first error is statistical and the second is systematic,
which was estimated to be close to $12\%$. The systematic error
is mainly determined by the
following contributions:

\begin{itemize}
\item the background subtraction error, which decreases almost linearly
with the increase of invariant mass  $m_{\pi\pi}$ and is  $5\%$ on average;
\item error in the detection efficiency estimation, which increases with
 $m_{\pi\pi}$ and is equal to $8\%$ on average;
 \item systematic error in $B(\phi\to\eta\gamma)$ is equal to $5\%$.
 \end{itemize}

 The $m_{\pi\pi}$ invariant mass spectrum was fitted with a smooth curve
 (Fig.~\ref{fig7}) according to Refs.\cite{IVN2,IVN11} 
and further used for simulation of the
 decay (\ref{p0p0g}). 
As a result the detection efficiency of the process (\ref{p0p0g}) was
 estimated
 $\sim 15\%$ for invariant masses within the $600\div 1000~MeV$.
 This efficiency was used in the fitting of the $\phi$-resonance excitation
 curve. The cross section was described as a sum of the processes
 (\ref{p0p0g}), (\ref{etagn}), 
and (\ref{omp0n}) with radiative 
corrections taken into account (Fig.~\ref{fig8}).
 The background due to the process (\ref{omp0n}) 
was estimated by fitting the detection
 cross section of the events of the $\omega\pi^0$ type. The background from the
 process (\ref{etagn}) was obtained from simulation. 
The beam energy spread and the
 accuracy of the beam energy determination were also taken into account
 during fitting \cite{Prep.97,Ivanchenko/H-97}.
 As a result, the following value was obtained:
 \begin{equation}
 {
 B(\phi\to\pi^0\pi^0\gamma)=(1.14\pm 0.10\pm0.12)\cdot10^{-4},
 }
 \label{EQ7}
 \end{equation}
 which, in contrast with (\ref{EQ5}) and (\ref{EQ6}), 
is valid for the whole mass
 spectrum. In the systematic error estimation the following considerations
 were taken into account. 
In comparison with the results (\ref{EQ5}) and (\ref{EQ6}) the
 accuracy of normalization  $(3\%)$ and efficiency estimation
 $(5\%)$ are higher here, the background subtraction error $(5\%)$  is
 the same, but an additional systematic error of $(6\%)$ exists, due to
 uncertainty in extrapolation of the invariant mass spectrum into the region
 $m_{\pi\pi}<600~MeV$.
 Smaller systematic uncertainty has a ratio of branching ratios:
 \begin{equation}
 {\frac{B(\phi\to\pi^0\pi^0\gamma)}{B(\phi\to\eta\gamma)}
 =(0.90\pm 0.08\pm0.07)\cdot10^{-2},
 }
 \label{EQ8}
 \end{equation}
 Assuming that the process (\ref{p0p0g}) is fully determined by  $f_0\gamma$
 mechanism, using the relation $B(f_0\to\pi^+\pi^-)=2B(f_0\to\pi^0\pi^0)$,
 and neglecting the decay $\phi\to KK\gamma$ 
\cite{IVN2}, we can estimate from (\ref{EQ7})
 $$B(\phi \to f_0(980) \gamma) = (3.42 \pm 0.30\pm0.36)\cdot 10^{-4}.$$

In the $\pi^0\pi^0$ invariant mass spectrum in Fig.~\ref{fig8} 
the $f_0$-meson peak
is clearly seen. The visible peak position is close to $960~MeV$, with the
width about $100~MeV$. 
The analysis of our preliminary data together with other already
known properties of the $f_0$ meson, performed by N.N.Achasov 
\cite{IVN22}, permits
to assume, that the 4-quark component constitutes a large part of the
$f_0$ meson. Actually, the observed decay probability about $10^{-4}$
together with the invariant mass spectrum (Fig.\ref{fig8}) can be only due to
$s$ quarks  constituting noticeable part of the $f_0$ meson. 
The results of the
fitting of the mass spectrum, performed using formulas from
Refs.\cite{IVN2,IVN11}, are following:
$$m_f = (971\pm6\pm5)~MeV,~\Gamma_f(m_f)=(188^{+48}_{-33})~MeV,$$
$$\frac{g^2_{fK^+K^-}}{4\pi}=2.10^{+0.88}_{-0.56} ~GeV^2,~
\frac{g^2_{f\pi^+\pi^-}}{4\pi}=0.51^{+0.13}_{-0.09} ~GeV^2,$$
$$g^2_{fK^+K^-}/g^2_{f\pi^+\pi^-}=4.1\pm0.9.$$
The systematic error in mass is determined by the detector resolution.
Although the values
of the constants are strongly model dependent, their ratio is almost
model independent.
The value of the coupling constant $g^2_{fKK}/4\pi$ obtained from the fit
agrees with the predictions of 4-quark model  $(2.3~GeV^2~\cite{IVN2})$, but
contradicts the value predicted by 2-quark model $(0.3~GeV^2~\cite{IVN2})$ and
by almost 3 standard deviations higher than molecular model prediction:
$(0.6~GeV^2~\cite{IVN11})$.

Of course the more complex mechanism of the decay under study cannot
be excluded, e.g. contribution of the heavy and broad 
$\sigma$ state \cite{IVN11}.
But such a state can probably produce a smooth invariant mass spectrum,
not masking resonance signal from the $f_0$ meson in the mass region 
$m_{\pi\pi}>900~MeV.$ 
To take into account all
mechanisms of the transition one have to perform a simultaneous analysis
of all $f_0$ data within the framework of one model.

In conclusion we would like to emphasize that the
$\phi \to\pi^0\pi^0\gamma$ radiative decay
was observed for the first time and its branching ratio was measured.
It was shown, that the  $f_0(980)\gamma$ transition mechanism
dominates in this decay. 
Invariant mass spectrum of $\pi^0\pi^0$ system and the rate
of 
the decay give grounds to the assumption of
4-quark structure of the $f_0$ meson.

\subsection{Search for the $\phi \to \eta \pi^0 \gamma $ decay}

The $\phi \to \eta \pi^0 \gamma$ decay can be described
as a radiative  electric dipole transition from a light
vector meson state into a scalar state like $a_0$(980) with emission of a
photon.
Although electric dipole transitions are common in radiative decays of
heavy quarkonia, only few were observed in the light quark
mesons and none of $\phi(1020)$-meson.
Since even lightest known scalar resonances are quite heavy, the
energy yield in the decay is low ($\leq 100~MeV$) leading to
a small phase space for a photon. A number of estimations exist on the
$\phi \to \eta \pi^0 \gamma$ branching ratio
in different models \cite{IVN2}.
Since the values strongly depend on the
quark structure of the scalar intermediate state, the decays of this type
could be a unique probe of the structure of light scalar mesons.
It was suggested, that
the decay proceeds mainly through the $a_0(980)$ intermediate state and the
decay probability is determined by its structure. At present 
the structure of $a_0(980)$ meson it is not well established and several
theoretical models exist, including simple two-quark model,
$K^0\overline{K}^0$ molecular model, different 4- quark schemes. The values of
$\phi \to \eta \pi^0 \gamma$ branching ratio calculated in these models
vary in the range of $10^{-6} \div 10^{-4}$, making the
$\phi \to \eta \pi^0 \gamma$ decay a good probe of the inner structure of
the $a_0(980)$ state.

Because of great significance of the $ \phi \to \eta \pi^0 \gamma $ decay, 
two independent analyses were made.  The results obtained are
described below.

\subsubsection{Analysis 1}

The main background sources for the process under study
(\ref{etap0g}) 
are the following $\phi$-meson decays:
\(e^+e^- \to \phi \to \pi^0 \pi^0 \gamma \to 5\gamma\)
(\ref{p0p0g}),
\(e^+e^- \to \phi \to \eta \gamma \to 3\pi^0 \gamma \to 7\gamma\)
(\ref{etagn}),
\(e^+e^- \to \phi \to K_S K_L \to neutrals\)
(\ref{kskln})
and non-resonant process 
\(e^+e^- \to \omega \pi^0 \to \pi^0 \pi^0 \gamma\)
(\ref{omp0n}).

The expected number of events of the process (\ref{etap0g}) at a branching
ratio of  $Br(\phi \to \eta \pi^0 \gamma)=10^{-4}$ is about 300, while
the number of background events (\ref{etagn}) is $3\cdot10^4$. 
Although this process does not
produce 5-$\gamma$ final states, the topology of the
process (\ref{etap0g}) could be faked 
due to merging of close photons and loss of
soft photons through openings in the calorimeter. 
Total statistics of the process (\ref{kskln}) is
$8\cdot10^5$ $K_SK_L$ events with $K_S\to\pi^0\pi^0$ decays.
The $K_L$-s produced in $\phi$-meson decays are slow and their decay length is
about 3 m, while the nuclear interaction length in NaI(Tl) is about 30 cm.
Characteristic feature of fake 5-$\gamma$ events produced by $K_SK_L$ decays
due to either nuclear interaction of $K_L$-s or their decays in flight
is an energy-momentum imbalance and poor quality of at least one
photon in the event.

Primary event selection was based on simple criteria
(\ref{5gamma}) for $5\gamma$- final state.
Such criteria greatly reduce background from the
processes (\ref{etagn}) and (\ref{kskln}), 
not affecting the decays 
(\ref{p0p0g}, \ref{etap0g}), and (\ref{omp0n}).

Next step in the event selection was based on photons quality
parameter $\zeta$ and kinematic fit parameter $\chi_E^2$.
The requirements were imposed that  $\zeta<0$, for all photons.
Then kinematic fit was performed under the assumption, that selected
events are $e^+e^-\to 5\gamma$ ones and corresponding parameter $\chi_E^2$,
describing the likelihood of this assumption was calculated. Events with
$\chi_E^2>10$ were also rejected. Study of simulated events of
true $5\gamma$ processes (\ref{p0p0g}, \ref{etap0g})
and (\ref{omp0n}) shows, that the  $\chi_E^2$
and  $\zeta$ cuts reject less than $15\%$ of true $5\gamma$ events
while suppressing the process (\ref{etagn}) by a factor of 3 and making the
expected background due to the process (\ref{kskln}) 
very small, of the order of 10 events.
It should be noted, that in contrast with other background processes the
simulation of the process (\ref{kskln}) 
is much less accurate due to nuclear interaction
of $K_L$, and data analysis may not exclusively rely on this estimation.

Characteristic feature of the
process (\ref{etap0g}) is that its events must contain 
two photon pairs with invariant
masses of $\eta$ and $\pi^0$ mesons. Simulation shows that if the energies
of photons in an event are enumerated
in descending order in energy, the photons from $\eta\to\gamma\gamma$ decay
have the numbers of either 1 and 2 or 1 and 3.
Corresponding experimental and simulated
distributions in $m_{12}$ and $m_{13}$ with an additional
requirement, that the rest of photons contain a pair with
$|m_{ij}-m_{\pi^0}|<20 MeV/c^2$ and more stringent photon quality requirements
$\zeta<-5$.
are shown in Fig.~\ref{INCLUSIVE}. 
Background estimations are based on PDG table
value for $\phi\to\eta\gamma$ decay branching ratio \cite{PDG},
our study of $e^+e^-\to\omega\pi^0$ reaction  
and $\phi\to\pi^0\pi^0\gamma$ decay \cite{Prep.97}.
Distribution of experimental events
in $m_{12} \oplus m_{13}$ shows an enhancement
centered at $\eta$-meson mass, while sum of background
processes is nearly flat in this region. The sum of all
simulated background processes, where each one was normalized to the
number of events expected for a given integrated luminosity and total number
$\phi$-meson decays, describes the spectrum outside the
enhancement quite well. If this enhancement is due to the decay 
(\ref{etap0g}), its branching ratio should be
of the order of $7\cdot10^{-5}$, but it is not possible
to extract more accurate result and error estimates from these inclusive
spectra, because of poor signal to background ratio and decay
dynamics remaining obscure.

Detailed study of the process (\ref{etap0g}) requires considerable
suppression of
background. It was done using kinematic fit with intermediate $\eta$
and $\pi^0$ mesons
taken into account. For each event $\eta\pi^0\gamma$ and $\pi^0\pi^0\gamma$
hypotheses were tried and corresponding
$\chi_{\eta\pi^0\gamma}^2$,
and $\chi_{\pi^0\pi^0\gamma}^2$ calculated.
To suppress the processes (\ref{p0p0g}) and 
(\ref{omp0n}) the following requirements were imposed:
$\chi_{\eta\pi^0\gamma}^2<20$,
$\chi_{\pi^0\pi^0\gamma}^2>20$, 
and for additional suppression the process 
(\ref{etagn}): $\zeta<-4$. With these
requirements, the contribution from the processes (\ref{p0p0g}) and 
(\ref{omp0n}), which are 
themselves relatively rare becomes negligible.

\begin{figure}[htb]
\epsfxsize=0.49\textwidth

\begin{minipage}[t]{0.47\textwidth}
   \centerline{\epsfig{figure=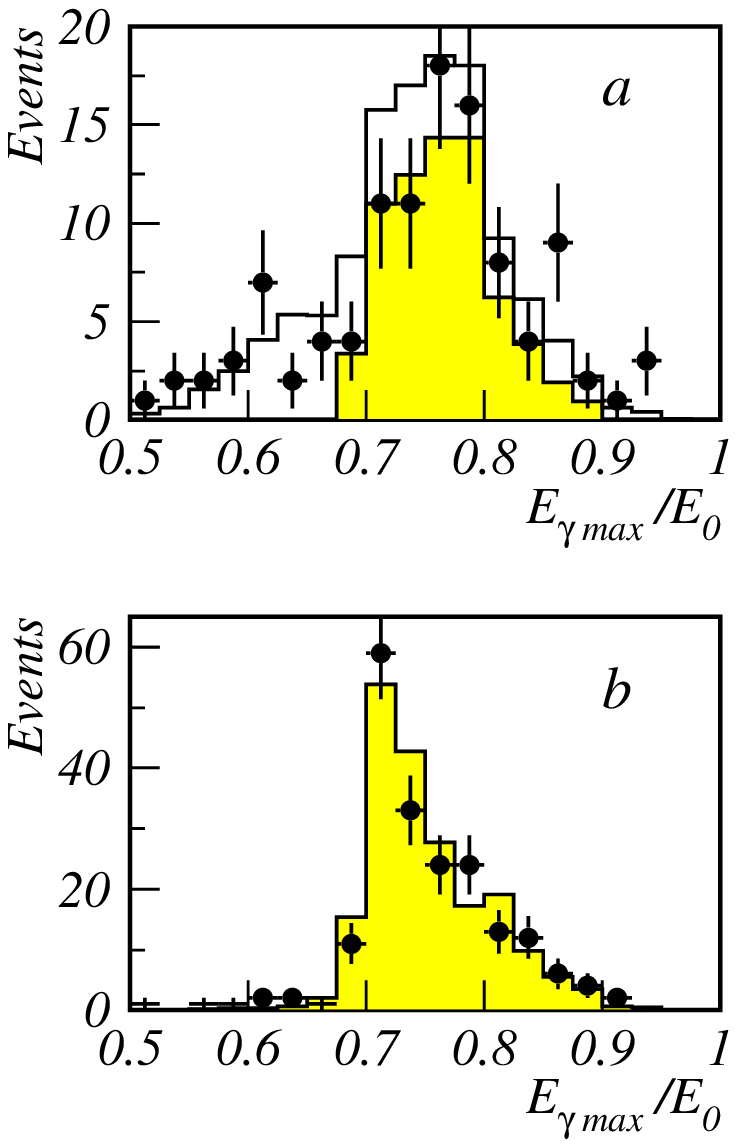,width=1.4\textwidth}} 
\vspace*{-5mm}
\caption{ Spectrum of the most energetic photon in an event
in search for $\phi\to\eta\pi^0\gamma$ decay:
a) photons quality $\zeta<-4$, b) $0<\zeta<10$. 
Circles with error bars depict experimental data,
shaded histogram --
 simulation of the process (\ref{etagn}) and clear histogram -- simulated sum
of (\ref{etagn}) and (\ref{etap0g}) with a $BR=10^{-4}$.}
\label{ER1N}
\end{minipage}
\hfill
\begin{minipage}[t]{0.47\textwidth}
   \centerline{\epsfig{figure=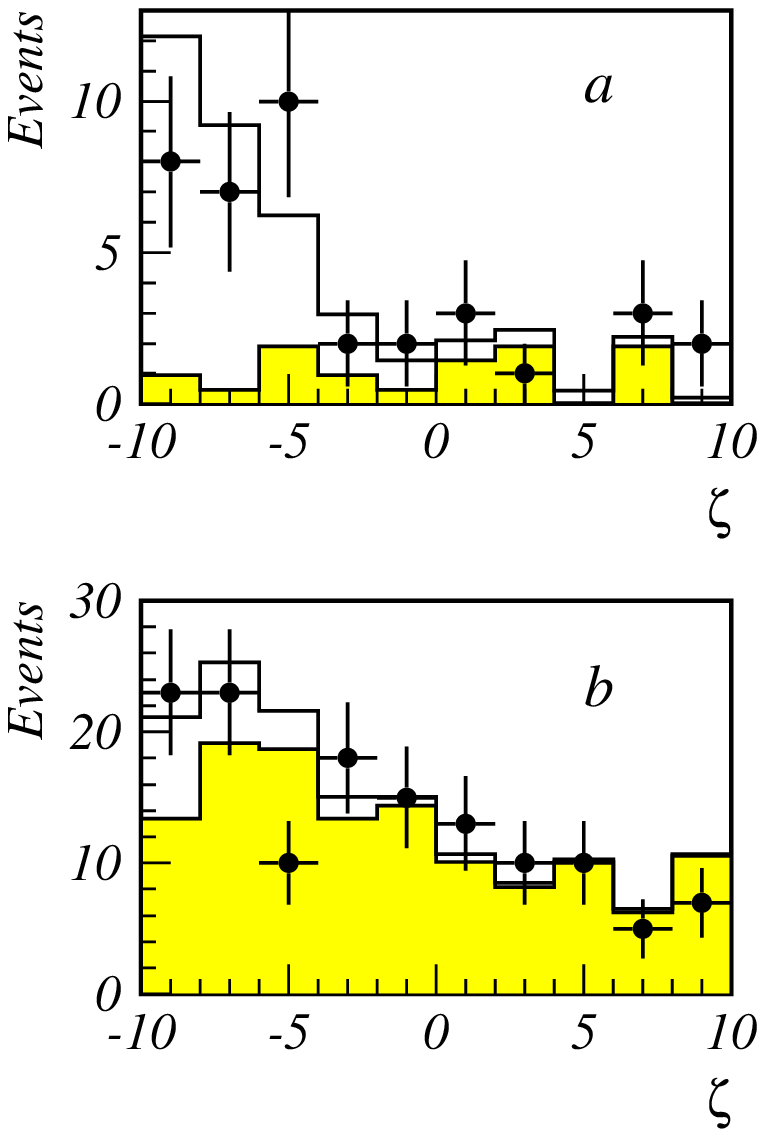,width=1.4\textwidth}} 
\vspace*{-5mm}
\caption{ Distribution of events over $\zeta$
in search for $\phi\to\eta\pi^0\gamma$ decay:
a) events with  $E_{\gamma max}/E_0<0.7$, b) events
with  $0.7<E_{\gamma max}/E_0<0.8$.
Circles with error bars -- experimental data,
shaded histogram -- simulation of the process (\ref{etagn}),
and clear histogram -- simulated sum
of (\ref{etagn}) and (\ref{etap0g}) with a $BR=10^{-4}$.}
\label{XINM}
\end{minipage}
\end{figure}

\begin{figure}[htb]
\epsfxsize=0.49\textwidth
\begin{minipage}[t]{0.47\textwidth}
\centerline{\epsfig{figure=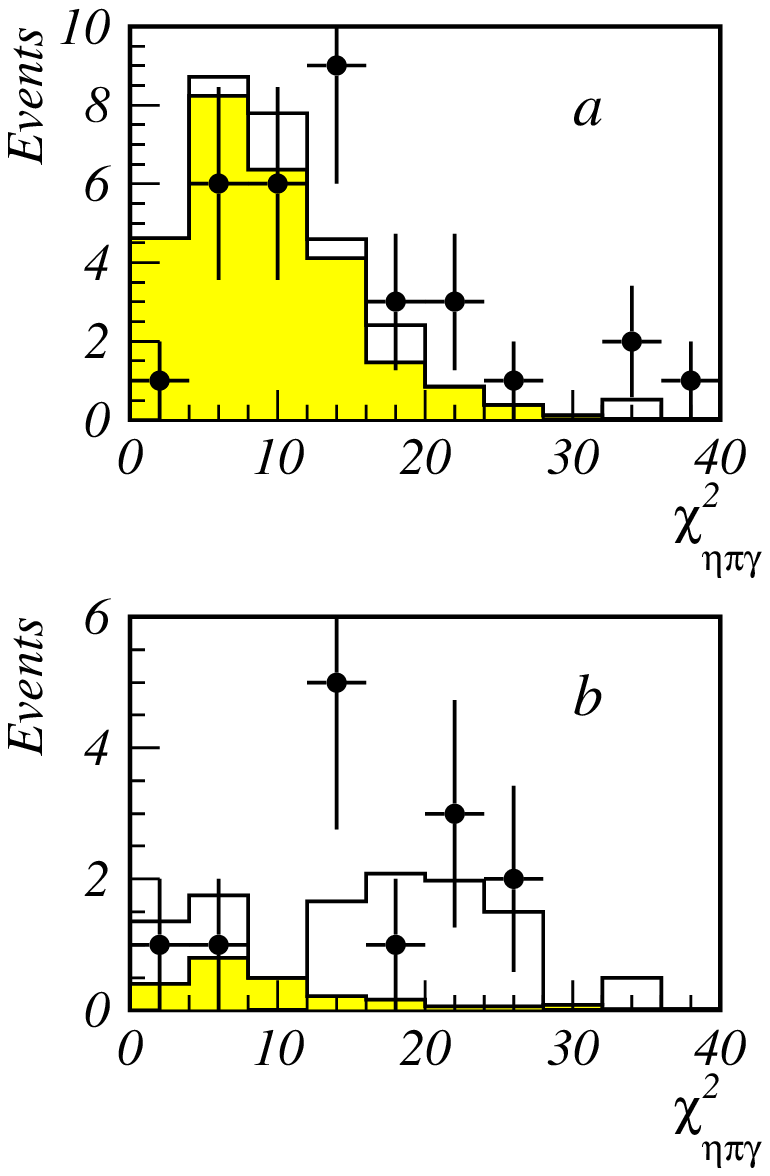,width=1.4\textwidth}} 
\vspace*{-5mm}
\caption{ Distribution of events over $\chi^2_a$
in search for $\phi\to\eta\pi^0\gamma$ decay:
a) events with  $\zeta<-4$, b) events
with  $0<\zeta<10$,
circles with error bars -- experimental data,
shaded histogram
depicts simulation of the process 
(\ref{etap0g}) and clear histogram -- simulated sum
of (\ref{etap0g}) and (\ref{etagn}) with a $BR=10^{-4}$.}
\label{XI2A}
\end{minipage}
\hfill
\begin{minipage}[t]{0.47\textwidth}
  \centerline{\epsfbox{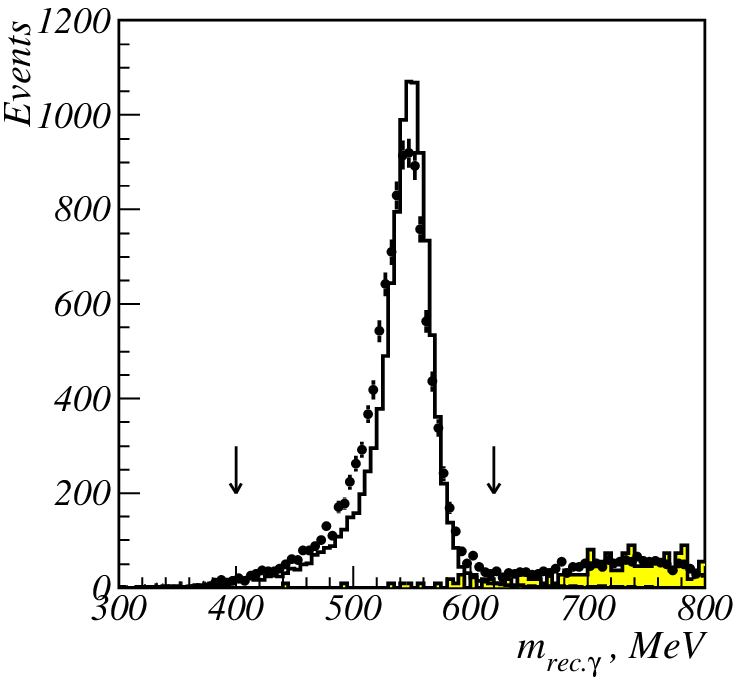}} 
\vspace*{-5mm}
\caption{The recoil mass distribution of the most energetic photon in an event.
Points -- data, histogram -- simulation, 
hatched histogram -- simulation of the background decay $\phi\to K_S K_L$.}
\label{etg98}
\end{minipage}
\end{figure}

The resulting spectra of the $E_{\gamma max}/E_0$, energy of the most
energetic photon, are shown in Fig.~\ref{ER1N}.
Since the recoil photon in the process (\ref{etagn})
has a narrow spectrum close to 360 MeV,
$E_{\gamma max}/E_0$ must be more than 0.7 in this decay. 
This can be seeing
in Fig.~\ref{ER1N}. While the spectrum in Fig.~\ref{ER1N}b 
is well reproduced
by simulation of the process (\ref{etagn}) alone, the spectrum in
Fig.~\ref{ER1N}a shows excess of events over the simulation 
of (\ref{etagn}).
Absence of corresponding excess in the Fig.~\ref{ER1N}b indicates, that
additional events are true 5-$\gamma$ ones, because the cut $0<\zeta<10$
suppresses the number of such events by an order of magnitude. In the
event sample with $\zeta<-4$, the background (\ref{etagn}) still dominates at
$E_{\gamma max}/E_0>0.7$, so
further analysis was conducted with events where $E_{\gamma max}/E_0$ was
less than 0.7, where contribution of (\ref{etagn}) is small. 
From Fig.~\ref{ER1N}a
one can see, that this roughly halves the detection efficiency for the process
(\ref{etap0g}) 
and enhances its dependence on the $\eta\pi^0$ invariant mass.
The detection efficiency varies from $1\%$ at 
$m_{\eta\pi^0}=970~MeV/c^2$ to
$5\%$ at $m_{\eta\pi^0}=700~MeV/c^2$. 

The distributions of the events over photon quality parameter $\zeta$ are
shown in Fig.~\ref{XINM}. Selection criteria here were the same except the
less stringent requirement on photon quality: $\zeta<10$. It can be seen,
that simulation well describes the background from the process 
(\ref{etagn}), while
the excess of events in Fig.~\ref{XINM}a at low $\zeta$ is compatible with
existence of the decay (\ref{etap0g}) with BR of the order of $10^{-4}$. 
Additional
information could be obtained from the distribution in 
$\chi^2_{\eta \pi^0 \gamma}$ 
(Fig.~\ref{XI2A}). The distributions of good events with $\zeta<-4$ and
background with $0<\zeta<10$ do not contradict the simulation. The
enhancement at low $\chi^2_{\eta \pi^0 \gamma}$ 
for good events is clearly seen. The number of
selected events is 25, from which 5 were estimated to be a background.
The corresponding branching ratio was calculated using the $\eta\pi^0$
invariant mass distribution of the selected events and detection
efficiencies, obtained from simulation. The resulting value is
$(8.3\pm2.3)\cdot10^{-5}$, where error is a statistical one. For such
stringent selection criteria the systematic error in detection efficiency
could be as high as $10\div15\%$, but at present level of 
experimental statistics the statistical error dominates.

\subsubsection{Analysis 2}

\begin{figure}[htb]
\epsfxsize=\textwidth
\begin{minipage}[t]{\textwidth}
  \centerline{\epsfig{figure=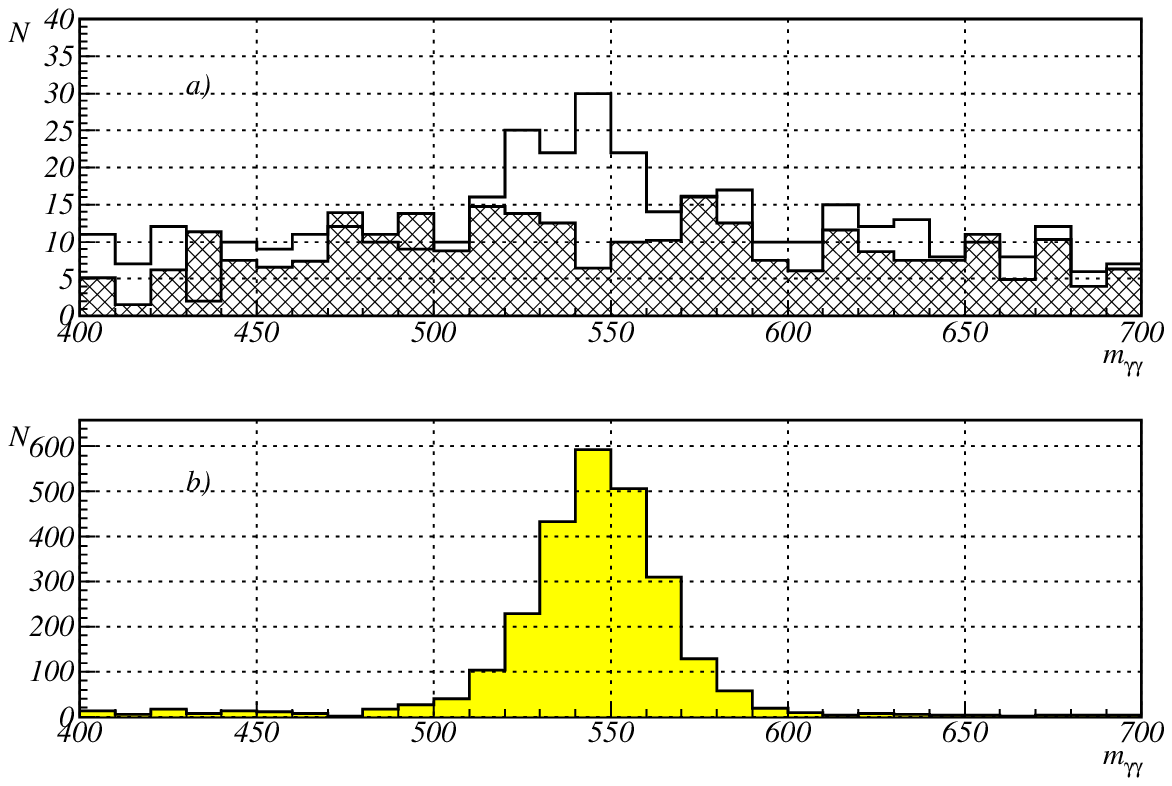,width=0.8\textwidth}} 
  \vspace*{-5mm}
\caption{\large Distribution of $\gamma\gamma$ invariant mass
closest to that of $\eta$-meson for photons not included into $\pi^0$
in search for the decay $\phi\to\eta\pi^0\gamma$.
a) histogram -- experimental data;
hatched histogram -- expected background;
b) $\phi\to\eta\pi^0\gamma$ simulation.}
\label{PEMETA}
\end{minipage}
\vfill
\begin{minipage}[t]{\textwidth}
   \centerline{\epsfig{figure=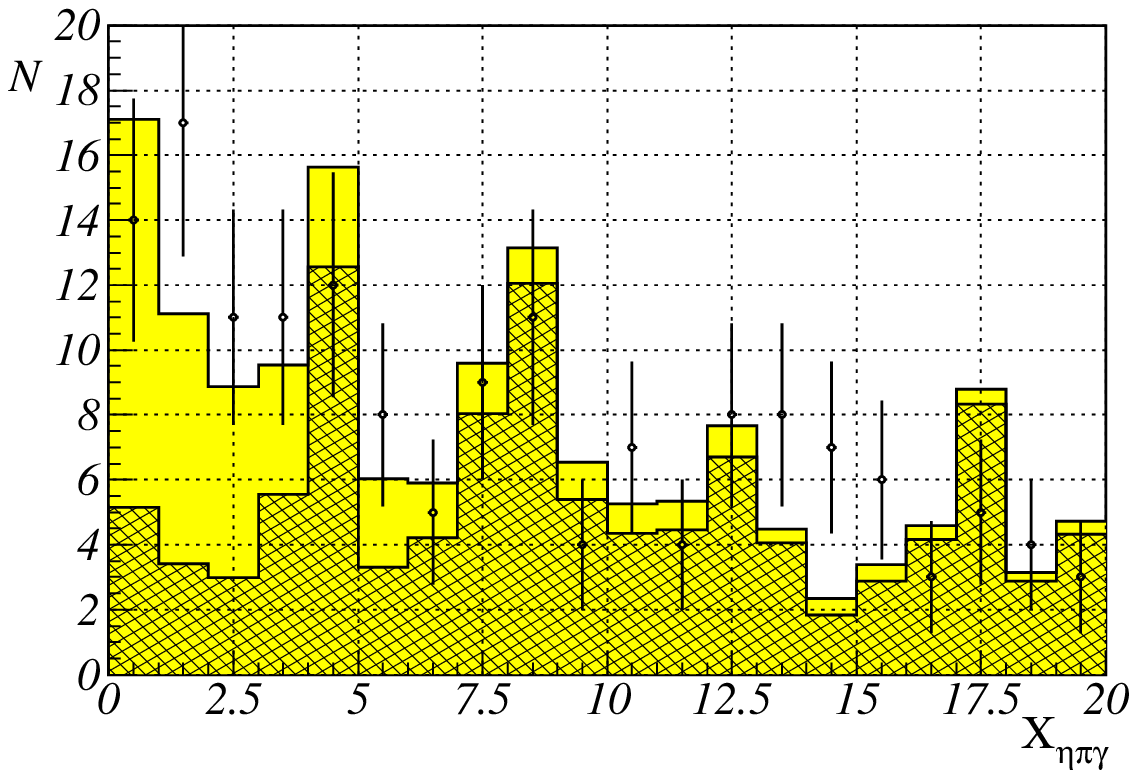,width=0.8\textwidth}} 
\vspace*{-7mm}
\caption{Distribution of the $\chi_{\eta\pi^0\gamma}^2$ parameter
in search for  $\phi\to\eta\pi^0\gamma$ decay,
with additional cut $\zeta <-4$.
Points with error bars -- experimental data;
hatched histogram -- expected background;
histogram -- sum of expected effect at
$B(\phi \to \eta\pi\gamma) = 10^{-4}$
and background contributions.}
\label{PEXA0G}
\end{minipage}
\end{figure}

In  the decay  $\phi\to\eta\pi^0\gamma$  we consider 5-$\gamma$ final
state with the following selection cuts :
\\$\bullet$ $N_{\gamma}=5$; $N_{cp}=0$; $0.8 < E_{tot}/2E_0 < 1.1$;
$P/2E_{tot} < 0.15$; $\theta_{min} > 27^0$;
\\$\bullet$ $\zeta <0$; $\chi_E^2 <25$.
\\ The main background processes (\ref{etagn}), (\ref{kskln}),
(\ref{omp0n}) and (\ref{p0p0g}) have 5 photons final state
with two $\pi^0$.  For  supressing this background
besides the cut $\zeta <0$ two additional cuts were applied:
1 -- $\chi_{\pi^0\pi^0\gamma}^2 > 50$ -- event is not consistent 
with $\pi^0\pi^0\gamma$ hypothesis,
2 -- $\chi_{\omega\pi^0}^2  > 30$ -- the kinematics of the event is
not consistent with the $\omega\pi^0$ hypothesis.
All experimental events, that passed the selection cuts above, contained
at least one $\pi^0$-meson candidate.
Among other three photons  not included into
$\pi^0$, the pair of
photons with invariant mass closest to that of $\eta$ meson, was chosen
(Fig.~\ref{PEMETA}). The visible excess of events close to  $\eta$-meson mass
could be considered as an evidence of the decay $\phi\to\eta\pi^0\gamma$.
For the background
estimation we used events in two intervals of $m_{\gamma\gamma}$:
$500~MeV \le m_{\gamma\gamma} \le 590~MeV$ and outside this interval:
$455~MeV <  m_{\gamma\gamma} < 500~MeV$ or
$590~MeV <  m_{\gamma\gamma} < 635~MeV$.

104  experimental events were found outside $\eta$-mass  interval and
96 simulated events from  background processes
(\ref{etagn}), (\ref{omp0n}). Their ratio is equal to $1.08 \pm 0.12$.
Taking into account this value and the detection efficiency of 7.2\%, 
we obtained $B(\phi \to \eta\pi^0\gamma) = (0.87 \pm 0.30)\cdot 10^{-4}$.

\begin{figure}[htb]
\epsfxsize=0.49\textwidth

\begin{minipage}[t]{0.47\textwidth}
  \centerline{\epsfbox{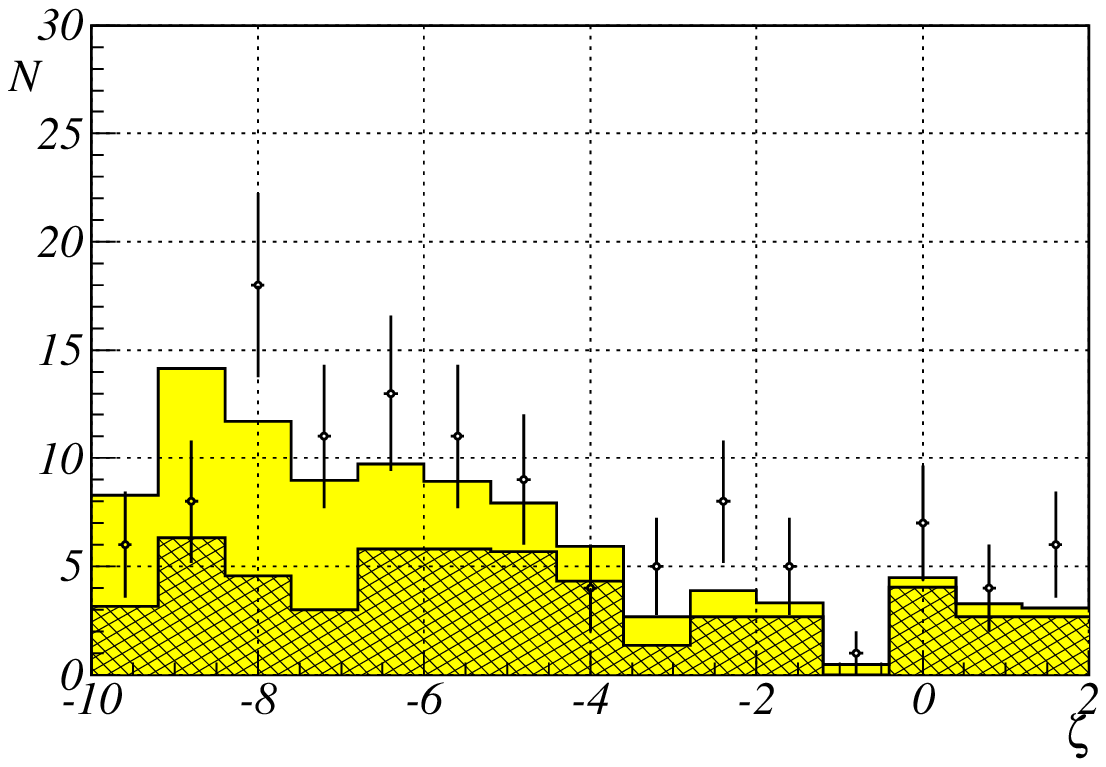}} 
\vspace*{-5mm}
\caption{Distribution of the parameter $\zeta $
in search for the decay $\phi\to\eta\pi^0\gamma$
with additional cut $\chi_{\eta\pi^0\gamma}^2 < 7$.
Circles with error bars -- experimental data;
hatched histogram -- expected background;
histogram -- sum of expected signal at
$B(\phi \to \eta\pi\gamma) = 10^{-4}$
and background contributions. }
\label{PEXINM}
\end{minipage}
\hfill
\begin{minipage}[t]{0.47\textwidth}
  \centerline{\epsfbox{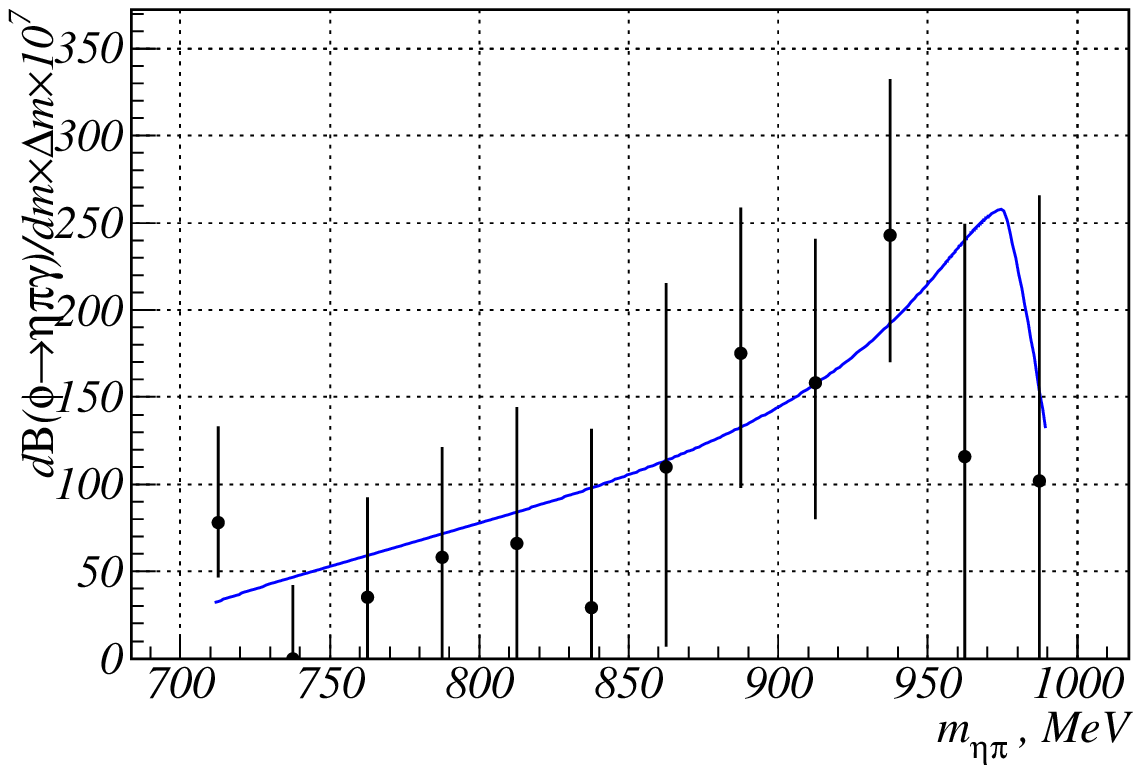}} 
\vspace*{-5mm}
\caption{Distribution of the $\eta\pi^0$ invariant mass
in a search for the $\phi\to\eta\pi^0\gamma$ decay. 
Solid line is a calculated distribution for the
optimal value of fit parameters.}
\label{PEMA0G}
\end{minipage}
\vfill
\begin{minipage}[t]{0.47\textwidth}
  \centerline{\epsfbox{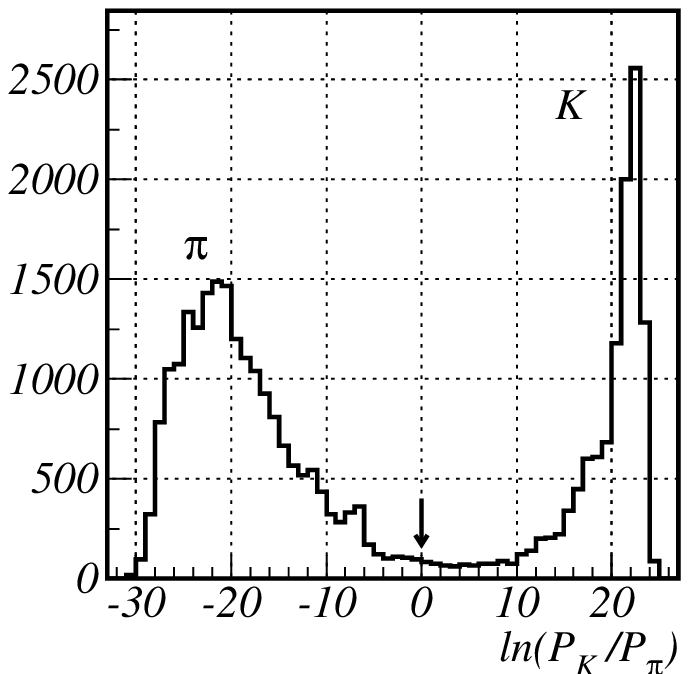}} 
\vspace*{-5mm}
\caption{Distribution of the $\pi/K$ separation parameter
in search for the process
$ e^+e^- \to \omega \pi^0 \to \pi^+ \pi^- \pi^0 \pi^0$.
Arrow shows a cut suppressing background from 
$\phi\to K^+K^-$ decay.} 
\label{vdcfig1}
\end{minipage}
\hfill
\begin{minipage}[t]{0.47\textwidth}
  \centerline{\epsfbox{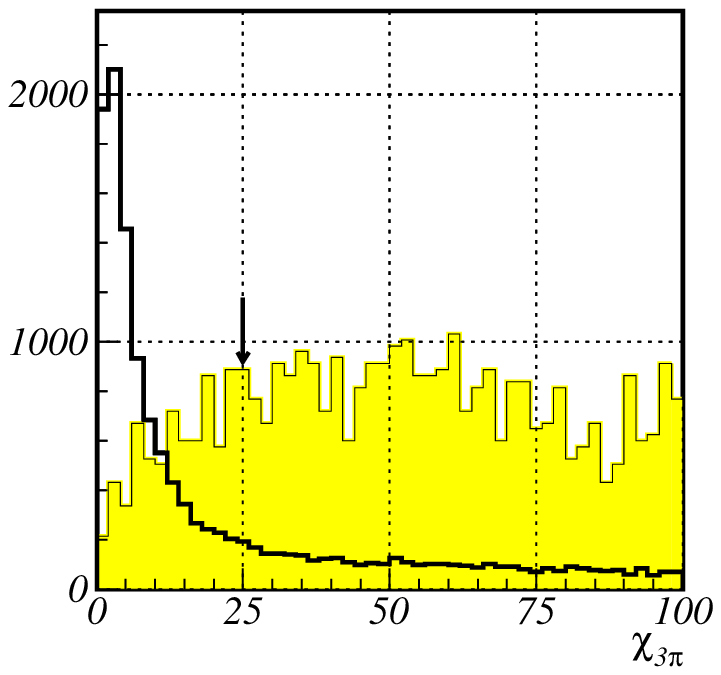}} 
\vspace*{-5mm}
\caption{Distribution of the parameter
$\chi_{3\pi}$ for experimental and simulated events of the process
$ e^+e^- \to\omega\pi^0\to\pi^+\pi^-\pi^0\pi^0$
(hatched histogram).
A peak in the experimental data distribution is due to the
the process $ e^+e^- \to \phi \to\pi^+\pi^-\pi^0$
(\ref{phi3pi}). 
Arrow shows the cut value for this parameter.}
\label{vdcfig2}
\end{minipage}
\end{figure}

Then the following cuts were applied:
 $\chi_{\pi^0\pi^0\gamma}^2 > 50$,
 $\chi_{\eta\pi^0\gamma}^2 <7$,
 $\zeta <-4$.
Fig.\ref{PEXA0G} and  \ref{PEXINM} show the distribution
over $\chi_{\eta\pi^0\gamma}^2$ for the cut $\zeta <-4$,
and over the parameter $\zeta$ for the cut
$\chi_{\eta\pi^0\gamma}^2 <7$ respectively.
  It is seen that at the small values of $\zeta$
and  $\chi_{\eta\pi^0\gamma}^2$ there is a considerable excess of
experimental events (78) over the expected background ( 34 for the
background process (\ref{etagn}) and 4 for  (\ref{omp0n}),
while in the region
$\zeta >-4$ and $\chi_{\eta\pi^0\gamma}^2 >7$ the experimental events
number is in agreement with the expected background. Taking into account
the selection efficiency of the process under study in the
region $\zeta <-4$ and $\chi_{\eta\pi^0\gamma}^2 <7$ (4.5\%) we
have:
$B(\phi \to \eta\pi\gamma) = (1.1 \pm 0.30)\cdot 10^{-4}$.

    For more accurate background estimation, the events, satisfying
the cut $\zeta <2$ and $\chi_{\eta\pi^0\gamma}^2 <15$,    
were divided into 4 classes (table \ref{Four_groups}).
\begin{table}[tbp]
\begin{center}
\caption{\label{Four_groups}  }
\begin{tabular}{|c|c|c|}
\hline
  & $\zeta <-4$ & $-4 \le \zeta < 2$     \\
\hline
  $\chi_{\eta\pi^0\gamma}^2 < 7$  & 78        &    38 \\
\hline
  $\chi_{\eta\pi^0\gamma}^2 \ge 7$  & 58        &    27 \\
\hline
\end{tabular}
\end{center}
\end{table}

The c.m. energy dependence of the cross section in each of
4 classes, was approximated by the sum of the cross section of
the process under study, background process (\ref{etagn})
with coefficient  $K_{\eta\gamma}$  and
nonresonant background.
The expected distributions over all main parameters ($\zeta $,
$\chi_{\eta\pi^0\gamma}^2$) in each class of events
were obtained from simulation. Only  $\zeta $ distribution
for the process under study
was derived from experimental events of the process (\ref{p0p0g}).
The fit results obtained are the following:
\begin{equation}
\begin{array}{l}
    K_{\eta\gamma} = 1.26\pm0.18,\\
    B(\phi \to \eta\pi^0\gamma)  = (0.96\pm0.32)\cdot 10^{-4},
\end{array}
\end{equation}
where $K_{\eta\gamma}$ is the ratio of found cross section
 of the process (\ref{etagn}) to the expected one. Nonresonant background
 was found to be negligible.
  Fig.\ref{PEMA0G} shows the distribution over $\eta\pi^0$ invariant
mass , corrected for detection efficiency dependence on the $\eta\pi^0$
invariant mass.
The errors indicated include statistical errors  and
error in background estimation. The fitted curve 
\cite{IVN2} is 
indicated by solid line. The accuracy of our measurements
is not enough to obtain
the coupling constant $g_{a_0\eta\pi}$. Its value was set to 
$g_{a_0\eta\pi}=0.85g_{a_0K\bar K}$ (\cite{IVN2}). As a result of the fit 
the following parameters were obtained:
\begin{equation}
\begin{array}{l}
      m_{a_0} = 986^{+23}_{-10}~MeV,
      g^{2}_{a_0K\bar K}/4\pi   = (1.5\pm0.5)~GeV^{2}.
\end{array}
\end{equation}

\subsection{ Evidence of the
$\phi \to \omega \pi^0 \to \pi^+ \pi^- \pi^0 \pi^0 $ decay}
\label{Sec.V.D.cha}

    Recently in experiments with SND and CMD-2 detectors at VEPP-2M the study
of $\phi$ meson rare decays with branching ratios of
$10^{-4} \div 10^{-5}$
became possible \cite{Serednyakov/H-97,Prep.97,CMD-2}. One of the decays
this kind of decays is OZI and G-parity forbidden $\phi\to\omega\pi^0$ decay.
The expected branching fraction of this decay is of order of
$5\cdot 10^{-5}$ \cite{theor1,theor2}. Predictions vary in wide limits
depending on the nature of $\phi$, $\omega$, $\rho$ mixing and existence of
direct  $\phi\to\omega\pi^0$ transition. Because of the significant 
cross-section  of the non-resonant $e^+e^-\to\omega\pi^0$ reaction
in the vicinity of $\phi$-resonance, the decay $\phi\to\omega\pi^0$ 
reveals itself as a narrow interference pattern in the  
cross section energy dependence. This allows to determine from the data
both the real and imaginary parts of the decay amplitude. 
The  $\phi\to\omega\pi^0$ decay was not observed by now.
In our preliminary study \cite{Serednyakov/H-97,Prep.97}
only an upper limit of the decay probability was established
at $5\cdot 10^{-5}$.
The present work is based on full statistics of PHI-96 run
and part of the data from PHI-97 run at the energy points 980,
1040 and 1060 MeV. The total integrated luminosity is
$\Delta L=3.7~pb^{-1}$, corresponding to $6.3\cdot 10^{6}$
$\phi$-mesons produced.

{\bf Events selection.} For search of the decay  $\phi\to\omega\pi^0$
we studied the cross section of the process 
\(e^+e^-\to\omega\pi^0\to\pi^+\pi^-\pi^0\pi^0\)
(\ref{omp0c}) in the vicinity of the $\phi$ resonance. The final state
consists of 2 charged particles and 4 photons.
Unfortunately, in addition to 4 produced photons, we found in
selected events  with the probability of $\sim$20\% one or two
faked photons. So, for analysis we selected events with 2 charged
particles and 4 or more photons. The reconstructed production point of charged
particles should be within $\pm$0.5 cm from the beam and
$\pm$7.5 cm from the center of the detector in the beam direction
(the interaction region length $\sigma_{z}$ is about 2 cm).  

    Because of the large probability of faked photons production
virtually all main $\phi$-meson decays are a potential source of background 
in the search for  $\phi\to\omega\pi^0$ decay. To suppress 
the background from the decays
\( e^+e^- \to\phi\to K^+K^-\)
(\ref{kpkn}),
\( e^+e^- \to\phi\to K_SK_L, \,\,\, K_S \to\pi^+\pi^-\)
(\ref{ksklc})
the following selection criteria were applied:
\\ $\bullet$ $\alpha _{\pi \pi} < 140^\circ$;
\\ $\bullet$  Ionization losses of charged particles 
are close to those of minimum ionizing particles.

The first condition suppressed events from the process (\ref{ksklc}),
where minimum angle between pions from the  $K_S$ decay is $150^\circ$,
and considerable number of events from the process (\ref{kpkn}),
where charged kaons produce two collinear tracks. Part of events of the
process (\ref{kpkn}) survives this cut due to decay or nuclear
interaction of one of the kaons in the material of the beam pipe or 
drift chamber shell. Remaining charged kaon due to its low velocity
$\beta\approx 0.25$ has high dE/dx in the SND drift chamber,
allowing to reject these events using $\pi /K$ separation parameter
$ln(P_K/P_\pi$) (Fig.~\ref{vdcfig1}).

  In the kinematic fit the energy-momentum balance and masses of intermediate
particles ($\pi^0,\eta ,\omega$) were included. Three following
hypotheses of the event origin were considered for each event:
\\$\bullet$
The event originates from the process  $e^+e^-\to\pi^+\pi^-\pi^0$.
The value of likelihood function is $\chi_{3\pi}$.
\\$\bullet$ 
The event is from the process  
$e^+e^- \to\pi^+\pi^-\pi^0\gamma$. 
The photon recoiled mass $m_{rec.}$
was calculated.
\\$\bullet$ 
The event is due to the process $e^+e^-\to\pi^+\pi^-\pi^0\pi^0$
The value of likelihood function is $\chi_{4\pi}$. The recoil mass of
$\pi^0$ mesons was calculated and one $m_{3\pi}$
closest to $\omega$ meson mass was chosen.

\begin{figure}[htb]
\epsfxsize=0.49\textwidth

\begin{minipage}[t]{0.47\textwidth}
  \centerline{\epsfbox{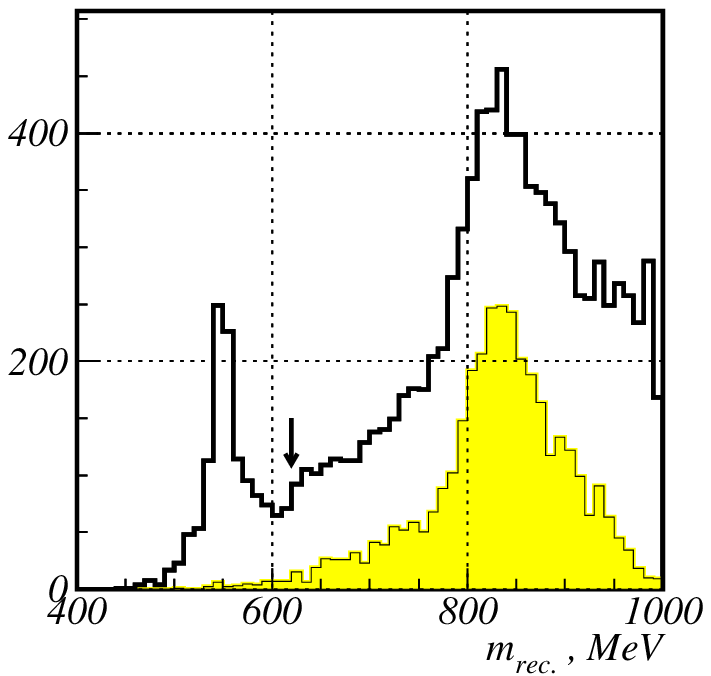}} 
\vspace*{-5mm}
\caption{Distribution of the parameter $m_{rec.}$
for experimental data and simulated process
$ e^+e^- \to\omega\pi^0\to\pi^+\pi^-\pi^0\pi^0$
(hatched histogram).
A peak at $\eta$-meson mass in experimental data is determined by
contribution from the process $ e^+e^- \to \phi \to \eta \gamma$,
$eta\to\pi^+\pi^-\pi^0$ (\ref{etagc}). 
Arrow shows the cut value for this parameter.}
\label{vdcfig3}
\end{minipage}
\hfill
\begin{minipage}[t]{0.47\textwidth}
  \centerline{\epsfbox{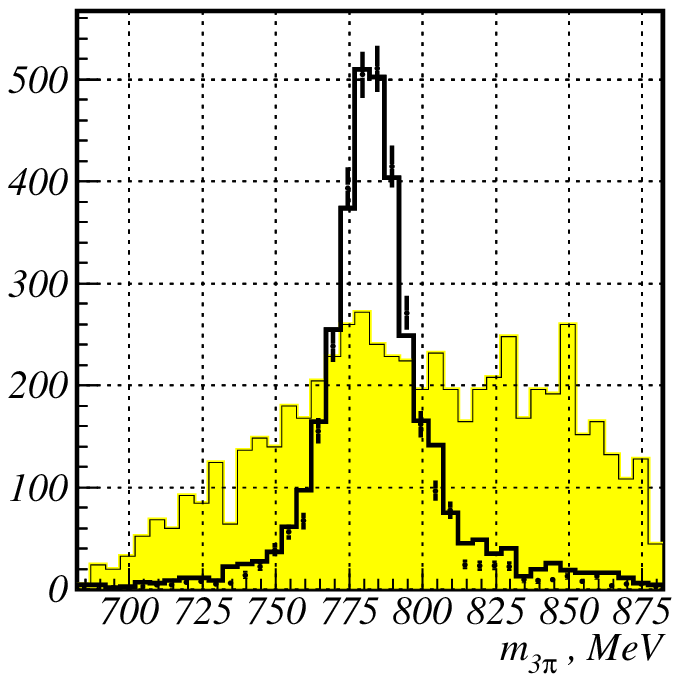}} 
\vspace*{-5mm}
\caption{Distribution of the parameter $m_{3\pi}$
for experimental (histogram) and simulated (hatched histogram)
events of the process
$ e^+e^- \to \pi^+\pi^-\pi^0\pi^0$.
Dots with error bars -- simulation of the process 
$ e^+e^- \to\omega\pi^0\to\pi^+\pi^-\pi^0\pi^0$
with 7\% contribution of the process
$ e^+e^- \to \pi^+\pi^-\pi^0\pi^0$.}
\label{vdcfig5}
\end{minipage}
\vfill
\begin{minipage}[t]{0.47\textwidth}
  \centerline{\epsfbox{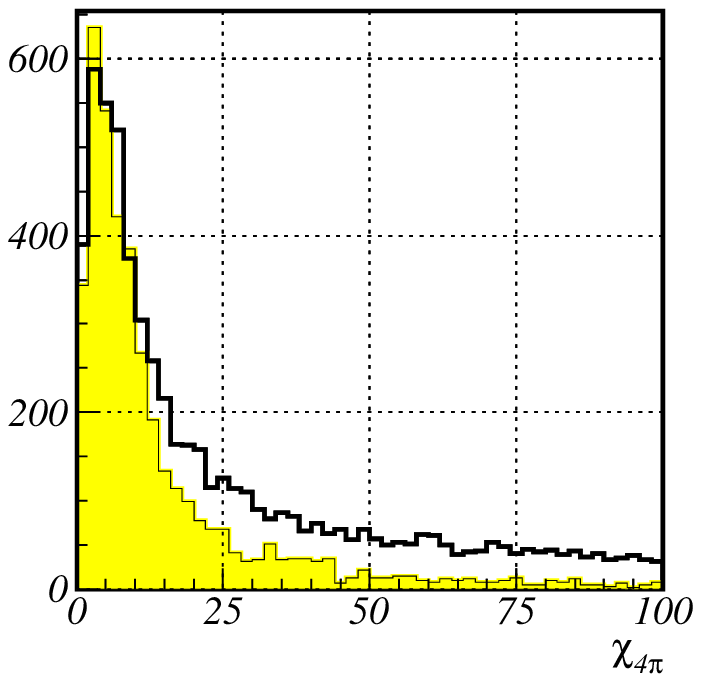}} 
\vspace*{-5mm}
\caption{Distribution of the parameter 
$\chi_{4\pi}$ for experimental (histogram) and simulated 
(hatched histogram) 
events of the process
$ e^+e^- \to\omega\pi^0\to\pi^+\pi^-\pi^0\pi^0$.}
\label{vdcfig4}
\end{minipage}
\hfill
\begin{minipage}[t]{0.47\textwidth}
  \centerline{\epsfbox{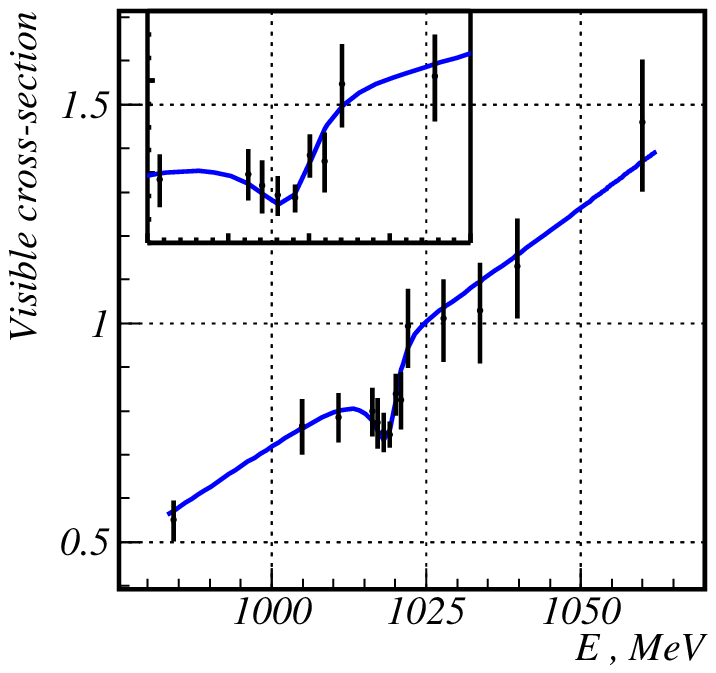}} 
\vspace*{-5mm}
\caption{Visible cross section of the process
$ e^+e^- \to\omega\pi^0\to\pi^+\pi^-\pi^0\pi^0$
as a function of energy and its optimal approximation.}
\label{vdcfig6}
\end{minipage}
\end{figure}

In case if number of found photons was greater than that in considered
hypothesis, extra photons were rejected as fake photons. To do that,
all combinations were considered and one was left with a minimum value of
likelihood function. The same approach was applied in search
for the best $\pi^0$ candidates among all possible photon pairs in the event.
The distribution of
experimental and simulated events from the process (\ref{omp0c})
over the parameter $\chi_{3\pi}$, is shown in Fig.~\ref{vdcfig2}.
One can see a considerable contribution from process (\ref{phi3pi})
as a peak at low values of  $\chi_{3\pi}$. Fig.~\ref{vdcfig3}
shows the experimental spectrum over  $m_{rec.}$, where clear peak
of $\eta$ meson from the reaction (\ref{etagc}) is seen.
To suppress the background from processes (\ref{phi3pi}) and
(\ref{etagc} the following cuts were used:
\\$\bullet~ \chi_{3\pi} > 25$,
\\$\bullet~ m_{rec.} > 620~MeV$.

{\bf Data analysis.} Fig.~\ref{vdcfig4} shows the distribution of
experimental and simulated events from the process (\ref{omp0c}) over
$\chi_{4\pi}$. The considerable difference between the ``tails'' of 
measured and simulated
spectra is a sign of background, which survived
the cuts. In Fig.\ref{vdcfig5} the
distribution over  $\pi^0$ recoil mass $m_{3\pi}$  is presented for 
experimental events with $\chi_{4\pi}<20$, simulated events 
of the process (\ref{ppp0p0}) without $\omega\pi^0$ 
intermediate state (the model $\rho\pi\pi$ was used)
and simulation of the process  (\ref{omp0c}) with 7\% addition 
of the process  (\ref{ppp0p0}). The last distribution 
is in a good agreement with experimental data.

For further analysis all events were divided into 4 classes:  
\begin{enumerate}
\item $\chi_{4\pi}<20$,    $ |m_{3\pi}-782|<30$;
\item $\chi_{4\pi}<20$,    $30<|m_{3\pi}-782|<60$;
\item $20<\chi_{4\pi}<40$, $|m_{3\pi}-782|<30$;
\item $20<\chi_{4\pi}<40$, $30<|m_{3\pi}-782|<60$.
\end{enumerate}

   The visible cross section for each class  was fitted using the
following formulae:
$$\sigma_{vis} = \alpha_i\sigma_{\omega\pi}+
\beta_i\sigma_{4\pi}+\sigma_{\phi i},$$
$$
\sigma_{\omega\pi} = \varepsilon B(\omega\to 3\pi) 
(\sigma_0+A(E-m_\phi))\cdot
\biggl |1-Z \frac{m_\phi\Gamma_\phi}
{D_\phi}\biggr |^2(1+\delta),
$$
where $E=2E_0$, $\sigma_0$ is a nonresonant cross section of the
process (\ref{omp0c}) at  $E=m_\phi$, $A$ -- its slope,
$\varepsilon$ -- the detection efficiency of the process (\ref{omp0c})
with $\chi_{4\pi}<40$ and  $|m_{3\pi}-782|<60 MeV$, $Z$ -- complex
interference amplitude, $m_\phi,\Gamma_\phi$ --
$\phi$ meson mass and width,
$D_\phi=m_\phi^2-E^2-iE\Gamma_\phi(E)$ --  $\phi$-meson propagator,
$B(\omega\to 3\pi )=0.888$ -- the branching ratio of the decay
$\omega\to 3\pi $\cite{PDG}, $\delta$ -- radiative correction \cite{RadCor},
$\sigma_{4\pi}$ -- the visible cross section of the process (\ref{ppp0p0}),
$\sigma_{\phi i}$ -- the visible resonant background
cross section for $i$-class selection, $\alpha_i$ and $\beta_i$ -- 
the probabilities for events from processes (\ref{omp0c}) 
or (\ref{ppp0p0}) to be found in the $i$-class.

   The fitting was performed for all 4 classes simultaneously.
The class 1 with a small resonant background,
was the most important for evaluation of $\sigma_0$, $A$, $Z$ parameters.
Classes 2-4 were used to determine
the background from $\phi$-decays $\sigma_{\phi 2}$, $\sigma_{\phi 3}$,
$\sigma_{\phi 4}$. It was suggested in the fit, that for resonant background
the distribution over  $m_{3\pi}$ does not depend on $\chi_{4\pi}$,
and for class 1 the background can be obtained from the expression:
$\sigma_{\phi 1}=\sigma_{\phi 2}\cdot(\sigma_{\phi 3}/\sigma_{\phi 4})$.
The cross section of the resonant background $\sigma_{\phi 1}$
in the resonance maximum was found to be $\sigma_{\phi 1}=(28\pm 17)~pb$,
what was about 4\% of the visible cross section (\ref{omp0c}) 
in class 1.

   The contribution of the process (\ref{ppp0p0})
was determined from the ratio of nonresonant cross section
for events in the intervals $\pm50~MeV$ and  $\pm100~MeV$ 
of the $\omega$-meson mass.
The value of the cross section (\ref{ppp0p0}) was found to be
$(6.9\pm3.8)$\% of the cross section (\ref{omp0c}).
With the chosen cut $\pm 60~MeV$ this contribution
was suppressed down to $(4.2\pm 2.3)\%$. 

  The chosen selection criteria suppress resonant background, but
decrease the detection efficiency of the process (\ref{omp0c}).
In addition, the tracking in the drift chambers
and nuclear interaction of pions
are simulated imprecisely. Therefore, the detection efficiency,
obtained from the simulation, was corrected in three ways.
First, we processed the data with softer cuts on the parameters
$\chi_{4\pi}$ and  $m_{3\pi}$, second,
we studied a class of events with 1 charged and 4 photons
outside the $\phi$-peak,
and third, we compared the efficiencies under different cuts for both data
and simulation. The total correction obtained from experimental data
was found to be 12\%. It was mainly due to inaccuracy of the simulation of
$\chi_{4\pi}$ parameter distribution and drift chamber reconstruction
errors.
The corrected detection efficiency
was $16.9\pm1.7\%$ at $E=m_\phi$. Its value is almost constant
in the energy range under study. The systematic error was estimated
as 10\%. The number of events in classes 1--4 ($\alpha_i$, $\beta_i$)
was obtained in part from simulation or from the fit results. The
parameters $\alpha_i$, $\beta_i$ of the energy dependence in 
linear approximation were obtained also from the fit with the constraint,
that detection efficiency is energy independent.

    The number of fit parameters, describing energy dependence of the
cross sections in 4 classes of selected events, was 12. In each class
the cross section was measured in 14 energy points. At  
$\chi^2/d.f.=35/44$ the following values of main fit 
parameters were obtained:
\begin{eqnarray}
\label{vdcres}
&\sigma_0=(8.6\pm0.9)~nb,&\nonumber \\
&A=(0.090\pm0.011)~nb/MeV,&\\
&Re(Z)=0.108\pm0.026,&\nonumber\\
&Im(Z)=-0.127\pm0.027.&\nonumber
\end{eqnarray}

  The visible cross section in class 1 and fit are shown in
Fig.~\ref{vdcfig6}. Despite 4\% resonant background, the
interference pattern is clearly seen. The interference amplitude
can be presented in the form
$Z=|Z|\cdot e^{\psi}$  with
$$|Z|=0.167\pm0.027,$$
$$\psi=(-50\pm9)^\circ\, .$$

The branching ratio of the decay  $\phi\to\omega\pi^0$ is
obtained in the following way:
$$B(\phi\to\omega\pi^0)=\frac{\sigma_0\cdot|Z|^2}{\sigma_\phi}=
(5.7^{+2.0}_{-1.8})\cdot 10^{-5},$$
where $\sigma_{\phi} =12\pi B(\phi\to e^+e^-)/m_{\phi}^2=4240~nb$ --
the cross section in the  $\phi$ maximum  \cite{PDG}.

   To estimate systematic errors caused by possible detector instability
during lengthy data taking runs, we processed  data from 1~--~3
runs separately from
4~--~6 runs and PHI-97 run. It was found, that all three data samples 
are well described by the fit obtained above from the whole data.

    To investigate reliability of the obtained results (\ref{vdcres}),
we changed the cuts: for the parameter $\chi_{4\pi}$ we used cuts
50 and 100 instead of 20 and 40, for the parameter $|m_{3\pi}-782|$
we used values 50 and 100 instead of 30 and 60. The detection
efficiency increased up to 20\%, the resonant background in the
maximum was  $\sigma_{b1}=0.206\pm0.069~nb$, that is 20\%
from the visible cross section (\ref{omp0c}). But the interference
amplitude was obtained:
$Re(Z)=0.117\pm0.025$,
$Im(Z)=-0.132\pm0.028$, 
which does not contradict to  (\ref{vdcres}) and thus confirms our 
procedure of the background subtraction. As a final result we prefer
(\ref{vdcres}), because it was obtained at lower resonant background.

{\bf Conclusion.} Obtained in this work non-resonant cross section
of the process $e^+e^-\to\omega\pi^0$ is in agreement with our old result
\cite{Ph.Rep} for the neutral
$\omega \to \pi^0 \gamma$ decay:
$\sigma(e^+e^- \to \omega \pi^0)=(8.7\pm1.0\pm0.7)~nb$. 
Our preliminary result in \cite{Prep.97}
is by 16\% lower, but this deviation is explained by systematic
uncertainty in simulation, discussed above.
The measured non-resonant cross section  $\sigma_0$
exceeds in two times the expected value, where only
$\rho(770)\to\omega\pi^0$ transition is taken into account.
The agreement should be significantly improved with radial excitations,
included into calculations.
The measured interference amplitude  $Z$ (\ref{vdcres})
is near lower edge of theoretical predictions \cite{theor2}.
But in \cite{theor2}  the known radial excitations of $\rho$ also were
not considered. Another important remark is small value of measured
real part of interference amplitude $Re(Z)$, which could hardly be
explained by well known $\phi-\omega$ mixing model \cite{theor2}.
For example, the predicted in \cite{theor1} branching ratio of
the decay $\phi\to\omega\pi^0$, is 
1.5 times higher than measured in this work.
The interference amplitude 
$$|Z|=0.166\pm0.027,$$
measured in this work, is of six
standard deviation significance. Thus, we claim the existence of the
decay $\phi\to\omega\pi^0$ with the branching ratio of
$$B(\phi\to\omega \pi^0)=(5.7^{+2.0}_{-1.8})\cdot 10^{-5}.$$

\subsection{Study of the process 
$e^+e^- \to \omega \pi^0 \to \pi^0 \pi^0 \gamma $}

       During the study of the reaction
$e^+e^-\to\omega\pi^0\to\pi^+\pi^-\pi^0\pi^0$ (\ref{omp0c}),
the $\phi\to\omega \pi^0$ decay
was observed for the first time with
the branching ratio  $5\cdot 10^{-5}$ \ref{Sec.V.D.cha}.
The decay reveals itself as an interference wave in non-resonant
cross section of the process  $e^+e^-\to\omega\pi^0$. In principle,
the similar picture should be observed in neutral channel
$e^+e^-\to\omega\pi^0\to\pi^0\pi^0\gamma$ (\ref{omp0n}). Really
the whole situation here looks more complicated because of other
$\phi$ meson neutral decays like  $\phi\to\rho^0\pi^0$,
$\phi\to f_0\gamma,\,\varepsilon\gamma$ \cite{IVN2}, which can create
the same final state and interfere with the (\ref{omp0n}) process.
The interference amplitude with the  $\phi\to\rho^0\pi^0$ decay is
about 10\%, which is close to the value 17\%, obtained in 
\cite{Prep.97}.
In our preceding study of the reaction  (\ref{omp0c})
\cite{Serednyakov/H-97,Prep.97}, we did not observe the interference
because of small statistics and non-resonant background.

    In the present work we studied the cross section of the
process (\ref{omp0n}) in the vicinity of  $\phi$ meson.
Besides this process, the same 5 photon final state have
processes 
\(e^+e^- \to \phi \to \pi^0 \pi^0 \gamma \to 5\gamma\)
(\ref{p0p0g}),
\(e^+e^- \to \phi \to K_S K_L \to neutrals\)
(\ref{kskln}),
\(e^+e^- \to \phi \to \eta \gamma \to 3\pi^0 \gamma \to 7\gamma\)
(\ref{etagn}),

   Some contribution to the background give cosmic rays and particles
from collider beams. The events from two photon annihilation 
with splitting showers also mimic 5 photon configuration, because
of its large cross section. The cut imposed on the total energy deposition 
and total momentum (\ref{5gamma}), strongly suppressed the process
(\ref{kskln}), the cosmic and beam background and considerably
reduced the contribution of the process (\ref{etagn}). 
Two photon events were suppressed by the cut on the the energy	
of two most energetic photons $E_2 < 0.8\cdot E_0$.

\begin{figure}[htb]
\epsfxsize=0.49\textwidth

\begin{minipage}[t]{0.47\textwidth}
  \centerline{\epsfbox{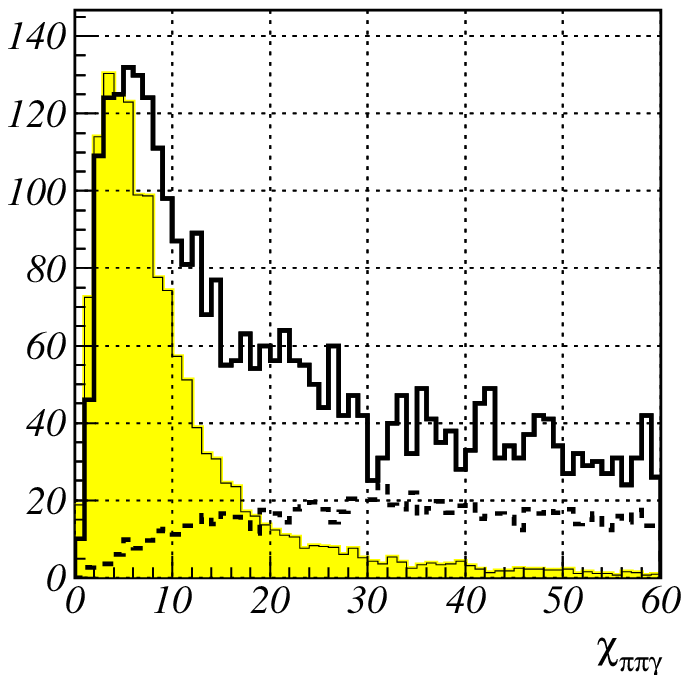}} 
\vspace*{-5mm}
\caption{Distribution of the parameter
$\chi_{\pi^0\pi^0\gamma}$
for experimental (histogram) and simulated events of the processes
$e^+e^- \to \omega \pi^0 \to \pi^0 \pi^0 \gamma $ 
(hatched histogram), 
$e^+e^- \to \phi \to \eta \gamma \to 3\pi^0 \gamma$ 
(dashed-line histogram). }
\label{f1}
\end{minipage}
\hfill
\begin{minipage}[t]{0.47\textwidth}
  \centerline{\epsfbox{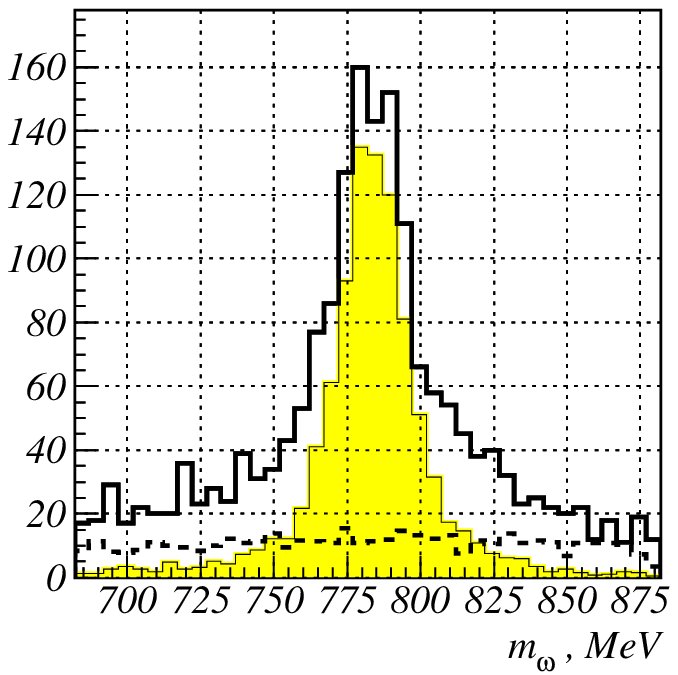}} 
\vspace*{-5mm}
\caption{Distribution of the parameter $m_{\omega}$ 
for experimental (histogram)
and simulated events of the processes
$e^+e^- \to \omega \pi^0 \to \pi^0 \pi^0 \gamma $ 
(hatched histogram),
$e^+e^- \to \phi \to \eta \gamma \to 3\pi^0 \gamma$ 
(dashed-line histogram). }
\label{f2}
\end{minipage}
\vfill
\begin{minipage}[t]{0.47\textwidth}
  \centerline{\epsfbox{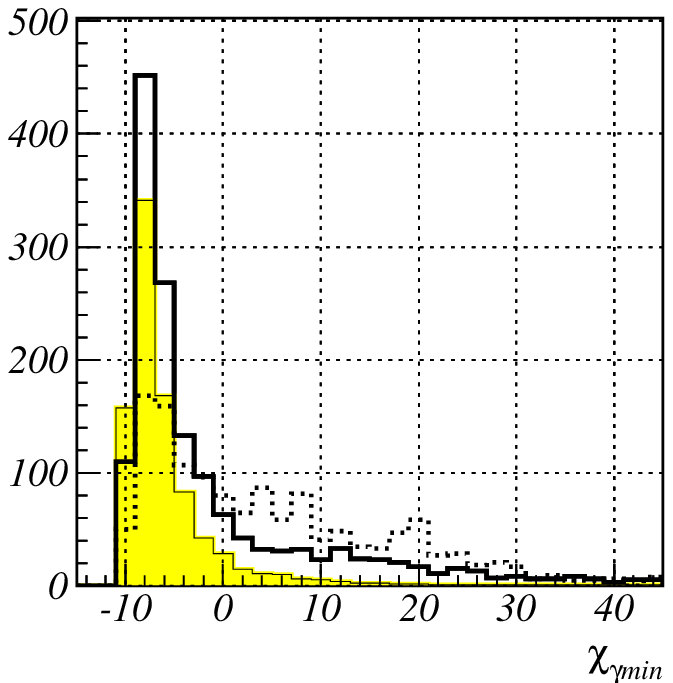}} 
\vspace*{-5mm}
\caption{Distribution on the parameter
$\chi_{\gamma max}$
for experimental (histogram)
and simulated events of the processes
$e^+e^- \to \omega \pi^0 \to \pi^0 \pi^0 \gamma $ 
(hatched histogram),
$e^+e^- \to \phi \to \eta \gamma \to 3\pi^0 \gamma$ 
(dashed-line histogram). }
\label{f3}
\end{minipage}
\hfill
\begin{minipage}[t]{0.47\textwidth}
  \centerline{\epsfbox{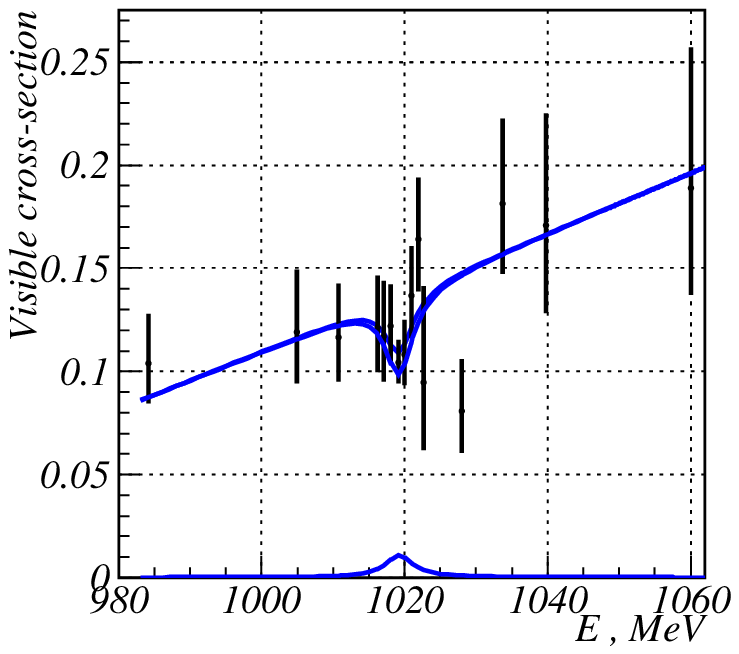}} 
\vspace*{-5mm}
\caption{Visible cross section of the process
$e^+e^- \to \omega \pi^0 \to \pi^0 \pi^0 \gamma $. 
The optimal fits of visible cross section  
and resonance background are shown.}
\label{f4}
\end{minipage}
\end{figure}

  In the kinematic fit the requirement was imposed, that two $\pi^0$
are found in 5 photon event. The corresponding parameter
$\chi^2_{\pi^0\pi^0\gamma}$ for experimental events and simulation
of the processes (\ref{omp0n}) and  (\ref{etagn}) is shown at the
Fig.~\ref{f1}. The cut  $\chi^2_{\pi^0\pi^0\gamma} < 40$ was
imposed for further analysis. Then the  $\pi^0$ recoil
mass was chosen  $m_{\omega}$, closest to  $\omega$ mass
(Fig.~\ref{f2}). The cut for this parameter $m_{\omega}-782| < 60$
was used later. The additional suppression of processes
(\ref{kskln}) and (\ref{etagn}) was obtained with the parameter
$\zeta$, describing the transverse profile of the electromagnetic
shower (Fig.~\ref{f3}).  For events selection the cut  $\zeta < 25$
was applied.

{\bf Analysis of data.}
The events selected by the criteria, described above, were
divided into 7 following classes:
$$
\begin{array}{llll}
1) & \chi^2_{\pi^0\pi^0\gamma} <20,   & |m_{\omega }-782|<30,
& \zeta < -5; \\
2) & \chi^2_{\pi^0\pi^0\gamma} <20,   & 30<|m_{\omega }-782|<60,
& \zeta < -5; \\
3) & 20<\chi^2_{\pi^0\pi^0\gamma} <40,& |m_{\omega }-782|<30,
& \zeta < -5; \\
4) & 20<\chi^2_{\pi^0\pi^0\gamma} <40,& 30<|m_{\omega }-782|<60,
& \zeta < -5; \\
5) & \chi^2_{\pi^0\pi^0\gamma} <40,   & |m_{\omega }-782|<60,
& -5<\zeta<5; \\
6) & \chi^2_{\pi^0\pi^0\gamma} <20,   & |m_{\omega }-782|<30,
& 5<\zeta<25; \\
7) & \chi^2_{\pi^0\pi^0\gamma} <20,   & |m_{\omega }-782|<30,
& N_\gamma=6.
\end{array}
$$
   The 6 photon events were put into the last class. It was done to
investigate, for example, a possible background from beams, where
sixth photon is superimposed on the event. In the fit such a photon
was supposed to be spare. In the Table~\ref{tb1} there are shown the 
probabilities for the process under study to be found in classes,
described above.
\begin{table*}[htb]
\small
\caption
{The probabilities $\alpha_i$  for events of (\ref{omp0n}) to be found
in $i$-class, the ratio $\beta_i$ of non-resonant cross section in $i$-class
to the $e^+e^-\to\phi\to\eta\gamma$ process cross section,
obtained by simulation and from the cross section fit.}
\label{tb1}
\begin{tabular}{ccccc}
\hline
Class &
$\alpha_i,exp$              &  $\alpha_i,MC$   &
$\beta_i,MC(\eta\gamma, f_0\gamma, K_{S}K_{L})\cdot 10^3$ &
$\beta_i,exp\cdot 10^3$     \\
\hline
1 & 0.591$\pm$ 0.031 & 0.633 & 0.18+0.20+0.00=0.38 & 0.25         \\
2 & 0.061$\pm$ 0.013 & 0.067 & 0.20+0.18+0.00=0.38 & 0.23$\pm$ 0.11 \\
3 & 0.069$\pm$ 0.015 & 0.049 & 0.43+0.02+0.00=0.45 & 0.49$\pm$ 0.12 \\
4 & 0.020$\pm$ 0.009 & 0.012 & 0.37+0.02+0.00=0.39 & 0.36$\pm$ 0.08 \\
5 & 0.196$\pm$ 0.023 & 0.202 & 1.33+0.11+0.09=1.53 & 1.74$\pm$ 0.23 \\
6 & 0.042$\pm$ 0.012 & 0.033 & 0.30+0.01+0.00=0.31 & 0.40$\pm$ 0.10 \\
7 & 0.021$\pm$ 0.012 & 0.006 &                 --- & 1.87$\pm$ 0.16 \\
\hline
\end{tabular}
\end{table*}
    The probabilities for events of the process under study to be
found  in classes 1--7 are shown in Table~\ref{tb1} as well as
the background resonant cross sections, normalized by the
$e^+e^-\to\phi\to\eta\gamma$ cross section. The process
(\ref{p0p0g}) was simulated for  $f_0\gamma$ intermediate state
with the branching fraction  of $0.9\cdot 10^{-4}$.
The simulation shows that the signal to noise ratio (S/N)
in  class 1 is the largest, which was used later for investigation
of the process (\ref{omp0n}). The contribution of  the process (\ref{omp0n})
in the class 2 is about 10 times lower, which allows
to extract from the data the resonant background from the processes
(\ref{p0p0g}), (\ref{kskln}), (\ref{etagn}). Using ratio of the
resonant background in classes 6 and 7 obtained from the simulation,
one can estimate the background in class 1. The ratio of the
background processes in the regions $|m_{\omega}-782|<30$ and
$30<|m_{\omega}-782|<60$ weakly depends on the cuts in
$\chi^2_{\pi^0\pi^0\gamma}$ and  $\zeta$ parameters. For the
process (\ref{p0p0g}) this ratio varies from
1.09 to 1.12, while for the process (\ref{etagn}) --- from 0.9
to 1.07. But in the latter case this range is determined by low
simulation statistics of the process  (\ref{etagn}).
The coefficient $1.1\pm 0.2$ was taken for background recalculation
from class 2 to class 1. The classes 3~--~7 were used for estimation
of systematic errors, depending on different selection cuts. 

       The visible cross section in each class was presented in the
following form:       
$$\sigma_{vis} = \alpha_i\sigma_{\omega\pi}+
\beta_i\sigma_{\phi\to\eta\gamma},$$
$$
\sigma_{\omega\pi} = \varepsilon 
(\sigma_0+A(E-m_\phi))\cdot
\biggl |1-Z \frac{m_\phi\Gamma_\phi}
{D_\phi}\biggr |^2(1+\delta),
$$
where $\sigma_0$ -- non-resonant cross section of the process
$e^+e^-\to\omega\pi^0\to\pi^0\pi^0\gamma$
at $E=m_\phi$, $A$ -- the slope of the cross section,
$\varepsilon$ -- detection efficiency
of the process  (\ref{omp0n}),
$Z$ -- complex interference amplitude,
$m_\phi,\Gamma_\phi$ -- mass and width of $\phi$ meson,
$D_\phi=m_\phi^2-E^2-iE\Gamma_\phi(E)$ -- $\phi$-meson propagator,
$B(\omega\to\pi^0\gamma)=0.085$ -- the decay
$\omega\to\pi^0\gamma$ branching ratio \cite{PDG},
$\delta$ -- radiative correction \cite{RadCor},
$\sigma_{\phi\to\eta\gamma}$ -- the cross section of the process
$e^+e^-\to\phi\to\eta\gamma$. The parameters
$\alpha_i$ mean the probability for the process (\ref{omp0n})
to be found in the class $i$, the parameters  $\beta_i$ --
non-resonant cross section in the class $i$, normalized to the process
$e^+e^-\to\phi\to\eta\gamma$ cross section. The $\phi$ meson
excitation curve was described by the main background process
$e^+e^-\to\phi\to\eta\gamma$. The differences in excitation curves
for different processes (\ref{kskln}), (\ref{etagn}),(\ref{p0p0g})
were neglected at the present level of accuracy. The fitting
of the visible cross section was done for all four classes.
The parameters
$\sigma_0$, $A$, $Re(Z)$, $Im(Z)$, $\alpha_i$, $\beta_i$ except
$\beta_1$ и $\alpha_1$ were set free. $\beta_1$ was found from
the expression $\beta_1=(1.1\pm 0.2)\cdot\beta_2$,
$\alpha_1$ from normalization $\sum\alpha_i=1$.
The detection efficiency $\varepsilon$=$39\%$,
obtained from the simulation, does not depend on the energy.
The coefficients  $\alpha_i$ are also energy independent.
The total number of fit parameters is 16. In each class the cross
section was measured in 15 points. The following values of fit
parameters were obtained ($\chi^2/d.f.=63/89$):
\begin{eqnarray}
\label{vdres1}
&\sigma_0=(0.65\pm0.04)~nb,&\nonumber \\
&A=(0.0065\pm0.0018)~nb/MeV,&\\
&Re(Z)=0.036\pm0.052,&\nonumber\\
&Im(Z)=-0.186\pm0.063.&\nonumber
\end{eqnarray}
    One can see from the Table~\ref{tb1}  rather good agreement
between experimental and simulated parameters  $\alpha_i$,
$\beta_i$. This allowed to put the total  systematic error to 6\%
for $\sigma_0$.  For other parameters  in  (\ref{vdres1})
the statistical error is much higher than  the systematic one.
The visible cross section for class 1 and fitting curve with
$\chi^2/d.f.=11.9/11$ are shown in Fig.~\ref{f4}.
Fitting curves for the visible cross section of the process  (\ref{omp0n})
and for non-resonant background
are also shown there. One could see,
that in spite of imposed strong cuts  in class 1, the resonant background
is about one third of interference amplitude wave. It remains the dominant
source of systematic error in  $Z$.

{\bf Conclusions.}
   The measured non-resonant cross section of the process
$e^+e^-\to\omega\pi^0$ $(7.7\pm0.5\pm0.5)~nb$
agrees with the result $(8.7\pm 1.0\pm 0.7)~nb$ from  \cite{Ph.Rep}
and with the result from \cite{Prep.97} in the decay channel with charged pions
$e^+e^-\to\omega\pi^0\to\pi^+\pi^-\pi^0\pi^0$. The measured interference
amplitude is three standard deviations above zero. The measured in \cite{Prep.97}
interference amplitude in the $\phi\to\omega\pi^0$ decay is
$Re(Z)=108\pm0.026$, $Im(Z)=-0.127\pm0.027$. Our calculation of the
interference amplitude for the decay $\phi\to\rho\pi^0\to\pi^0\pi^0\gamma$
gives $Re(Z)=-0.079$, $Im(Z)=-0.053$. The sum of these contributions
$Re(Z)=0.029$, $Im(Z)=-0.180$ agrees with our measurement. We did not
make a theoretical estimations of the contributions of the processes
$\phi\to f_0\gamma,~\varepsilon\gamma\to\pi^0\pi^0\gamma$ into
interference amplitude. Finally we list parameters
of the $e^+e^-\to\omega\pi^0\to\pi^0\pi^0\gamma$
cross section, obtained in the present work: 
\begin{eqnarray}
&\sigma_0=(0.65\pm0.04\pm0.04)~nb,&\nonumber \\
&A=(0.0065\pm0.0018)~nb/MeV,&\nonumber \\
&Re(Z)=0.036\pm0.052,&\nonumber\\
&Im(Z)=-0.186\pm0.063.&\nonumber
\end{eqnarray}

\subsection{$K_S \to 3\pi^0$ decay search}

This decay is a pure CP-violating one,  similar to those of
$K_L \to \pi\pi$ and
unlike $K_S \to \pi^+ \pi^- \pi^0$ mode, where C-conserving amplitude
could be present.
Considering the model, where CP-violation originates from the mixing
of CP-even and CP-odd states $K_1^0$ and $K_2^0$,  
it leads to physical state $K_S \simeq K_1^0 + \varepsilon K_2^0$,
and the branching of $K_S$ depending on 
$\varepsilon$:
\( B(K_S \to 3\pi^0) \simeq |\varepsilon |^2\cdot
\frac{\tau_S}{\tau_L} B(K_L \to 3\pi^0) \simeq 10^{-9}. \)
Very small magnitude of the branching ratio and 
experimental problems with
separation between this decay and background from the process
$K_S \to 2\pi^0$, which has a 9 orders of magnitude higher probability,
result in rather high existing experimental upper limit:
\( B(K_S \to 3\pi^0) < 3.7\cdot 10^{-5} \).

{\bf Analysis.}
  In the present work the $\phi \to K_SK_L$ decay
with a total $N_{\phi}=7.5\cdot10^6$ is used as a source of $K_S$.
Our analysis is applied to those events, where
$K_L$ leaves the detector without either interaction or decay.
$10^4$ simulated events of $K_S\to 3\pi^0$ and
$10^4$ simulated events of $K_S\to 2\pi^0$ were
processed in the same way.
Because the initial decay $\phi\to K_S K_L$ was not simulated,
the calculations take into account the branching ratio
$B(\phi\to K_S K_L) = 0.343\pm0.007$ \cite{PDG} and the
probability $W_{K_L}$ for $K_L$ to leave the detector undetected.
The latter value cannot be obtained from simulation due to its low accuracy
in nuclear interactions of $K_L$ mesons, so it was
estimated to be $18\pm 1\%$ from the experimental data on 
$\phi\to K_S K_L, K_S\to 2\pi^0$ decay.

  The following criteria were applied to the experimental
data and simulation sample for $K_S\to 3\pi^0$ decay search:
\\$\bullet$
$N_\gamma \geq 6$;
$N_{cp}=0$;
positive neutral trigger;
$0.38<E_{tot}/2E_0<0.5$;
\\$\bullet$
$0.15<P/E_{tot}<0.3$;
$E_{\gamma min}>30~MeV$;
$\zeta<10$.
\\The hits in the first two layers of the calorimeter
are required.
The kinematic fit was done for the selected events,
in assumption of \mbox{$K_S\to 3\pi^0$} kinematics.
The output parameters within the limits shown below 
were the following:
$\chi^2_{K_S\to 3\pi^0} < 15$ for the found combination satisfying
$K_S\to 3\pi^0\to 6\gamma$ decay hypothesis;
minimal angle between photons of the $K_S$ decay
and beam axis $\alpha_{K_S\to 3\pi^0} > 30^\circ$;
polar angle of reconstructed $K_S$ 
$30<\vartheta_{K_S\to 3\pi^0}<150$;
raw invariant masses of three $\pi^0$'s:
$110<m_{\pi^0_{1,2,3}, K_S\to 3\pi^0}<160~MeV$.

For the selection of the process
$K_S\to 2\pi^0$ the following criteria were applied:
\\$\bullet$
$N_{\gamma}=4$;
positive neutral trigger;
$0.38<E_{tot}/2E_0<1.1$;
$P/E_{tot}<0.6$;
\\$\bullet$
$E_{\gamma min}>20~MeV$;
$E_{\gamma max}<300~MeV$.
\\ The cut
$0.05<E_{tot}/2E_0-P/E_{tot}<0.4$ selects a
slanting stripe in the
($E_{tot}/2E_0$, $P/E_{tot}$) plane, containing most of the
$\phi\to K_S K_L$ events with tagged $K_L$ (Fig.~\ref{dbfig01}).
The kinematic fit has been carried out also
in assumption of the $K_S\to 2\pi^0$ event kinematics.
The selection criteria after the kinematic fit were the following:
$\chi^2_{K_S\to 2\pi^0}<25$ for the found combination of
photons satisfying $K_S\to 2\pi^0\to 4\gamma$ decay hypothesis;
minimal angle between photons of the
$K_S$ decay and the beam axis
$\alpha_{K_S\to 2\pi^0}>27.5^\circ$;
the raw invariant masses of the two $\pi^0$ mesons
$110<m_{\pi^0_{1,2},K_S\to 2\pi^0}<160$;
the reconstructed $K_S$ momentum $50<P_{K_S\to 2\pi^0}<200$;
the reconstructed impact parameter
$-2.5<Z_{K_S\to 2\pi^0}<2.5$.

{\bf The results.} 
One candidate event of the  \mbox{$K_S\to 3\pi^0$} decay was found. 
The detection efficiency of this  process is 
$\varepsilon_{K_S\to 3\pi^0} \simeq 14\%$.
The detection efficiency  of the process \mbox{$K_S\to 2\pi^0$} 
is  $\varepsilon_{K_S\to 2\pi^0} \simeq 46\%$. 
The total statistics of found events of this process
after all cuts is 6869 events.

Using the formulae
\vspace{-3mm}
\begin{equation}
  N_{K_S\to 2\pi^0} = N_{\phi 2}\cdot B(\phi\to K_S K_L)\cdot
                      W_{K_L}\cdot B(K_S\to 2\pi^0)\cdot
		      \varepsilon_{K_S\to 2\pi^0},
\end{equation}
\vspace{-3mm}
\begin{equation}
  N_{K_S\to 3\pi^0} = N_{\phi 3}\cdot B(\phi\to K_S K_L)\cdot
                      W_{K_L}\cdot B(K_S\to 3\pi^0)\cdot
		      \varepsilon_{K_S\to 3\pi^0},
\end{equation}
\vspace{-3mm}
where $N_{\phi 2}, N_{\phi 3}$ are the numbers of produced $\phi$ mesons,
we found:
$$
  B(K_S\to 3\pi^0) = 1.8\cdot 10^{-5}
$$
the corresponding upper limit was evaluated to be
$$
B(K_S\to 3\pi^0) < 6.9\cdot 10^{-5},~~ C.L.~ 90\%.
$$
Our result is  close to the table limit of $<3.7\cdot 10^{-5}$ \cite{PDG}.

\subsection{ Study of  $ e^+ e^- \to  e^+ e^- \gamma $ and
          $ e^+ e^- \to  e^+ e^- \gamma \gamma$  processes} 

The process  
\(e^+e^- \to e^+e^- \gamma\)
(\ref{eeg}) is a QED process of the order of $\alpha^3$.
Its study is interesting not only for  QED testing,  
but also because it is important as an integrated luminosity check.
It is also a source of background in searches for rare
hadron processes.
To compare experimental results  with the QED the simulation was used
from the work \cite{TDUM}.

The process
\(e^+e^- \to e^+e^- \gamma \gamma\)
(\ref{eegg}) is a QED process of the 4-th order in $ \alpha$. 
Its investigation is also useful as a QED test, especially
because it was studied in only one experiment \cite{Ph.Rep}
with  ND detector,  where 223 events were seen.
The measurement  of its cross-section is important for study of
$\phi \to \eta e^+e^-$ и $ \eta \to e^+e^- \gamma$ decays,
where this process is a main source of background.

The study of the processes is based
 on the data sample with a total integrated luminosity of $ 4.3 pb^{-1}$ in
the energy range  $985 \div 1040$ MeV, while the production cross section
measurements are based on the data subset, corresponding
to $\sim 1 pb^{-1}$ integrated luminosity.

\begin{figure}[p]
\epsfxsize=\textwidth
\epsfbox{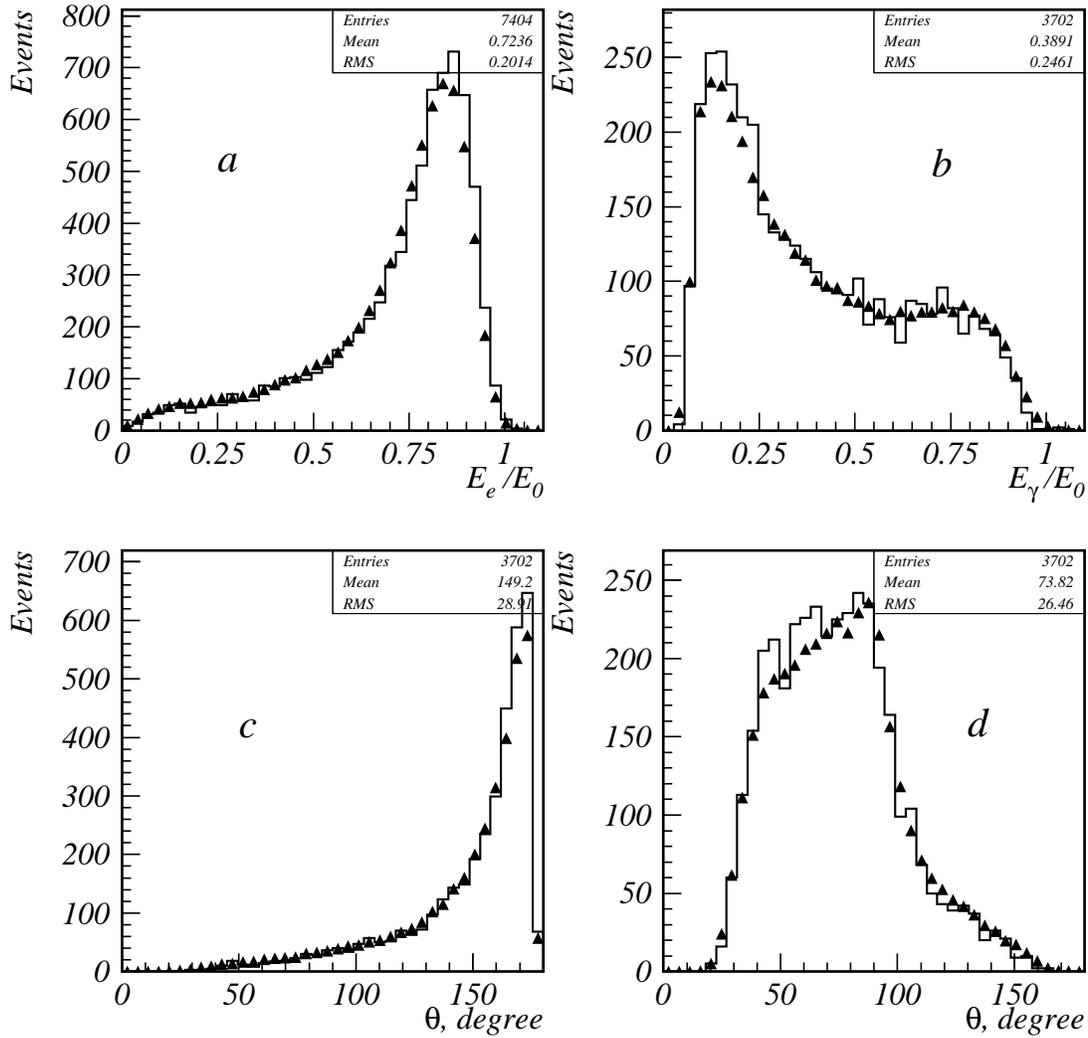} 
\caption{
Comparison of experimental data and simulation for 
the process $e^+e^- \to e^+e^- \gamma$:
a) normalized energy of the charged particle;
b) normalized photon energy;
c) angle between charged particles;
d) angle between photon and nearest charged particle.
Triangles -- experimental data, 
histogram -- simulation.}
\label{TDpic1}
\end{figure}

{\bf Analysis of  $e^+ e^- \to e^+ e^- \gamma$ (\ref{eeg}).}
 To study the process
(\ref{eeg})
the following selection criteria were used:
\\$\bullet$ positive charged trigger;
$N_{cp}=2$; $1 \leq N_{\gamma} \leq 3$;
$ R_1,~R_2<0.5 cm $; $ |Z_1|, |Z_2|<10 cm $;
$\theta _{min} > 36^\circ$;  $|\Delta \phi _{ee}| > 5^\circ$;
\\$\bullet$
no hits in muon system;
$ E_{tot}/2E_0>0.8 $, $P/E_{tot}<0.15$;
$\chi^2_E < 15$.\\
Extra photons were allowed to prevent the
events loss due to accidental signals in calorimeter
due to beam background.
For events passing these selection criteria, the 
experimental spectra and QED simulation \cite{TDUM} are
shown in Fig.~\ref{TDpic1}.
The events number in simulation  corresponds to  $190~nb^{-1}$
integrated luminosity. The cross section values were obtained 
separately for each scan. Resulting values are
lieing between 15.5 и 17.5~nb, giving  an average cross section 
$\sigma = 16.7 \pm 0.1$~nb , with $ \chi^2 = 24/5$d.f..  
After that all scans were processed again 
with the same configuration of trigger,
the event losses due to trigger inefficiency was taken into account and
luminosity based on $e^+e^-$ events was used.
As a result the difference of measured cross sections in different scans was
reduced to  17.1~$\div$~17.9~nb, that is less
than 5\% .
The average cross section was found to be $\sigma = 17.6 \pm 0.1$~nb ,
with $ \chi^2 = 7/5 $.
But the simulated cross-section was significantly higher
$\sigma_{sim.} =19.5 \pm 0.3~nb$,
so the difference between the experiment and simulation
was  $ \sim 10\% $.
It was assumed  that this difference appears because of different
angle $\theta$ cuts in our search and in luminosity calculations
(in the first case angle is determined using drift
chamber, in the second -- using
calorimeter). To study this problem in two scans
PHI\_9603 and PHI\_9604 the collinear events  were processed. 
The  ``$e^+e^-$'' events were selected with the same criteria
on  $\theta$,
$ E_{tot}/2E_0$, $P/E_{tot}$ as events ``$e^+e^-\gamma$''.
Then the value  of $ N_{ee\gamma} / N_{ee} $ was calculated for experiment and 
simulation. The results are shown below:
 $$ N_{ee\gamma} / N_{ee} (\mbox{PHI\_9603}) =   0.0138 \pm 0.0002, $$
 $$ N_{ee\gamma} / N_{ee} (\mbox{PHI\_9604}) =   0.0141 \pm 0.0001, $$
 $$ N_{ee\gamma} / N_{ee} ( Simulation ) =  0.0142 \pm 0.0003.  $$
One can see that in this case the results in both scans 
are in good agreement with each other and with simulation.
For 2 scans the integrated luminosities were recalculated and
the visible cross section was obtained:
$ \sigma_{exp.} = 19.0 \pm 0.2~nb $,
$\sigma_{sim.} = 19.5 \pm 0.3~nb $. Taking into account the detection
efficiency $80.5\%$, we obtained  the cross
section in the polar angle larger than  $36^\circ$ and acollinearity angle
larger than $5^\circ$:
$ \sigma_{exp.} = 23.6 \pm 0.3~nb $, $\sigma_{sim.} = 24.2 \pm 0.4~nb$.
The systematic error is determined  mainly  by the error in the integrated
luminosity and is equal to $5\%$.
The simulation and experiment are in good agreement
and the statistical error is at a level of $1\%$.
So, to increase the accuracy it is necessary to reduce the
systematic error.
 
{\bf Analysis of the process  
$e^+ e^- \to e^+ e^- \gamma\gamma (\ref{eegg})$.}
To study the process
(\ref{eegg})
the following selection criteria were used:
\\$\bullet$
charged trigger
$N_{cp}=2$; $2 \leq N_{\gamma} \leq 3$;
$ R_1,~R_2<0.5~cm $; $ |Z_1|, ~|Z_2|<10~cm $;
$\theta _{min} > 36^\circ$; $|\Delta \phi _{ee}| > 5^\circ$;
\\$\bullet$
no hits in muon system;  
$ E_{tot}/2E_0>0.8 $, $P/E_{tot}<0.15$;
\\$\bullet$
to cut events with  $\pi^0$ decays,
the invariant mass of two photons $m_{\gamma \gamma}$ was required to be
outside the interval from 110 to 170~MeV;
$E_{\gamma ~min}) > 50~MeV$; 
$\chi^2_E < 15 $.

\begin{figure}[p]
\epsfxsize=\textwidth
\epsfbox{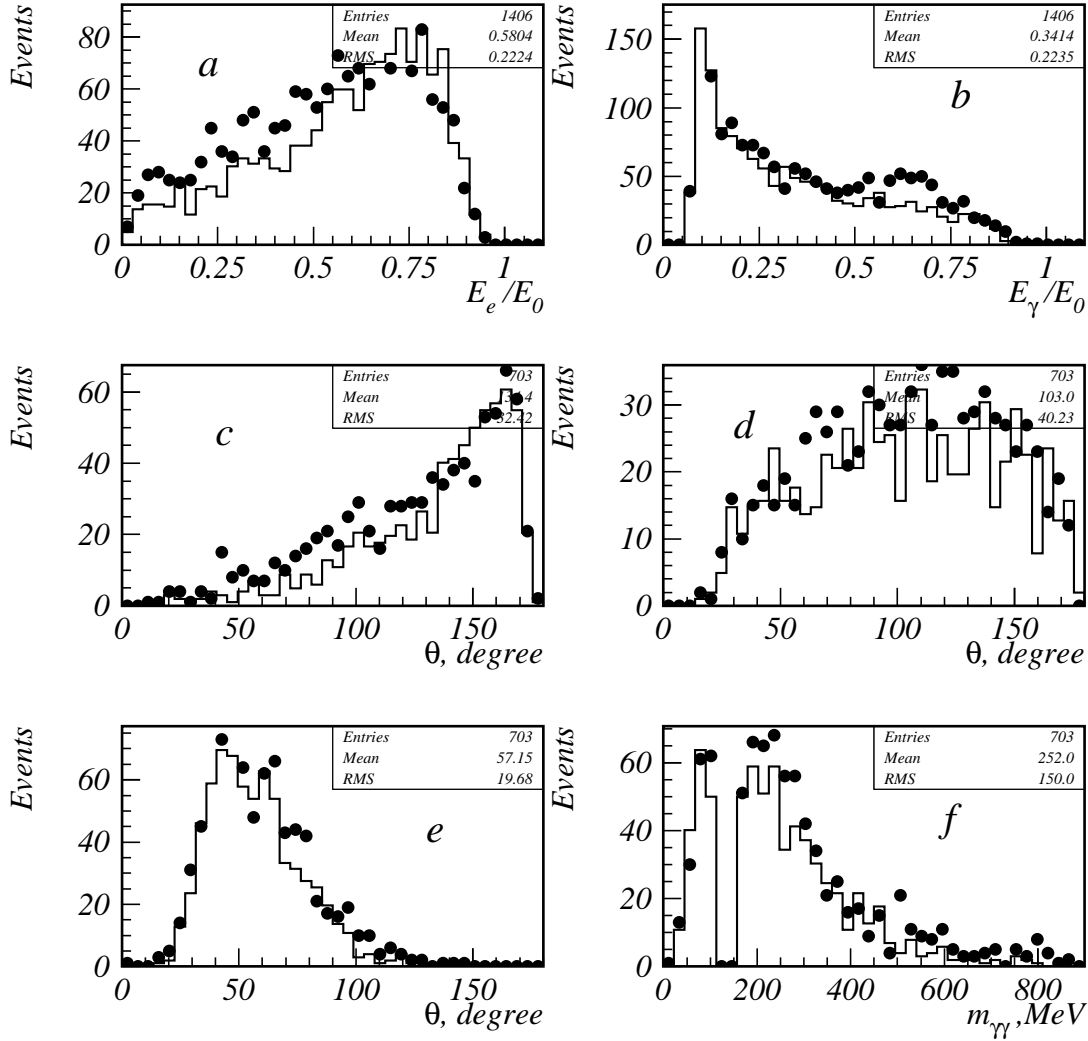} 
\caption{ 
Comparison of experimental data and simulation of the process
$e^+e^- \to e^+e^- \gamma \gamma$:
a) normalized charged particle energy;
b) normalized photon energy;
c) angle between charged particles;
d) angle between photons;
e) minimal angle between photon and charged particle;
f) invariant mass of photon pair.
Circles -- experimental data, histogram -- simulation.}
\label{TDpic2}
\end{figure}

For selected events the spectra are shown in Fig.\ref{TDpic2}
together with simulation\cite{TDEEGG}.
The number of events in simulation corresponds to integrated luminosity of
$4297~nb^{-1}$.
The main background for this process comes from the  reaction 
$e^+e^- \to e^+e^-\gamma$ (\ref{eeg})
with additional soft photon from the beam background.
To suppress these events the threshold in minimum photon energy was placed at
50~MeV.
For this process the ratio $ N_{ee\gamma\gamma} / N_{ee}$ 
in separate experimental runs and simulation was obtained:
 $$ N_{ee\gamma\gamma}/N_{ee}(\mbox{PHI\_9603})=
 (0.10 \pm 0.01)\cdot 10^{-3},$$
 $$ N_{ee\gamma\gamma}/N_{ee}(\mbox{PHI\_9604})=
 (0.12 \pm 0.01)\cdot 10^{-3},$$
 $$ N_{ee\gamma\gamma}/N_{ee}(Simulation)=(0.10 \pm 0.01)\cdot 10^{-3}.$$
Using corrected luminosity the visible cross section
was obtained:
$ \sigma_{exp.} = 0.14 \pm 0.03~nb, \sigma_{sim}=0.135\pm 0.006~nb$. 
Taking into account the detection efficiency $ 32.6 \% $, the 
cross section for polar angle greater than $36^\circ$ and
acollinearity angle larger than $5^\circ$ was obtained:
$ \sigma_{exp.} = 0.43 \pm 0.09~nb $,
and $ \sigma_{sim.} = 0.42 \pm 0.02~nb $.
The systematic error like in previous analysis is determined mainly
by error in luminosity and is equal  approximately to  $5\%$.
As one can see, the simulation and experiment are in good agreement,
but the statistical error in experiment is at a level of $20\%$.
So it is possible to improve accuracy with more statistics.
The energy dependence of cross section (\ref{eeg}) and (\ref{eegg}) 
is $1/E^2$  in agreement with QED predictions.

\subsection{Study of QED process  $e^+e^- \to 3 \gamma $} 

Earlier the three photon annihilation process was studied
in high energy region $2E_0\sim 10~GeV$ \cite{3g1,3g2}. 
In our previous experiment with  ND detector
this process was studied  as a QED background \cite{Ph.Rep}.
Investigation of (\ref{QEDggg}) process at low energy near
$\phi$-meson peak has an advantage, because its 
cross-section energy dependence $\sim\alpha^3/E_0^2$ provides us with
high statistics. This gives us a possibility for more detailed 
QED test in the third order of perturbation theory. Besides, the 
\(e^+e^- \to 3 \gamma\) process is important as a background source 
for $\phi$-meson radiative decays 
\(e^+e^- \to \phi \to \eta\gamma \to 3\gamma\) 
(\ref{etag3g}) and 
\(e^+e^- \to \phi \to \pi^0\gamma \to 3\gamma\)
(\ref{p0g3g}).

\begin{figure}[htb]
\epsfxsize=0.49\textwidth

\begin{minipage}[t]{0.47\textwidth}
  \centerline{\epsfbox{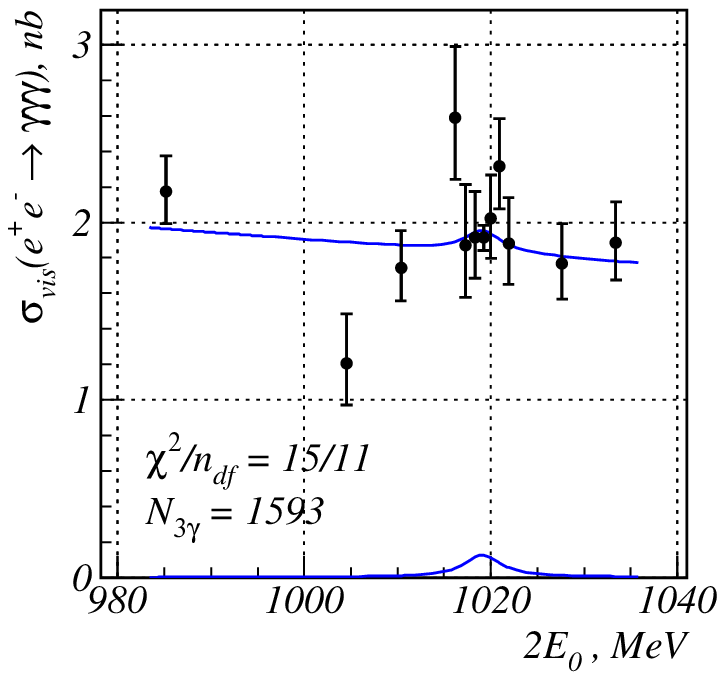}} 
\vspace*{-5mm}
\caption{Visible cross section of the process
$e^+e^- \to \gamma\gamma\gamma$. 
Upper curve -- optimal fit. Lower curve -- resonance background.}
\label{bavcrs}
\end{minipage}
\hfill
\begin{minipage}[t]{0.47\textwidth}
  \centerline{\epsfbox{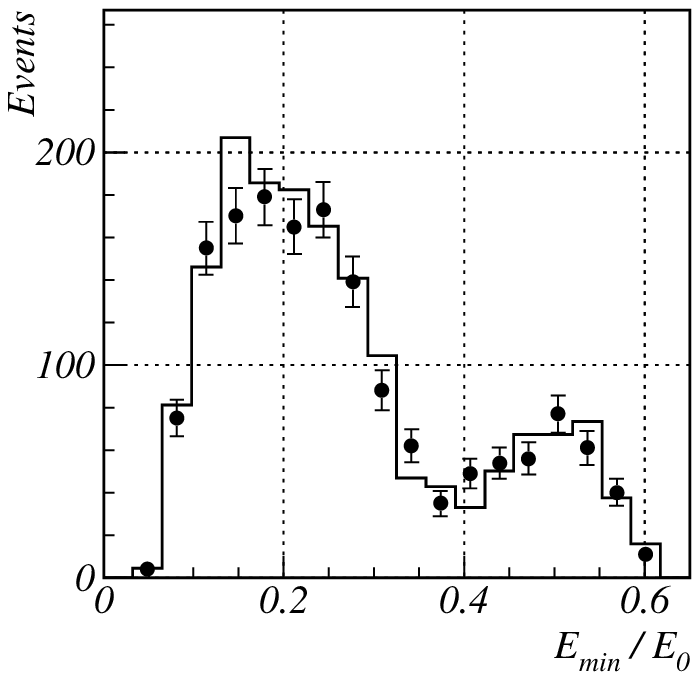}} 
\vspace*{-5mm}
\caption{Distribution of normalized minimal photon energy
in the process
$e^+e^- \to \gamma\gamma\gamma$.
Circles with error bars -- experimental data;
histogram -- simulation.}
\label{bavemin}
\end{minipage}
\vfill
\begin{minipage}[t]{0.47\textwidth}
  \centerline{\epsfbox{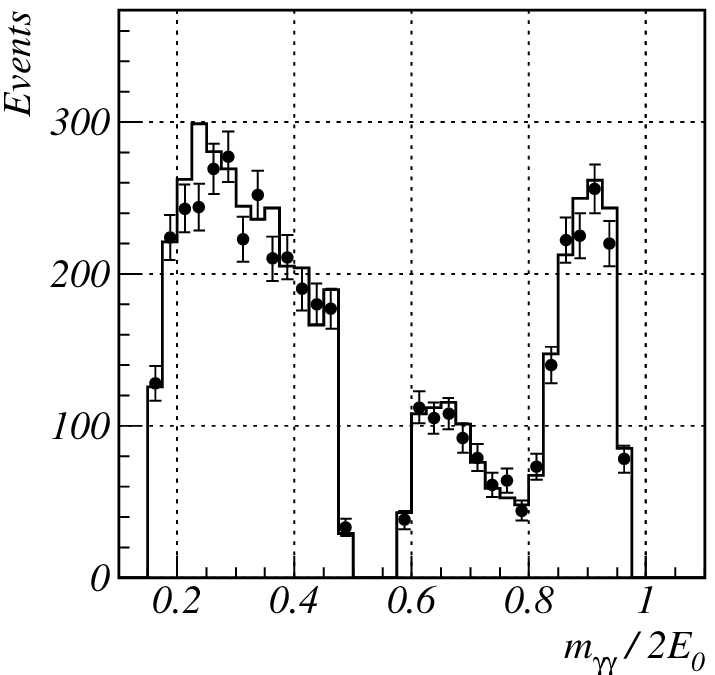}} 
\vspace*{-5mm}
\caption{Normalized spectrum of invariant mass of photon pairs
in the process $e^+e^- \to \gamma\gamma\gamma$, three entries per event.
Circles with error bars -- experimental data;
histogram -- simulation.}
\label{bavmgg}
\end{minipage}
\hfill
\begin{minipage}[t]{0.47\textwidth}
  \centerline{\epsfbox{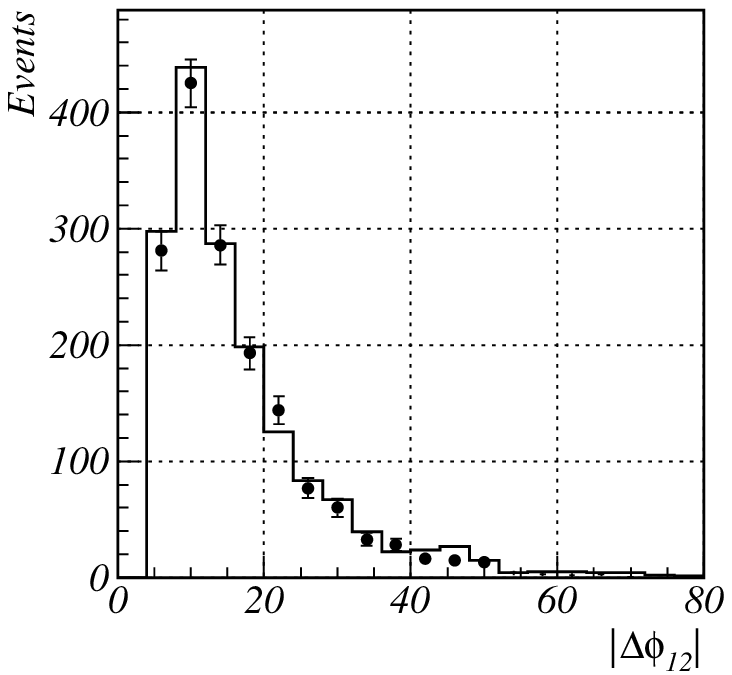}} 
\vspace*{-5mm}
\caption{Distribution of
azimuth acollinearity angle between 1-st and 2-nd photons
in the process $e^+e^- \to \gamma\gamma\gamma$.
Circles with error bars -- experimental data;
histogram -- simulation.}
\label{bavdphi}
\end{minipage}
\end{figure}

In the present study, the experimental data
of  PHI\_9603 scan were  analyzed. The total integrated luminosity is 
$\Delta L=822~nb^{-1}$.
The following selection cuts were applied:
$N_{\gamma}=3$,
$\theta_{min}>27^\circ$,
$0.7 < E_{tot}/2E_0 < 1.1$,
$\chi_E^2 < 40$.
The main background comes from the processes (\ref{etag3g}),
(\ref{p0g3g}) and 
\(e^+e^- \to \gamma \gamma\)
(\ref{2g}).
Taking into account presence of a quasi-monochromatic recoil photons
in radiative decays, we imposed the cut
$E_{\gamma , max}/E_0 < 0.974$ to suppress (\ref{p0g3g}) process.
To suppress  (\ref{etag3g}) process, the photons with energy
in the $0.65<E_{\gamma}/E_0<0.77$ interval were rejected.
The background from (\ref{2g}) process,
was significantly reduced by condition imposed
on acollinearity angle in the
azimuth direction between two most energetic photons
$|\Delta\phi_{12}|>6^\circ$.
The detection efficiency was obtained
by MC simulation, and for described selection criteria it equals
$(10.3\pm 0.3)\%$.
Fig.~\ref{bavcrs} shows the energy dependence of visible
cross-section of the $e^+e^- \to 3 \gamma $ process.
The value of the cross section at
$E_0=510~MeV$ was found to be:
$\sigma _{exp.} = (1.82 \pm 0.21 \pm 0.25)~nb$.
The first error is statistical, the second one -- systematic. The
simulation with the standard QED matrix element gives:
$\sigma _{sim.} = (2.01 \pm 0.05)~nb$.
The cross-section of background processes at $\phi$ peak
does not exceed $5\%$ of that of the process 
(\ref{QEDggg}). The value $\simeq 5\%$ is due to resonant background.
The detection efficiency for all background processes
(\ref{etag3g}), (\ref{p0g3g}), (\ref{2g}), obtained
from simulation was not higher than $0.25\%$. 

Figs.\ref{bavemin}, \ref{bavmgg}, \ref{bavdphi} show the softest photon
energy spectrum, distributions over photon pair invariant mass and
acollinearity angle in the azimuthal direction; points correspond to the
experimental data, histogram -- to the MC simulation. These
distributions show good agreement between experiment and simulation, 
confirming the validity of QED at our level of accuracy.

\section{Physical results from MHAD-97 experiment}

\subsection{Investigation of
\protect\( e^+e^-\to \pi ^+\pi ^-\pi ^0\pi ^0\protect \), 
\protect\( \pi ^+\pi ^-\pi ^+\pi ^-\protect \) reactions}

Processes of $e^+e^-$-annihilation into four pions attract attention
due to the following reasons. First, in the
$2E_0=1\div 2~GeV$ energy region these processes dominate and
determine the main part of the hadronic contributions into the anomalous
magnetic moment of muon and into QCD sum rules.
Second, the processes
\(e^+e^- \to \phi \to \pi^+ \pi^- \pi^+ \pi^- \)
(\ref{pppp}) and
\( e^+e^- \to \pi^+\pi^-\pi^0\pi^0\)
(\ref{ppp0p0})
are an important source of information for hadron  spectroscopy, 
in particular, for study  $\rho$-meson radial excitations. 

According to the existing data, in the $2E_0=1\div 2~GeV$
energy region there exist two radial excitations of
the $\rho$ meson: $\rho (1450)$ and $\rho (1700)$ \cite{PDG}. 
Determination of parameters of these states and
their interference with the $\rho$  meson, can be found,
for example, in work \cite{VSClegg1}, \cite{VSAchasov1}. 
A possible mixing of these excited $\rho$-states 
with the exotic ones (for example, 4-quark
states) is discussed in \cite{VSDonnachie2},\cite{VSDonnachie3}.
Recently some experimental evidence appeared in favor of 
$\rho _{x}(1300)$ state existence \cite{VSAston}, 
which possibly is not a conventional quark-antiquark 
meson \cite{VSDonnachie1}.

In the past these processes were studied at
VEPP-2M \cite{Ph.Rep,VSOLYA_1,VSOLYA_2}, DCI\cite{VSDM2_01}, 
ADONE\cite{VSYY2_1} $e^+e^-$ colliders.
Statistical accuracy achieved in these experiments 
is $ \sim 5\%$, with a systematic error of $\sim 15\% $,
and the discrepancy between different experiments is sometimes
as large as $\sim 20\%$ \cite{Ph.Rep, VSOLYA_1}. 
Therefore, measurements with smaller systematic errors are
needed to clarify  situation with multi-hadron production.

\subsubsection{ \protect\( e^+e^- \to \pi ^+\pi ^-\pi ^+\pi ^-
\protect \) process}

\begin{figure}[htb]
\epsfxsize=\textwidth
  \centerline{\epsfbox{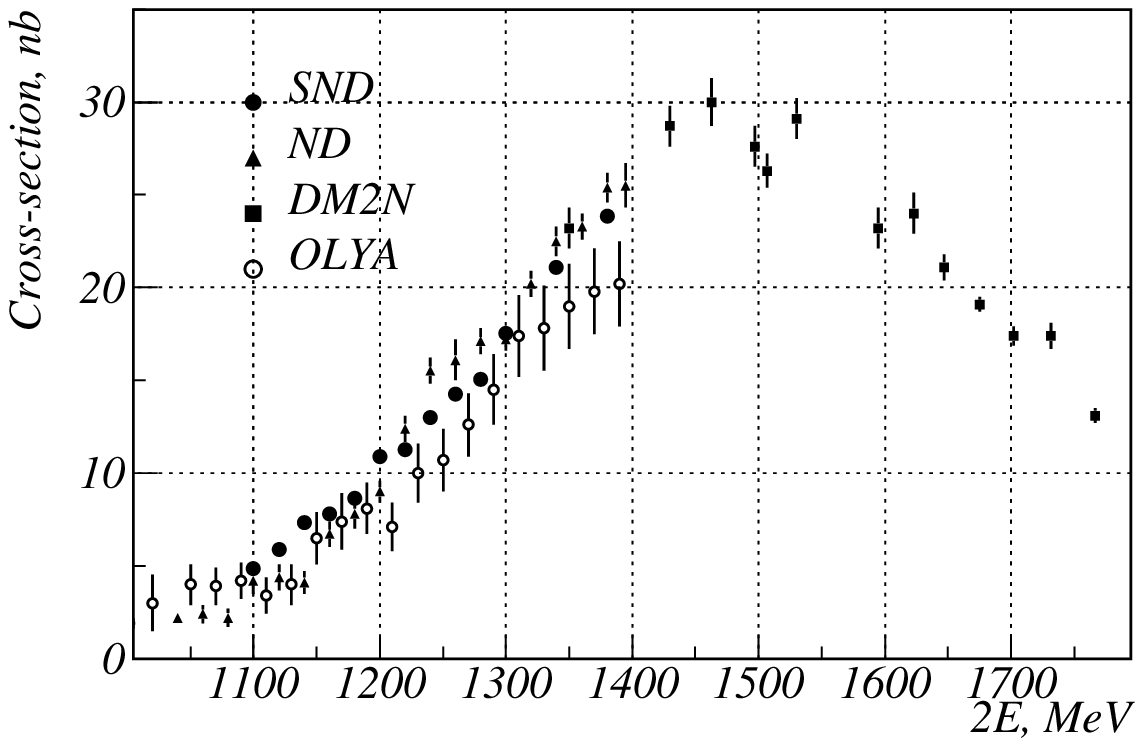}} 
\vspace*{-7mm}
\caption{
Experimental cross section of the process
$e^+e^-\to \pi ^+\pi ^-\pi ^+\pi ^-$.
Only statistical errors are shown.}
\label{sharycrs}
  \centerline{\epsfbox{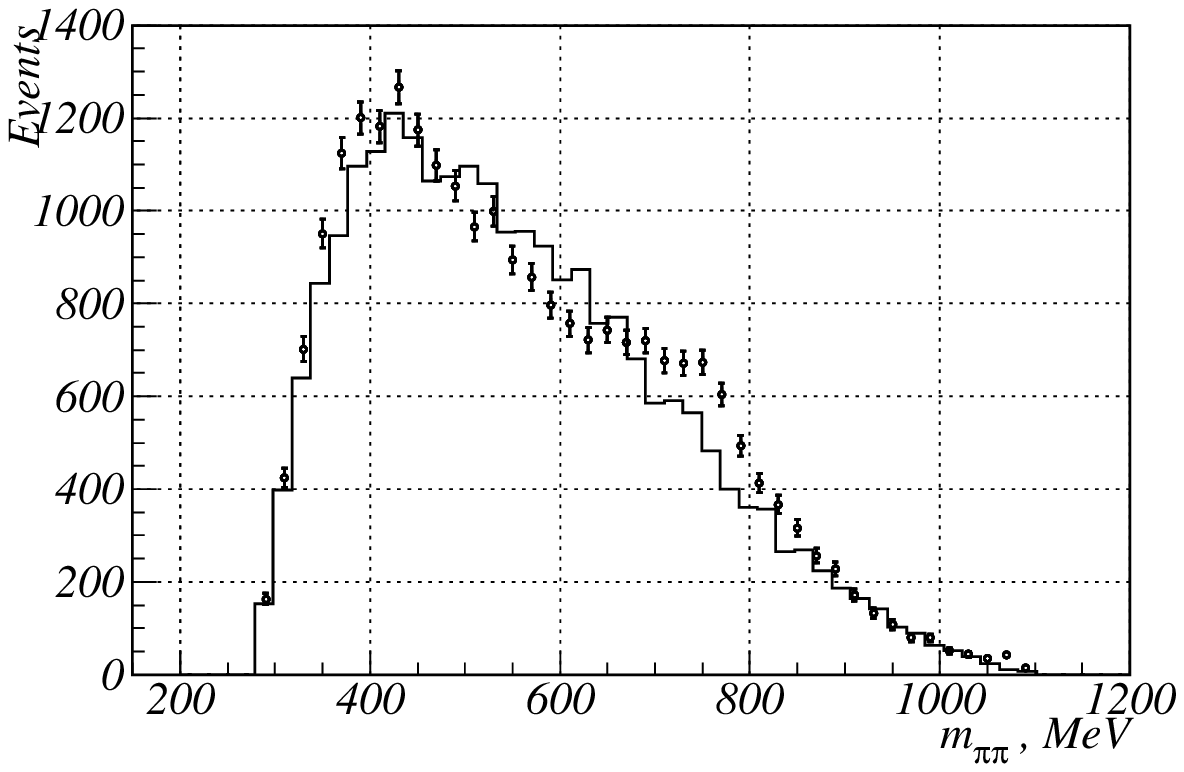}} 
\vspace*{-7mm}
\caption{Distribution of invariant mass of
$\pi^+\pi^-$ system in the process
$e^+e^- \to \pi ^+\pi ^-\pi ^+\pi ^-$.
Circles with error bars  -- experimental data; histogram -- LIPS simulation.}
\label{sharymp} 
\end{figure}

To select events of the process 
$e^+e^- \to \pi ^+\pi ^-\pi ^+\pi ^-$
(\ref{pppp})
the following selection criteria were applied:
\\$\bullet$  
$N_{cp} \geq 4$, $N_{\gamma} \geq 0 $,
$N_{wire} <30 $,
$E_{tot}/2E_0 < 0.8 $.
The energy dependence of visible cross section, obtained under
these conditions, is shown in Fig.~\ref{sharycrs} with error bars
indicating only statistical errors.
The following sources of systematic errors were taken into account: 
charged particles reconstruction errors, systematic
errors caused by the use of the selection cuts, i.g. cuts on the
recalculated trigger and the number of hit wires, inaccuracy in
luminosity determination. 
As a result, the systematic error was estimated as
$ \sim 15\%$ (preliminary).
The detection efficiency was determined from the Lorentz-invariant
phase space simulation (LIPS).
As it is seen from two-pion invariant mass distribution 
in Fig.~\ref{sharymp}, 
the experimental distribution
has a bump around 750 MeV, which is absent in LIPS simulation. 
This can be considered as a manifestation of $\rho$ meson 
in the four charged pions final state.
In future, simulation with the $\rho \pi \pi$ intermediate state will be
made to compare experimental data with the simulation. Refinement of the 
systematic error and the use of the whole MHAD9702 experiment statistics
is also planned.

\subsubsection{ \protect\( e^+e^-\to \pi ^+\pi ^-\pi ^0\pi ^0
\protect \) process }

The process (\ref{ppp0p0}) was studied in the final state 
with 2 charged particles and 4 photons.
In the energy range above $\phi$ meson this process dominates.
Background comes from $e^+e^- \to K^+K^- $ process,
from QED processes $ e^+e^- \to e^+e^-e^+e^-, e^+e^-\gamma \gamma $,
and also some background comes from cosmic rays and collider beams. 
To select events of the reaction (\ref{ppp0p0}), the following criteria were
applied:
\\ $\bullet$
$N_{cp} = 2$;
$N_{\gamma } \geq 4 $;
$E_{np}/2E_0 > 0.3$;
$R_1, R_2 <1~cm$;
$Z_1, Z_2 <10~cm$;
$|Z_1-Z_2| <5~cm$;
The calculated trigger corresponds to the experimental trigger for charged
particles with refined thresholds. 32195 events survived the cuts.
\\ $\bullet$
For energies $E_{np}<550~MeV$ a background from the
$ e^+e^-\to K^+K^- $ process is significant.
To suppress this background, additional cuts were used, 
related to ionization losses in drift chambers and acollinearity angle:
$\alpha _{\pi^+\pi^-}<160^\circ$,
$dE/dx_{\pi^{\pm}} + dE/dx_{\pi^{\mp}} < 5\cdot dE/dx_{m.i.p.}$.
\\ $\bullet$
The kinematic fit was made under assumption, that charged
particles are $ \pi $ mesons and 4 photons originate from two $ \pi^0$.
To suppress the remaining background, the following cuts were applied:
$\chi^2_E < 50$,
$\chi^2_{\pi^+\pi^-\pi^0\pi^0} < 50$,
$N_{\pi^0} = 2$.

\begin{figure}[htb]
\epsfxsize=\textwidth
  \centerline{\epsfbox{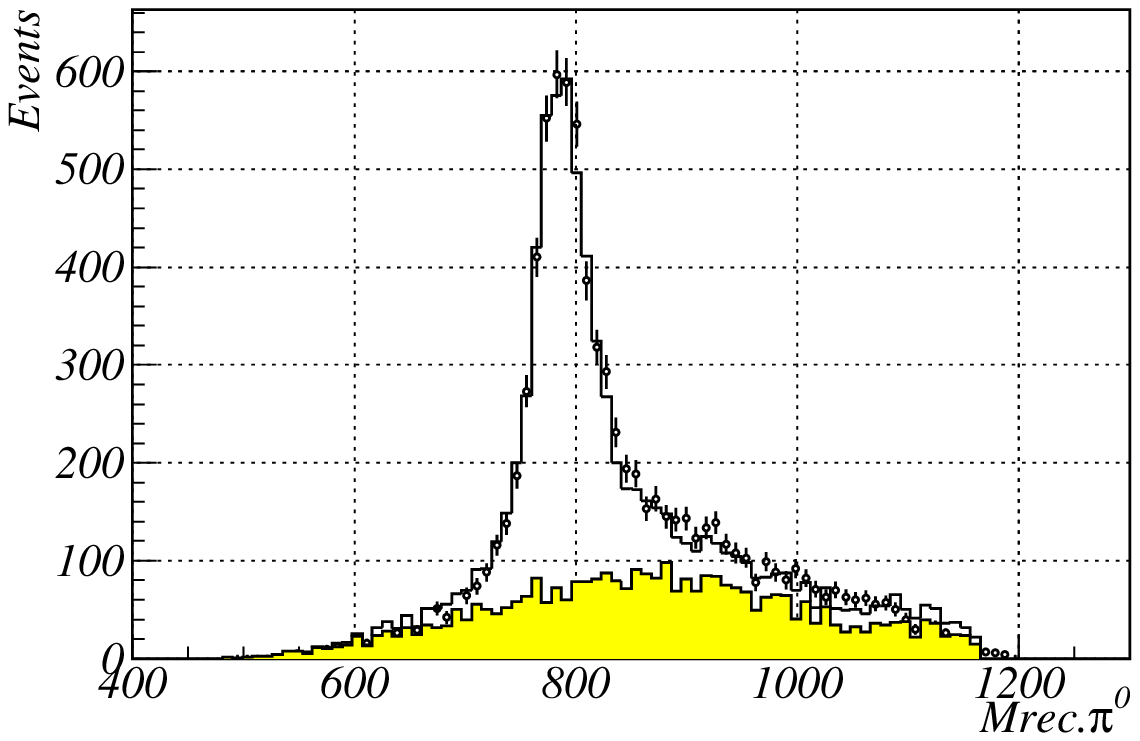}} 
\vspace*{-7mm}
\caption{Distribution of $\pi^0$-meson recoil mass
in the process $e^+e^- \to \pi^+\pi^-\pi^0\pi^0$.
Circles with error bars -- experimental data;
hatched histogram -- LIPS simulation; 
histogram -- simulation of the process
$e^+e^-\to \omega \pi ^0\to \pi ^+\pi ^-\pi ^0\pi ^0$.}
\label{dspic1}
  \centerline{\epsfbox{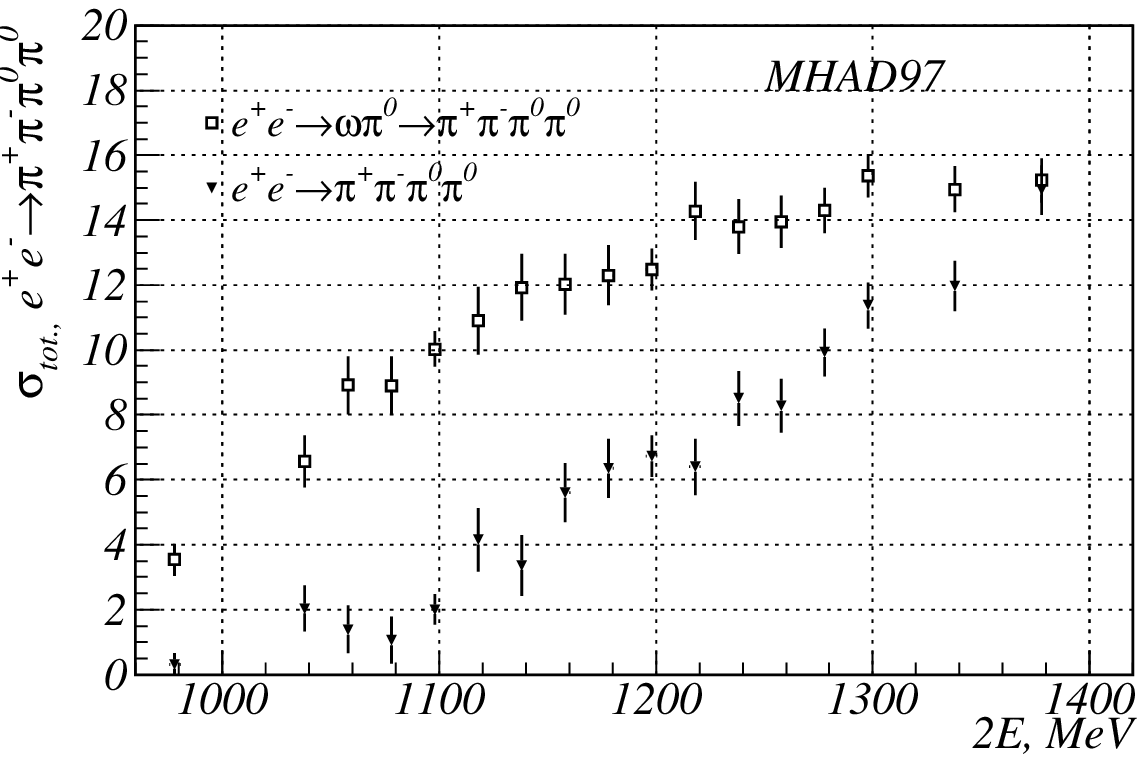}} 
\vspace*{-7mm}
\caption{Measured cross section of the process 
$ e^+e^-\to \pi ^+\pi ^-\pi ^0\pi ^0$ .}
\label{dspic3}
\end{figure}

{\bf Data analysis.}
The detection efficiency for the (\ref{ppp0p0}) process,
obtained from simulation, is practically energy 
independent and equals to  $ \varepsilon = 0.30 \pm 0.01 $
 Various intermediate states such as 
 $ \omega \pi^0 $,
$f_2  \rho^0 $, $ a_1^{\pm } \pi^{\mp } $ and others
can contribute to the reaction under study.
The $ \pi^0 $-meson recoil mass distribution in  Fig.\ref{dspic1}
shows that it is possible to separate the (\ref{omp0c}) process.
It is not possible to separate other intermediate states,
because they are very broad and distorted by interference effects.

To separate the (\ref{omp0c}) process, the selected  $ 4\pi $ events
were divided into two classes
\\ $\bullet$
$750~MeV < M_{rec. \pi^0} < 820~MeV $ ( $ \omega \pi^0 $, class 1),
\\ $\bullet$
$M_{rec. \pi^0} < 750~MeV $ or
$M_{rec. \pi^0} > 820~MeV $ (not $ \omega \pi^0 $, class 2). 

Using the detection efficiencies obtained from simulation with the
$ \omega \pi^0 $ intermediate state for the class 1, and with Lorenz
Invariant Phase Space (LIPS) model for the class 2, 
the total cross section of the (\ref{omp0c}) process was calculated. 
In Fig.\ref{dspic3} the cross section of the process (\ref{omp0c}) 
is shown together with the process
(\ref{ppp0p0}) with all other intermediate states without 
interference  between them.
Fig.\ref{dspic2} shows the cross section of the process (\ref{omp0c}), in
comparison with existing experimental results. It is seen  that the measured
cross section is lower that obtained in ND experiment \cite{Ph.Rep}, 
but in good agreement with earlier OLYA detector data \cite{VSOLYA_2}. 

Simulations of different intermediate states like $ \omega \pi^0$,
$\rho^0 \pi^0 \pi^0 $ and LIPS shows, that the detection efficiency 
varies within 5\%
range, which might be a source of systematic error.  
Another sources of systematic error are errors in event
reconstruction and inaccuracy in simulation of $\pi$-meson nuclear interaction.
In total, the systematic error is estimated still to be about 15\%.

\subsection{ The study of the process $e^+e^- \to \pi^+ \pi^- \pi^0 $}

      The process \( e^+e^- \to \pi^+ \pi^- \pi^0\)
(\ref{ppp}) in the energy range $2E_0$=1.1-1.4 GeV is interesting for several
reasons. It is well known, that near $\omega$ and $\phi$ resonances
the cross section $e^+e^-\to \omega, \phi \to \rho \pi \to \pi^+\pi^-\pi^0$
is well described by the Vector Dominance Model (VDM). New precise
measurements
in non-resonant region will allow to investigate the limitations of
VDM and determine possible contribution from
heavier states like $\omega(1420)$ or $\omega(1600)$ \cite{PDG}.
These states, decaying into $\pi^+ \pi^- \pi^0$ are considered now
 as radial excitations of $\omega(782)$. Due to existence of
$\omega$-meson decay into $\pi^+ \pi^-$ with a 2\% probability,
the process (\ref{ppp}) can proceed via
$e^+e^-\to \rho \to \omega \pi^0 \to \pi^+\pi^-\pi^0$ mechanism.
As a result, the  $\rho$--$\omega$ interference can be observed
in  $\pi^+\pi^-$ mass spectrum. According to \cite{3pi}, the radial
excitations $\rho(1450)$, $\rho(1700)$ \cite{PDG} can increase
the  $\rho$--$\omega$ interference significantly.

            The process  (\ref{ppp}) was studied earlier in the energy
range up to 1.4 GeV with ND detector at VEPP-2M	    
\cite{Ph.Rep}. The results and the VDM prediction are shown in
Fig.~\ref{vand}. One can see, that the measured cross section is
significantly higher than the predicted one.  

         In analysis of the process (\ref{ppp}) the events with two charged
particles and two photons were selected. To suppress the cosmic and beam
background and contribution from the process   $e^+e^-\to K_sK_l$
the cuts were imposed on the location of the production point
with respect to collision center: $R_1, R_2 < 0.3~cm$,
$Z_1, Z_2 < 6~cm$.
     The main background processes for (\ref{ppp}) are:
\begin{eqnarray}
&e^+e^-\to e^+e^-(\gamma)& \label{vaee} \\
&e^+e^-\to \pi^+\pi^-(\gamma)& \label{vappg} \\
&e^+e^-\to e^+e^-\gamma\gamma& \label{vaeeg} \\
&e^+e^-\to \pi^+ \pi^- 2\pi^0& \label{va4pi} 
\end{eqnarray}

     The process (\ref{vaeeg}) has the same final state as (\ref{ppp}):
The processes (\ref{vaee},\ref{vappg}) can mimic the process under study
(\ref{ppp}) because of the showers splitting in the calorimeter.
To reduce the contribution of the process  (\ref{vaee}) the cut
$E_{\gamma ~max}/E_0 < 0.8$ was imposed. Because the processes
(\ref{vaee},\ref{vappg},\ref{vaeeg}) have collinear tracks in
drift chamber, the cut $\alpha _{\pi ^+ \pi ^-} < 160^\circ$ was
added to suppress their contributions. The main background
comes from  (\ref{va4pi}), where two photons from  $\pi^0$-decays
are lost. We applied the cut  $\alpha _{\gamma \gamma} < 100^\circ$,
to decrease the contribution of this process.
     Then we applied kinematic fit to the selected events. To suppress the
process  (\ref{va4pi}), we used strong cut $\chi^2_E<10$.
The two-photon  invariant mass
$m_{\gamma\gamma}$ distribution for events which survived the
latter cut, is shown in Fig.~\ref{vams}.
For further analysis the events with $m_{\gamma\gamma}$
close to  $\pi^0$ mass within the interval
$105~MeV<m_{\gamma\gamma}<165~MeV$ were selected.   
The cross section can be  described as a sum of two terms: 
\begin{equation}
\sigma_{e^+e^- \to \pi^+\pi^-\pi^0}(s) = \sigma(s) + \sigma_{\gamma}(s), 
\label{vasum} 
\end{equation}
where $\sigma$ is a cross section (\ref{ppp}) with photon energy
less than 40~MeV, $\sigma_{\gamma}$ -- is the cross section with photon
energy larger than 40~MeV. The latter value, obtained from known
cross section of the processes $e^+e^-\to\omega,\phi \to 3\pi$,
can be calculated with the accuracy of 5\%.
The value  $\sigma (s)$ is the investigated cross section with the
radiative correction not more than 10\%.
The detection efficiency
of the process with $E_{\gamma}>40~MeV$ sharply decreases with energy,
because the selection criteria suppress events with hard
photons, emitted by initial particles.

      The total cross section  (\ref{ppp}) was presented in VDM form: 
\begin{equation}
\sigma(s)= 12\pi\frac{|A|^2}{\sqrt{s^3}}, \label{vacfit}  
\end{equation}
$$A=\sum_{V=\omega,\phi,\omega',\omega''}
\frac{\sqrt{\Gamma_{V\to3\pi}(s)\Gamma_{V\to ee}m_V^3}e^{i\theta}}
{s-m_V^2+im_V \Gamma_V(s)},$$
where $M$, $\Gamma$, $\theta$ -- mass, width and relative phase
of vector meson respectively. The values
$\Gamma_{\omega',\omega''\to3\pi}(s)\Gamma_{\omega',\omega''\to ee}$
and $\theta_{\omega',\omega''}$ are free fit parameters. Other values
were taken from the Tables  \cite{PDG}. The fit results are the following:

\begin{eqnarray*}
&B_{\omega'\to3\pi}B_{\omega'\to ee}=(0.14\pm0.02)\cdot 10^{-4}&  \\
&B_{\omega''\to3\pi}B_{\omega''\to ee}=(0.46\pm0.23)\cdot 10^{-7}&  \\
&\theta_{\omega'}=10^\circ\pm 10^\circ,\theta_{\omega''}=170^\circ\pm 24^\circ.&
\end{eqnarray*}

  Disagreement with PDG data might indicate the existence of
additional contributions besides
$\omega,\phi,\omega',\omega''$. In particular, in
the cross section (\ref{vacfit}) the interference with processes
$e^+e^-\to\rho,\rho'\to\omega\pi\to 3\pi$ was not taken into account.
Nevertheless, the function, which was used, describes the measured cross
section rather well. This allows to account for radiative corrections and
to calculate the cross section in Born approximation.
The values of Born cross section
in different energy points were calculated in the following way:
\(
\sigma=(\sigma_{exp.}^{vis.}-\sigma_{\gamma}^{vis.})/\delta\varepsilon.
\)
The results are shown in table ~\ref{vatab}.

\begin{center}
\begin{table}[htb]
\caption{Experimental cross section 
$\sigma_{e^+e^- \to \pi^+\pi^-\pi^0}$.}
\label{vatab}
\begin{tabular}{|c|c|c|c|}
\hline\
 Total energy   & Cross section &
 Total energy   & Cross section \\
$2E_0$, MeV & $\sigma_{e^+e^- \to \pi^+\pi^-\pi^0}$, nb &
$2E_0$, MeV & $\sigma_{e^+e^- \to \pi^+\pi^-\pi^0}$, nb \\
\hline 
1080 & $2.41\pm0.44 $ & 1220 & $4.36\pm0.41 $ \\
1100 & $2.71\pm0.31 $ & 1240 & $3.56\pm0.35 $ \\
1120 & $2.78\pm0.49 $ & 1260 & $3.62\pm0.34 $ \\
1140 & $2.69\pm0.49 $ & 1280 & $3.66\pm0.28 $ \\
1160 & $3.66\pm0.43 $ & 1300 & $2.89\pm0.25 $ \\
1180 & $3.22\pm0.40 $ & 1340 & $2.99\pm0.26 $ \\ 
1200 & $4.05\pm0.42 $ & 1380 & $3.04\pm0.18 $ \\ 
\hline
\end{tabular}
\end{table}
\end{center}
    
   The systematic error in the cross section is about 12\%.
It is determined by the detection efficiency error (10\%),
background subtraction error (5\%),
error in luminosity (5\%).
Fig.~\ref{vacomp} shows good agreement between old ND data \cite{ND} 
and new more accurate results, obtained in the present work. 

   To study intermediate state in  (\ref{ppp}), we measured invariant
masses of $\pi$-meson pairs in the final $3\pi$  state.
The intermediate state might be $\rho\pi$ and much less probable
$\omega\pi$ with the decay  $\omega\to 2\pi$. The interference between
these two intermediate states can be observed if one compare
mass spectra of $\pi^+\pi^-$ and  $\pi^0\pi^{\pm}$.
Experimental data in Fig.~\ref{vaspectr} show clear peak in
$\pi^+\pi^-$ mass spectrum, which proves the existence
of  $\rho$--$\omega$ interference in  $3\pi$ final state.
The relative number of events in the peak area in Fig.~\ref{vaspectr}
leads to the result
$$  N(\omega \to \pi^+\pi^-)/N(\rho \to \pi^+\pi^-) = 0.047\pm 0.011.$$
which is 4 standard deviations above zero.
To obtain the  $\rho$---$\omega$   interference  phase,
the simulation was done with different values of the phase. The experimental
data agrees the best with the phase value of $0^\circ$, which is predicted
in VDM model.

\subsection{Study of the reaction $e^+e^-\to K_SK_L$}

The cross-section of the process
\(e^+e^-\to K_SK_L\)
(\ref{ksklmhad})
is known with a high accuracy only in the energy range close 
to the $\phi$-meson peak.
In 1982 the reaction (\ref{ksklmhad}) was studied
with the DM1 detector (Orsay) in the energy range
$2E_0 = 1400 \div 2200~MeV$ \cite{bkMane}.
At the same time this process was measured
in the energy range $2E_0 = 1060 \div 1400~MeV$
with OLYA detector (Novosibirsk) \cite{olya}. 
In both experiments the achieved accuracy was not high,
so new measurements are desirable.

\begin{figure}[ht]
\epsfxsize=\textwidth
  \centerline{\epsfbox{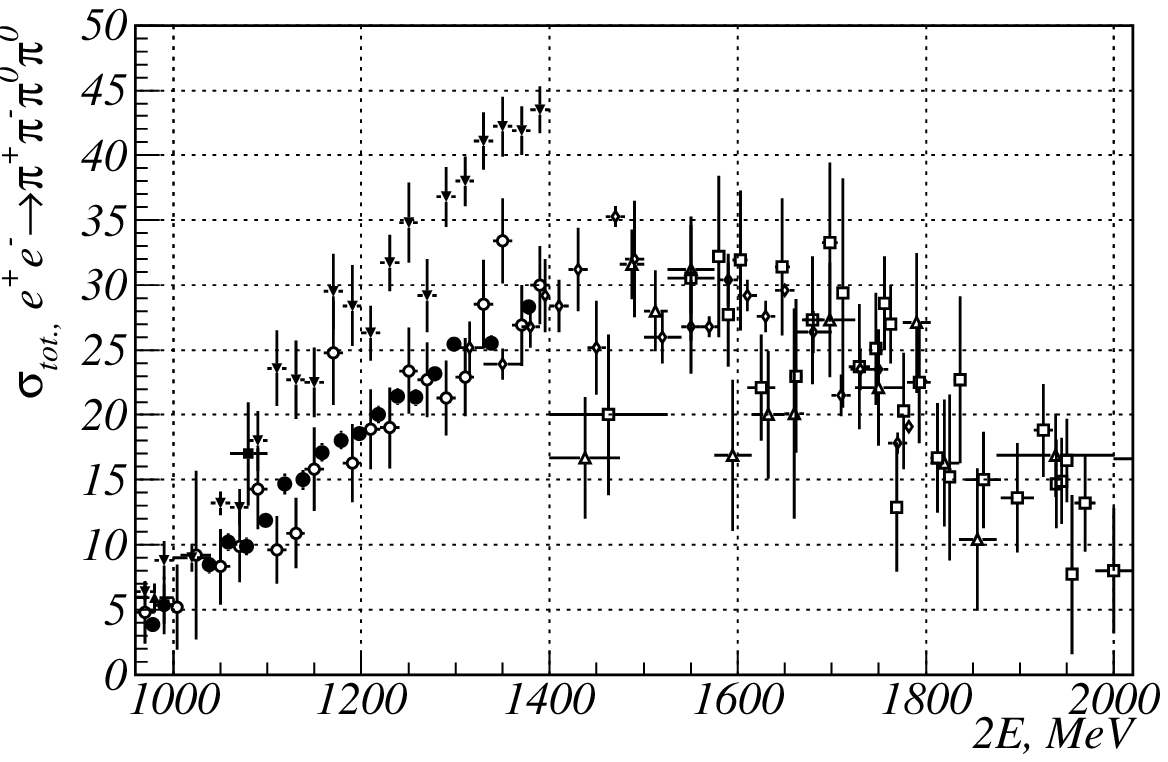}} 
\vspace*{-7mm}
\caption{Measured cross section of the process
$e^+e^-\to \pi ^+\pi ^-\pi ^0\pi ^0$. }
\label{dspic2}

\epsfxsize=0.49\textwidth
\begin{minipage}[t]{0.47\textwidth}
  \centerline{\epsfbox{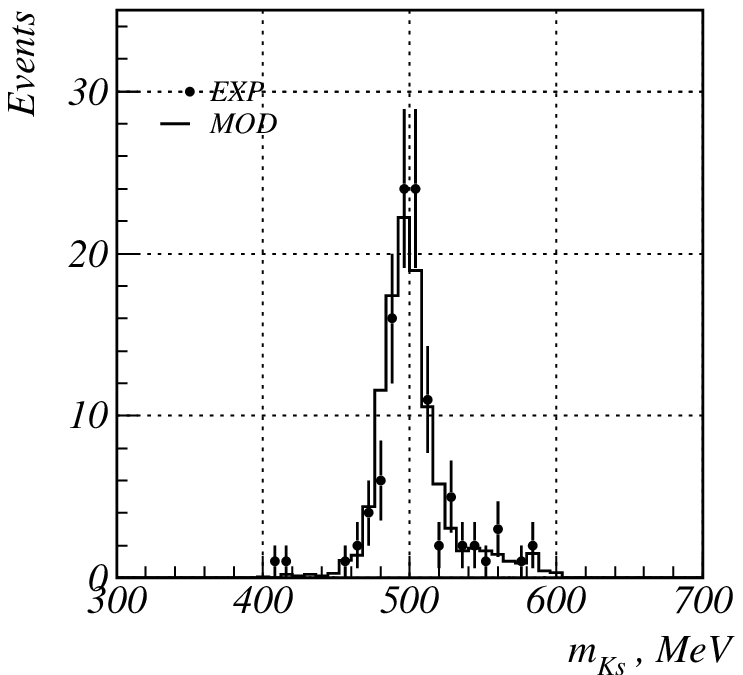}} 
\vspace*{-5mm}
\caption{Distribution of invariant mass of $\pi^0$ pairs 
in study of the process $e^+e^-\to K_SK_L$.
Points with error bars -- experimental data;
histogram -- simulation.}
\label{bkmks}
\end{minipage}
\hfill
\begin{minipage}[t]{0.47\textwidth}
  \centerline{\epsfbox{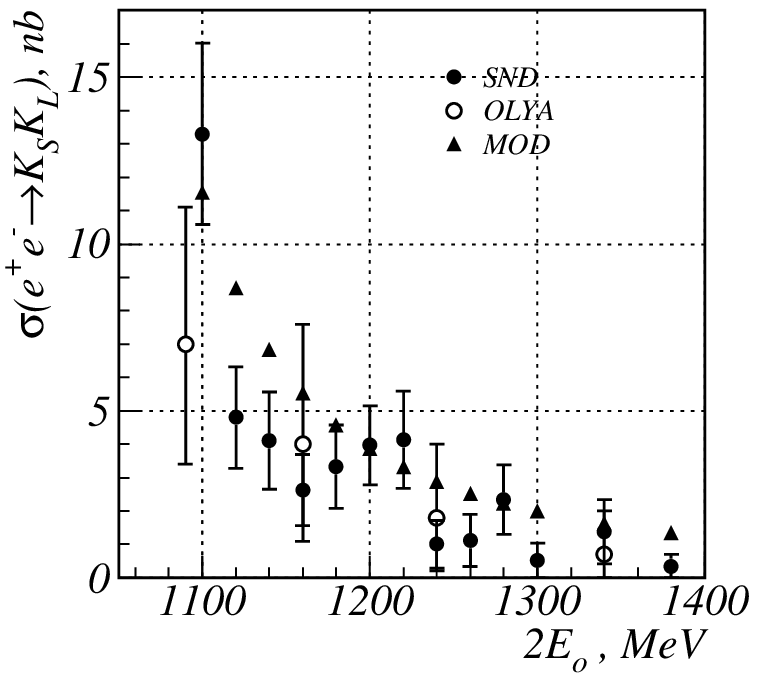}} 
\vspace*{-5mm}
\caption{Cross section of the process
$e^+e^-\to K_SK_L$:
$\bullet$ -- SND experimental data,
$\circ$   -- OLYA experimental data, 
triangles --- calculation.}
\label{bkcs}
\end{minipage}
\end{figure}

\begin{figure}[htb]
\epsfxsize=0.49\textwidth

\begin{minipage}[t]{0.47\textwidth}
  \centerline{\epsfbox{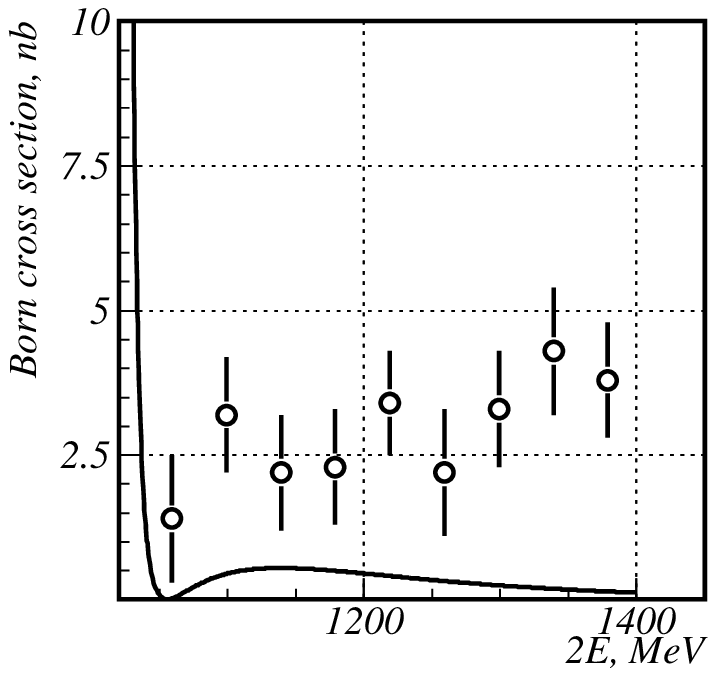}} 
\vspace*{-5mm}
\caption{Energy dependence of the total cross section
of the process $e^+e^-\to \pi ^+\pi ^-\pi ^0$
(\ref{ppp}) measured with ND detector.
Smooth curve -- vector dominance model.}
\label{vand}
\end{minipage}
\hfill
\begin{minipage}[t]{0.47\textwidth}
  \centerline{\epsfbox{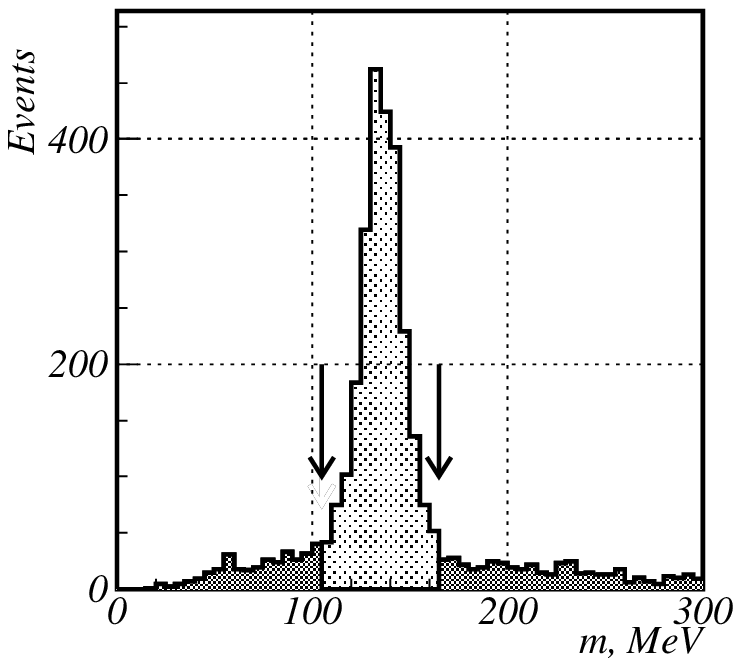}} 
\vspace*{-5mm}
\caption{Distribution of
$m_{\gamma\gamma}$ for selected experimental events
of the process $e^+e^-\to \pi ^+\pi ^-\pi ^0$.
Arrow shows the cut-off for this parameter.}
\label{vams}
\end{minipage}
\vfill
\begin{minipage}[t]{0.47\textwidth}
  \centerline{\epsfbox{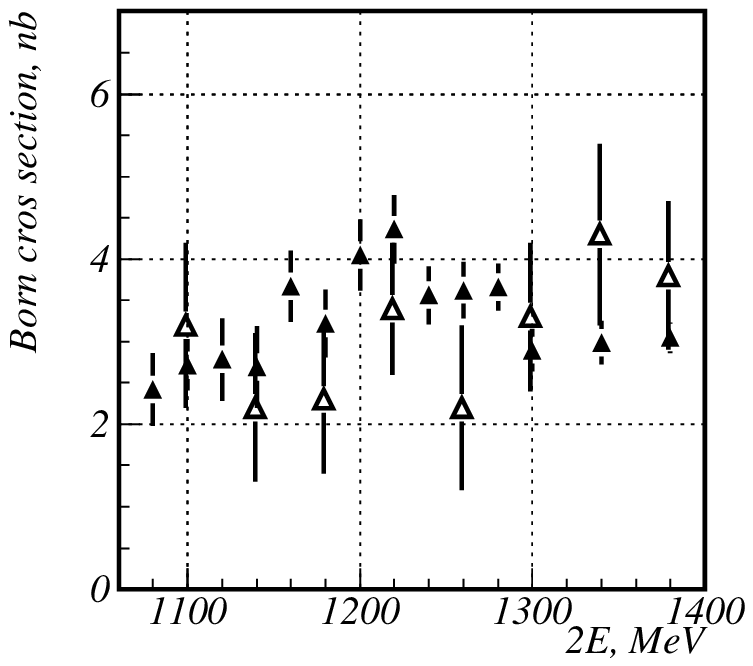}} 
\vspace*{-5mm}
\caption{Comparison of experimental results obtained
with the ND (clear triangles) and SND
(black triangles) detectors for the cross section of the
process $e^+e^-\to \pi ^+\pi ^-\pi ^0$.}
\label{vacomp}
\end{minipage}
\hfill
\begin{minipage}[t]{0.47\textwidth}
  \centerline{\epsfbox{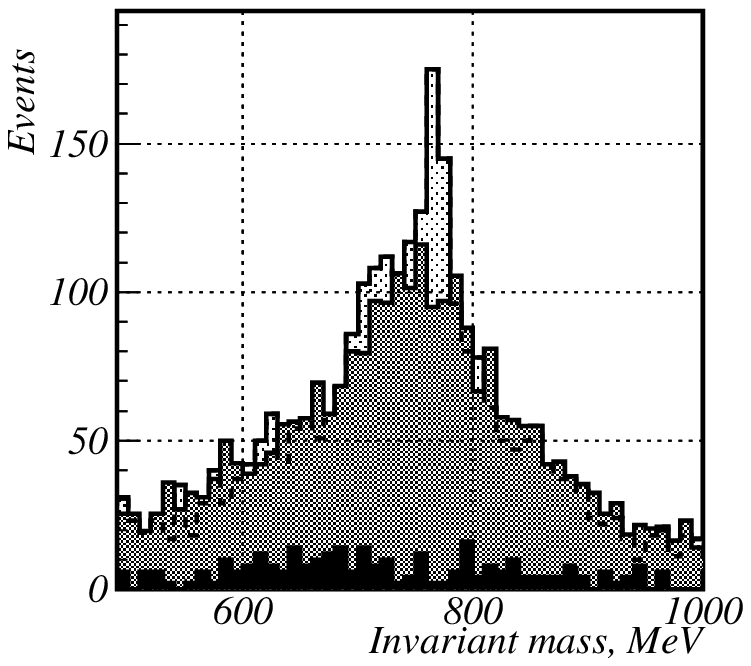}} 
\vspace*{-5mm}
\caption{Observation of the $\rho$--$\omega$ interference
in invariant mass spectrum of charged pion pairs 
in the reaction $e^+e^-\to \pi ^+\pi ^-\pi ^0$.
Histogram -- experimental distribution of $\pi^+\pi^-$  invariant mass;
hatched histogram -- experimental distribution of $\pi^0\pi^{\pm}$
invariant mass;
black histogram -- simulated $\pi^+\pi^-$ invariant mass distribution
for the background process $e^+e^- \to \pi^+\pi^-\pi^0\pi^0$.}
\label{vaspectr}
\end{minipage}
\end{figure}

   In the present work we analyzed a part of statistics of the scan 
MHAD\_9702 with integrated luminosity of $\sim 1.8~pb^{-1}$.
The process $e^+e^-\to K_SK_L$ was studied in the decay mode
$K_S\to \pi^0\pi^0$ (\ref{kspp}).
The main background comes from the non-resonant process
\(e^+e^- \to \omega \pi^0 \to \pi^0 \pi^0 \gamma\)
(\ref{omp0n}). Cosmic and beam background are also present.
To select events of the reaction (\ref{kspp}) the 
following cuts were applied:
\\$\bullet$ $N_{\gamma} \geq 4$, $N_{cp} = 0$.
\\$\bullet$
Cosmic events were rejected using SND calorimeter for
reconstruction of cosmic muon tracks.
The procedure was tested on simulated events of the process (\ref{ksklmhad})
at $2E_0 = 1100~MeV$. From 1733 simulated events
only 3 events were erraneously recognized as cosmic tracks,
while in the experimental data
1716 such tracks were found among 2319 events.
\\$\bullet$ The remaining events should have two 2$\pi^0$.
The invariant masses of photon pairs must be within the range:
$115 < m_{gamma\gamma} < 155~MeV$.
\\$\bullet$
Invariant mass of $\pi^0$-meson pair
$400 < m_{\pi^0\pi^0} < 600~MeV$ is compatible with $K_S$ mass.
The cut $\zeta < 0$ \cite{XINM}
was used to suppress possible background from the $K_L$ decay or
its nuclear interaction in the detector material.
  Events passing these cuts belong either to the process
(\ref{ksklmhad}) or to the process \(e^+e^-\to K_SK_L\gamma \)
with emission of photon by initial particles
and subsequent return back to the $\phi$ peak
(\ref{ksklg}), (\ref{omp0n}).
At $2E_0>m_{\phi}$ the radiative photon in the process (\ref{ksklg})
carries the energy up to $\sim 300~MeV$,  leading to
a significant difference in kinematics
with respect to the process (\ref{ksklmhad}).
In particular, the energy of $K_S$ meson decreases and as a result
the detection efficiency  with the additional cut 
$E_{K_S}/E_0 > 0.94$ goes down.
The cut $E_{K_S}/E_0 < 1.06$ rejects significantly the
process (\ref{omp0n}). The rest of the events of the process
(\ref{omp0n}) is rejected by cuts
$E_{tot}/2E_0 > 0.8$ and $P/E_{tot} < 0.15$.
Beam background events have photons located mainly in the 
small polar angle region. So, the cut 
$\theta_{\gamma ~min(K_S)} > 31.5^\circ$ was applied.
108 experimental events passed all cuts listed above.
Distribution over $m_{\pi^0\pi^0}$, invariant mass of $\pi^0$-mesons pairs
is shown in Fig.~\ref{bkmks}. Clear peak at $K_S$ mass is seen.
The detection efficiency to the investigated process (\ref{ksklmhad})
depends on $E_0$, decreasing from 3.4\%
at $2E_0=1100~MeV$ to 1.8\%
at $2E_0=1380~MeV$.
Efficiency to the process (\ref{ksklg}) varies from 1.9\% to 0.4\%
in the same energy interval.
Efficiency of the process (\ref{omp0n}) is $\leq$ 0.1\%.

The  visible cross section of the process (\ref{ksklmhad})
is shown in Fig.~\ref{bkcs} as well as expected  cross section, calculated
taking into account $\phi$-meson resonance and radiative corrections.
Experimental data from the  OLYA detector are also shown in Fig.~\ref{bkcs}.
One can see, that the data obtained by both detectors are in good
agreement and within experimental errors do not contradict 
theoretical estimations. 

\subsection{Search for the process $e^+e^- \to K_SK_L\pi^0$}

    $K\overline{K}\pi$ state could appear in  $e^+e^-\to \phi \pi^0$ and
$e^+e^-\to KK^{\ast}$ processes. The search for the process
$e^+e^- \to \phi \pi^0$ was carried out earlier in ND experiment
in the decay mode $\phi \to K_SK_L \to2\pi^0K_L$ \cite{Ph.Rep,ND},
in CMD experiment  \cite{cmd} and in DM1 experiment at DCI \cite{dm1_1}
in the process $e^+e^- \to \phi \pi^0 \to K^+K^- \pi^0$.
The study of the process $e^+e^- \to KK^{\ast}$ in the channel
$KK^{\ast} \to K_SK^{\pm}\pi^{\mp}$ was conducted in DM1-DCI
experiment  \cite{dm1_2} in the energy range $1400 \div 2180~MeV$ and in
the energy range $1350 \div 2400~MeV$ in DM2-DCI experiment \cite{dm2}.
The channel $e^+e^- \to KK^{\ast} \to K^+K^-\pi^0$ was studied in
DM2-DCI as well. 

New measurements, carried out with SND detector,
can significantly improve our knowledge of these
processes. In particular, new measurements could clarify the nature
of $C$(1480) state \cite{c1480_1}, found in IHEP (Protvino).
With quantum numbers $J^{PC}=1^{--}$, 
and decay into $\phi\pi^0$, this state 
is a possible candidate for exotic  hybrid
or 4-quark state. The model exists, where $C$(1480) is identical to
$\rho$(1450)-meson \cite{c1480_2}. The electron width of $C$-state depends on
its structure, so the study of  $C$(1480) production in $e^+e^-$
collisions could reveal its contents. Although the $C$ mass is
larger than VEPP-2M maximum center of mass energy, 
its left slope could be observed
with SND in the process  $e^+e^- \to \phi \pi^0 \to K_SK_L\pi^0$.

In our search for the process $e^+e^- \to K_SK_L\pi^0$ (\ref{ksklp0})
the decay channel  $K_S \to 2\pi^0$ was chosen. The following
multi-photon reactions can be a source of background:
\(e^+e^- \to \omega \pi^0 \to \pi^0 \pi^0 \gamma\)
(\ref{omp0n}),
\(e^+e^-  \to \eta \gamma \to 3\pi^0 \gamma\)
(\ref{etagnmhad}),
\(e^+e^- \to K_S K_L\)
(\ref{ksklmhad}),
\(e^+e^- \to \omega \pi^0 \pi^0 \to 3 \pi^0 \gamma\)
(\ref{op0p0}),
\(e^+e^- \to 4 \gamma, 5\gamma\)
(\ref{QED45}).

\begin{figure}[p]
\epsfxsize=\textwidth
  \centerline{\epsfbox{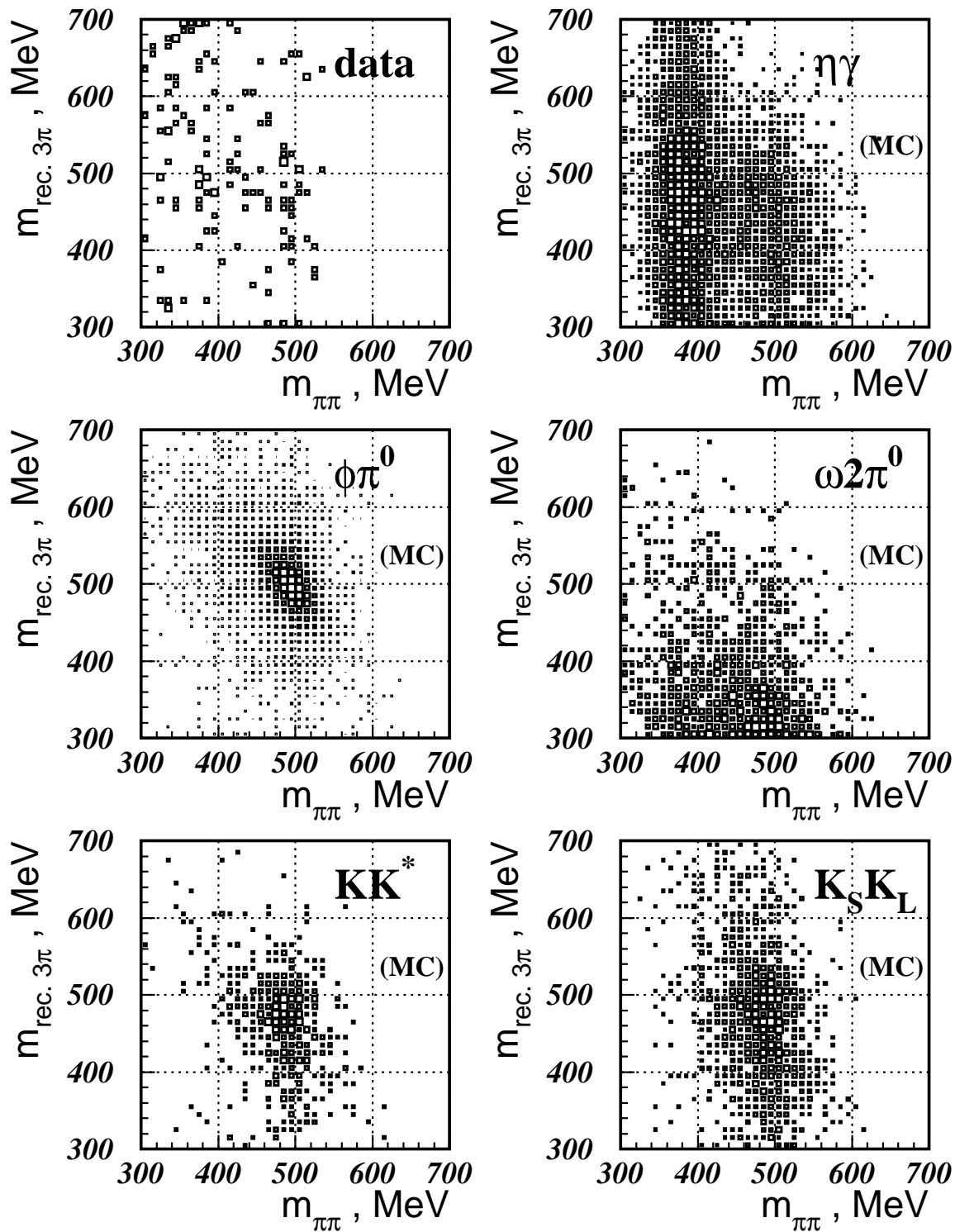}} 
\vspace*{-7mm}
\caption{\label{bermksl}
Distribtion of selected events over
recoil mass of the $\pi^0\pi^0\pi^0$ system versus
invariant mass of $\pi^0\pi^0$ pair in the search for
$e^+e^- \to K_SK_L\pi^0$ decay.}
\end{figure}

   The main selection cuts were $E_{tot}/2E_0>0.38$,
$N_{\gamma}\ge6$, $N_{cp}=0$. The number of found $\pi^0$-s should be
not less than 3 with two of them forming $K_S$ meson, the recoil mass
of 3  $\pi^0$'s should be compatible with
$K_L$ mass. Fig.\ref{bermksl} presents
the distribution over invariant mass for experiment and simulation.
The contribution of QED background from processes  (\ref{QED45})
and from the process (\ref{omp0n}) was found to be negligible.
To suppress possible background from the process (\ref{etagnmhad})
with unknown cross section we used a set of harder cuts on ETON, PTRT, and
$\pi^0$ triples with invariant mass close to that of $\eta$ meson.
 We estimated from VDM
model, that under these cuts the expected  contribution from
this process should be small. The same hard cuts allow to suppress
the process (\ref{op0p0}) \cite{dm2}.
Main background comes from the reaction
$e^+e^-\to\phi\gamma$  with subsequent decay $\phi\to K_SK_L$.
The number of such events was estimated \cite{olya} to be
$(6 \pm 3)$, which is close to observed 5 events.
So, on the basis of the available data only upper limits of the cross
sections in the energy range $1300\div1400~MeV$ can be placed:
\begin{center}
\begin{tabular}{cc}
$\sigma(e^+e^- \to \phi\pi^0) < 0.2~nb$ & (90\% C.L.), \\
$\sigma(e^+e^- \to KK^{\ast}) < 1.6~nb$ & (90\% C.L.), \\
$\Gamma(C \to e^+e^-) \cdot B(C \to \phi\pi^0) < 36$~eV & (90\% C.L.), \\
\end{tabular}
\end{center}
The obtained limits are lower than existing values \cite{Ph.Rep,ND,cmd}.

\subsection{Upper limit for electron width of $f_2(1270)$ meson}

\begin{figure}[htb]
\epsfxsize=0.49\textwidth
\begin{minipage}[t]{0.47\textwidth}
  \centerline{\epsfbox{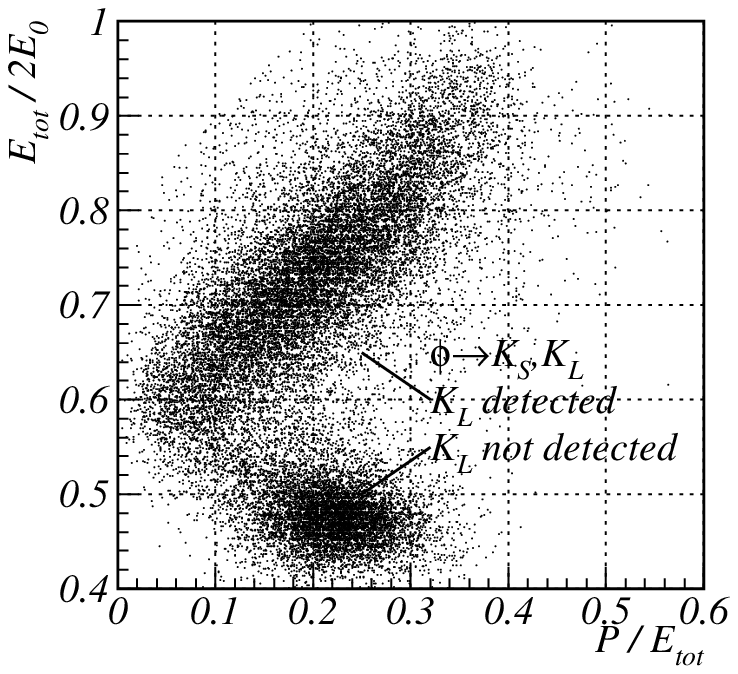}} 
\vspace*{-5mm}
\caption{ \label{dbfig01}
Distribution of experimental events over
$E_{tot}/2E_0$ versus $P/E_{tot}$,
in the search for
$K_S \to 3 \pi^0$ decay.}
\end{minipage}
\hfill
\begin{minipage}[t]{0.47\textwidth}
  \centerline{\epsfbox{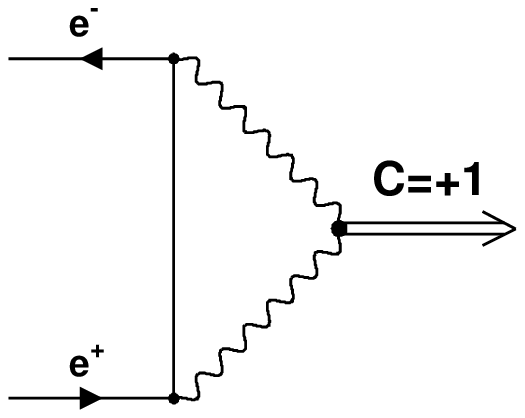}} 
\vspace*{-5mm}
\caption{Feinman graph for direct production of C-even resonance 
in $e^+e^-$ collision.}
\label{vslfig01}
\end{minipage}
\vfill
\begin{minipage}[t]{0.47\textwidth}
  \centerline{\epsfbox{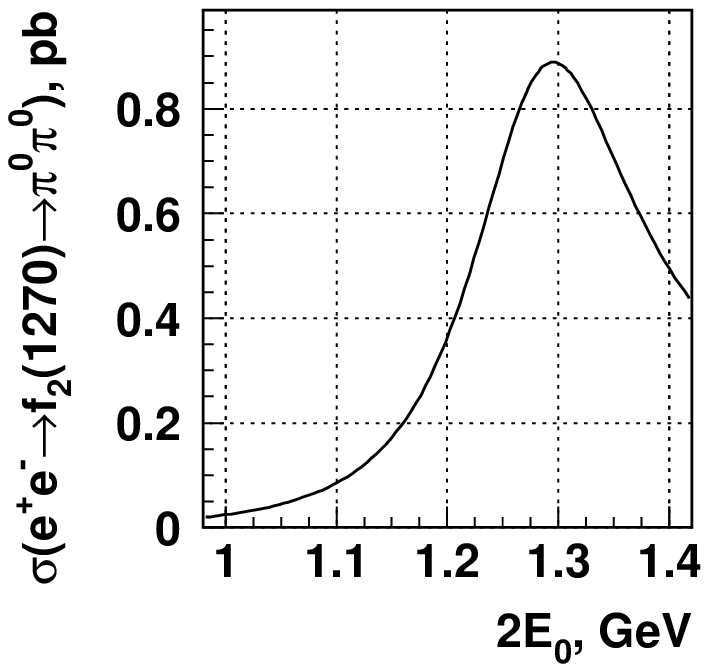}} 
\vspace*{-5mm}
\caption{Calculated total cross section of the process
$e^+e^- \to f_2(1270) \to \pi^0\pi^0$.}
\label{vslfig02}
\end{minipage}
\hfill
\begin{minipage}[t]{0.47\textwidth}
  \centerline{\epsfbox{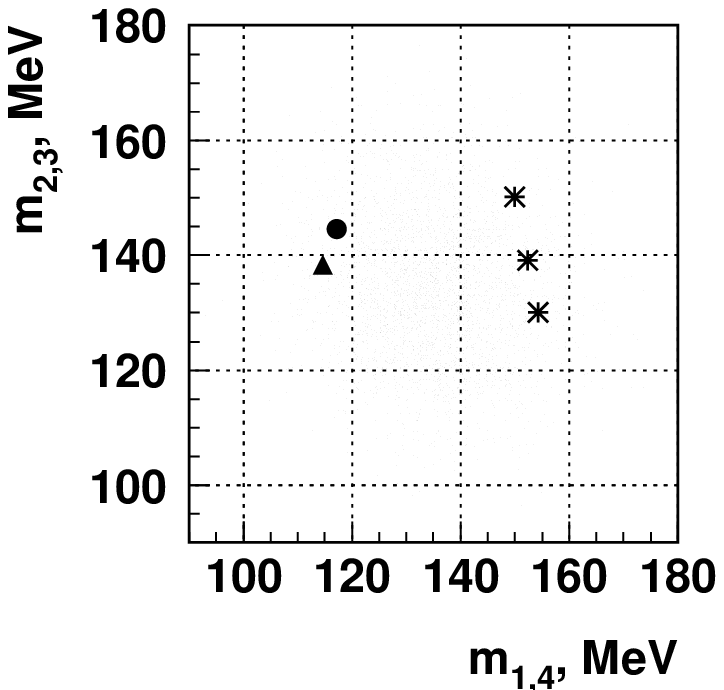}} 
\vspace*{-5mm}
\caption{Scatter plot  
of invariant masses of photon pairs in search for the process
$e^+e^- \to f_2(1270) \to \pi^0\pi^0$. 
Small dots -- simulation of the process (\ref{vslequ02});
triangles -- simulation of the background process (\ref{QED45});
stars -- simulation of the background process (\ref{omp0n});
circle -- the only
experimental event, which passed all the selection cuts.}
\label{vslfig03}
\end{minipage}
\end{figure}

C-even resonance production in $e^+e^-$ collisions is described 
in the lowest order in $\alpha$ by a Feinman diagram shown in
(Fig.~\ref{vslfig01}).
The unitary limit of
the electron width of the $f_2(1270)$ meson can be expressed
through its two-photon
width and a factor of $\sim \alpha^2$ \cite{vslyaf13}, \cite{vslyaf21}.  
It can be estimated using the table value of 
$\Gamma_{f_2\to \gamma \gamma}$ \cite{PDG}:
\begin{equation}
\Gamma_{unit.lim.}(f_2(1270) \to e^+e^-) \sim 3 \cdot 10^{-2}~eV.
\label{vslequ01}
\end{equation}
Actual cross section can be several times larger due to 
transition form factor.
The only experimental result was obtained at VEPP-2M collider
with ND detector \cite{ND}. 
The $e^+e^- \to f_2(1270) \to \pi^0 \pi^0$ reaction
(\ref{vslequ02}) was studied in a $4\gamma$ final state 
\cite{vslyaf48,Ph.Rep} and
the following upper limit was obtained:
$\Gamma (f_2(1270) \to e^+e^-) < 1.7~eV$ at 90\%
 confidence level.

It this work, like in \cite{vslyaf48}, the reaction (\ref{vslequ02})
was  studied again. The cross section was opproximated according to
\cite{vslyaf40}:
\begin{equation}
\frac{d\sigma}{d\Omega}(e^+e^-\to f_2(1270)\to\pi^0 \pi^0)= \\
12.5\cdot\Biggl(\frac{2E_0}{m}\Biggr)^6\cdot \frac{\Gamma^2
\cdot B_{ee}\cdot B_{\pi^0\pi^0}}{(m^2-s)^2 + m^2 \Gamma^2}
\cdot sin^2(2\theta),
\label{vslequ03}
\end{equation}
where $s=4E_0^2$, $m$, $\Gamma$,
$B_{ee}$ и $B_{\pi^0 \pi^0}$ -- are the $f_2(1270)$ meson
mass, full width, branching  ratios of the decays into $e^+e^-$ and
$\pi^0\pi^0$ respectively.
In the unitary limit the total cross section  of the
reaction (\ref{vslequ02}) (Fig.~\ref{vslfig02}) is about $1~pb$
at $\sqrt{s}=m_{f_2}$. This value is about 30 \%
larger than that
extracted from Breit-Wigner  formula, used  in \cite{vslyaf48}.

{\bf Event selection}
For analysis of the reaction (\ref{vslequ02}), events with four
photons in the final state where selected. The main background 
processes are the following: 
\(e^+e^- \to 4 \gamma, 5\gamma\)
(\ref{QED45}) and
\(e^+e^- \to \omega \pi^0 \to \pi^0 \pi^0 \gamma\)
(\ref{omp0n}).
The cross section each of these processes is approximately
three order of magnitude larger than that of the process (\ref{vslequ02}).
To choose selection criteria and estimate detection
efficiency the Monte Carlo simulation was carried out
for the processes (\ref{vslequ02}, \ref{QED45}, \ref{omp0n}).
In order to suppress background the following selection criteria were applied:
\\$\bullet$ $N_{\gamma}=4$; $N_{cp}=0$; 
$\theta _{min} >27^\circ$; 
$E_{tot}/2E_0 > 0.85$;
$P/E_{tot} < 0.1$; 
\\$\bullet$ $\zeta < 0$; $\chi^2_{\pi^0\pi^0} < 15$;
\\$\bullet$ $85~MeV < m_{14},~m_{23} < 185~MeV$ ;
$E_2 < 0.8 E_0$; $E_4 > 0.1 E_0$;

The shown above  choise of certain combinations of photon pairs
$m_{14}$, $m_{23}$ is based on specific
kinematics of the process (\ref{vslequ02}).
The probability for other  combinations to form $\pi^0$-s is about 20 \%.
The restrictions on the second and fourth photons energies -- 
$E_2$, $E_4$ essentially reduced background from the 
process (\ref{QED45}).

As a result of all described cuts, 3 experimental events left.
The expected background is about 10 events mainly from the process 
(\ref{omp0n}). 
To reduce background, the kinematic fit was applied for remaining events.
The fit required the energy-momentum balance, presence of
two $\pi^0$ meson in event. Events with  
($\chi^2_{\pi^0\pi^0} < 15$) were selected. The  experimental events
are shown in Fig.~\ref{vslfig03}.
The expected background from the processes(\ref{QED45}, \ref{omp0n}),
estimated from simulation was found to be $1.0 \pm 0.7$~event.

{\bf Results}
According to simulation the detection efficiency
for the process (\ref{vslequ02}) is 21\%.
Based on the single found event the following upper limit 
could be placed:
\begin{equation}
\Gamma(f_2(1270)\to e^+e^-)<\frac{N_{observed}}{N_{expected}}
\cdot k_1\cdot \Gamma_{unit.lim.}(f_2(1270) \to e^+e^-)=0.4~eV,
\label{vslequ05}
\end{equation}
 C.L.~90\%,
where $k_1=3.89$ is the Poisson coefficient for upper limit at 90\%
confidence level for one detected event.
Obtained result in four times lower than previous one, obtained 
with ND \cite{Ph.Rep},
but still 15 times higher than unitary limit:
$\Gamma(f_2(1270) \to e^+e^-)$.


\begin{table}[tbp]
\parbox{\textwidth}{\caption{ \label{BRAN} \large Measured branching ratios of particle
decays in $\phi$- 96 experiments. }}
\begin{center}
\hspace{5mm}
\begin{normalsize}
\begin{tabular}{|p{28mm}|p{32mm}|p{20mm}|p{30mm}|p{25mm}|}
\hline
\parbox[t]{25mm}{Decay mode}                  &
\parbox[t]{30mm}{Branching ratio (BR) (this work)} &
\parbox[t]{20mm}{Expected BR, ref.}           &
\parbox[t]{30mm}{PDG,1996 or recent result}  &
Comments \\
\hline
$\phi \to \pi^0 \pi^0 \gamma$        &
   $(1.14\pm 0.10 \pm 0.12)\cdot 10^{-4}$ &
   $10^{-4} \div 10^{-5}$                 &
   $< 10^{-3}$ \cite{IVN1}                & 
   First observation \\
\hline
$phi \to \eta \pi^0 \gamma$ &
   $(8.3\pm 2.3)\cdot 10^{-5}$    &
   $10^{-6} \div 10^{-4}$         &
    ---  & First observation \\
\hline
$\phi \to \eta' \gamma$                         &
   \parbox[t]{30mm}{
   $(6.7^{+3.4}_{-2.9})\cdot 10^{-5}$;\\
   $ < 1.1 \cdot 10^{-4}$ (in 7-$\gamma$ mode)     } &
   $0.6\cdot 10^{-4} \div 10^{-4}$    \cite{Q-model} &
   $(1.2^{+0.7}_{-0.5})\cdot 10^{-4}$ \cite{CMD-2}   & 
Second measurement \\
\hline
 $\phi \to \omega \pi^0$                          &
   $(5.7^{+2.0}_{-1.8})\cdot 10^{-5}$                 &
   $  5 \cdot 10^{-5}$ \cite{theor1,theor2}           &
   $< 5 \cdot 10^{-5}$ \cite{Prep.97,Serednyakov/H-97}&
   First observation \\
\hline
$\phi \to \eta \pi^0 \pi^0 \gamma$ & 
   $<2\cdot 10^{-5}$                    &
   $1.5 \div 2\cdot 10^{-5}$ \cite{Serednyakov/H-97} & 
   $<2\cdot 10^{-4}$ & First attempt \\
\hline
 $\phi \to \eta \gamma$                               &
   $(1.21\pm 0.03 \pm 0.05)\%$                            &
   --- &
   $(1.26\pm0.06)\%$ \cite{PDG} & Most precise measurement
   \\
\hline
 $\phi \to \eta e^+e^-$                               &
   $(1.42\pm 0.39 \pm 0.23)\cdot 10^{-4}$                 &
   $1.1\cdot 10^{-4}$ (for unit form.f.) \cite{Landsberg} &
\parbox[t]{30mm}{
   $(1.3^{+0.8}_{-0.6}) \cdot 10^{-4}$ \cite{NDETAEE} \\
   $(1.1\pm 0.5\pm 0.2) \cdot 10^{-4}$ \cite{CMD96}
                } &
--- \\
\hline
$\eta \to e^+e^- \gamma$          &
   $(6.8\pm 1.1 \pm 0.7)\cdot 10^{-3}$ &
   $6.3\cdot 10^{-3}$                  &
   $(4.9\pm 1.1)\cdot 10^{-3}$         & 
--- \\
\hline
 $K_S \to \pi^0\pi^0\pi^0$     &
   $ 6.9\cdot 10^{-5}$             &
   $ 1.9\cdot 10^{-9}$  \cite{IVN7}&
   $<1.9\cdot 10^{-5}$             & 
--- \\
\hline
 $f_2(1270)\to e^+e^-$&
\parbox[t]{30mm}{
   $\Gamma < 0.4~eV$\\  
   $(BR < 2.2 \cdot 10^{-9})$ 
                } &
   $ \Gamma = 0.03~eV$ (Unit.limit) &
   $ \Gamma < 1.7~eV$ \cite{Ph.Rep,vslyaf48}& 
   C.L. 90\%, the best upper limit \\
\hline
\end{tabular}
\end{normalsize}  
\end{center}
\end{table}

\begin{table}[tbp]
\parbox{\textwidth}{\caption{\label{CROSS} \large
The $e^+e^-$ annihilation processes under study
in the energy range $2E_0=985\div 1400~MeV$.}}
\begin{center}
\begin{normalsize}
\begin{tabular}{|c|c|c|}
\hline
The reaction & 
\parbox[t]{25mm}{The average \\ cross-section\\} &
Comments \\
\hline
 $e^+e^- \to \omega \pi^0 \to \pi^+ \pi^- \pi^0 \pi^0$ &
   $7.6 \pm 0.8~nb$ & 
   $2E_0 = m_\phi $\\
\hline
$e^+e^- \to \omega \pi^0 \to \pi^0 \pi^0 \gamma$ &
   $0.65 \pm 0.04 \pm 0.04~nb$ & 
   $2E_0 = m_\phi $\\
\hline
$e^+e^- \to e^+e^-\gamma$ &
   $23.6 \pm 0.3~nb$ &
   $2E_0 = 985 \div 1040~MeV $; $\theta_{min}>36^\circ$;
   $\Delta \psi > 5^\circ$\\
\hline
 $e^+e^- \to e^+e^-\gamma\gamma$ &
   $0.43 \pm 0.09~nb$ &
   $2E_0 = 985 \div 1040~MeV $; $\theta_{min}>36^\circ$;
   $\Delta \psi > 5^\circ$\\
\hline
$e^+e^- \to \gamma\gamma\gamma$ &
   $1.82 \pm 0.21 \pm 0.25~nb$ &
   $2E_0 = m_\phi$; $\theta_{min}>27^\circ$; $\Delta \psi > 6^\circ$\\
\hline
$e^+e^- \to \pi^+ \pi^- \pi^+ \pi^-$  &
   $3 \div 24~nb$ & 
   $2E_0 = 1000 \div 1380~MeV $ \\
\hline
 $e^+e^- \to \omega \pi^0 \to \pi^+ \pi^- \pi^0 \pi^0 $ &
   $3 \div 15~nb$ & 
   $2E_0 = 985 \div 1380~MeV $ \\
\hline
$e^+e^- \to \pi^+ \pi^- \pi^0 \pi^0 $ excl. $\omega \pi^0$&
   $0 \div 15~nb$ & 
   $2E_0 = 985 \div 1380~MeV $ \\
\hline
$e^+e^- \to \pi^+ \pi^- \pi^0$  &
   $2.4 \div 4.4~nb$ & 
   $2E_0 = 1100 \div 1380~MeV $ \\
\hline
 $e^+e^- \to K_S K_L $  & 
   $13 \div 0.5~nb$ & 
   $2E_0 = 1040 \div 1380~MeV $ \\
\hline
$e^+e^- \to \phi \pi^0$  & 
   $ < 0.2~nb$ & 
   $2E_0 = 1040 \div 1380~MeV $; C.L. 90\% \\
\hline
 $e^+e^- \to KK^{\ast}$  & 
   $ < 1.6~nb$ & 
   $2E_0 = 1040 \div 1380~MeV $; C.L. 90\% \\
\hline

\end{tabular}
\end{normalsize}  
\end{center}
\end{table}

\section{Conclusion}
In Tables~\ref{BRAN} and ~\ref{CROSS} the SND physical results, 
obtained up to now, are shown. Let us list the most important of them:
\\1. The existence of $\phi$ meson  radiative electric dipole decays
$\phi \to f_0\gamma, a_0\gamma$, announced in our previous
work \cite{Prep.97}, is confirmed. The measured branching ratios
and mass spectrum of $\pi^0\pi^0$ and $\eta\pi^0$
system support the 4-quark model structure of $f_0$ and
$a_0$ mesons.
\\2. The decay $\phi \to \eta'\gamma$ is observed, confirming
the 1997 CMD-2 first observation of this decay, but SND branching 
ratio almost two times less, unless the errors is still high.
\\3. The decay $\phi \to \omega \pi^0$ is observed for the first time.
\\4. The measurements of $\eta \to e^+e^-\gamma$ and
$\phi \to \eta e^+e^-$ decay branching ratios
agree well with existing data.
\\5. The cross-sections of $e^+e^- \to \pi^+\pi^-\pi^0,~
\pi^+\pi^-\pi^0\pi^0\,~\pi^+\pi^-\pi^+\pi^-$ reactions
were measured in agreement with previous results.
\\6. In analysis of  $e^+e^- \to \pi^+\pi^-\pi^0$ process
the effect of $\rho -\omega$-interference was observed
in final state $\pi^+\pi^-$ mass spectrum.
\\7. We continued study of high order QED processes
$e^+e^- \to e^+e^- \gamma,~e^+e^- \gamma \gamma,~\gamma\gamma\gamma$.

\section{Acknowledgment}
This work is supported in part by Russian Foundation of
Basic Researches, grants
No.96-02-19192; No.96-15-96327; 
No.97-02-18561; No.97-02-18563;
STP ``Integration'' (Grant No 274);
INTAS Foundation, grant No.94-763.


\clearpage

\clearpage

\clearpage

\clearpage

\clearpage

\clearpage

\clearpage

\clearpage

\clearpage

\clearpage

\clearpage

\clearpage

\clearpage

\clearpage
\onecolumn

\clearpage

\clearpage

\clearpage

\clearpage

\clearpage

\clearpage

\clearpage

\clearpage

\clearpage

\clearpage

\begin{thebibliography}{10}
\large

\bibitem{SND}
V.M.Aulchenko et al., Proc.  
Workshop on Physics and Detectors for DA$\Phi$NE,
Frascati, Italy, April 9--12 (1991), p.605.

\bibitem{Prep.87} V.M.Aulchenko et al., Preprint INP 87-36, in Russian.

\bibitem{ND} V.B.Golubev et al., Nucl. Instrum. Meth. 227 (1984), p.467.
V.M.Aulchenko et al., Pisma v JETP, 43 (1987) 118 (in Russian). 

\bibitem{VEPP} G.M.Tumaikin, Proceedings of the 10-th International 
Conference on High Energy Particle Accelerators, Serpukhov, 
v.1 (1977) p.443.

\bibitem{Calibr.}
M.N.Achasov et al., Nucl. Instrum. Meth. A441(2--3) (1998) p.337--342;
M.N.Achasov et al., Nucl. Instrum. Meth. A401 (1997) p.179.

\bibitem{STUB} V.M.Aulchenko et al., 
Preprint INP 85-122, Novosibirsk, 1985.

\bibitem{Prep.97} M.N.Achasov et al., 
Preprint Budker INP 97-78, Novosibirsk, 1997;
e-Print Archive: hep-ex/9710017.

\bibitem{Wiggler} V.V.Anashin et al.,
Preprint INP 84-123, Novosibirsk, 1984.

\bibitem{XINM} A.V.Bozhenok, V.N.Ivanchenko, Z.K.Silagadze,
Transverse energy profile of electromagnetic
shower, Nucl. Inst. and Meth., A379 (1996) pp.507--508.

\bibitem{Ivanchenko/H-97}
V.N.Ivanchenko, Parallel talk given at HADRON-97 Conference, Brookhaven,
August 25-30, 1997, to be published in Phys. Atom. Nucl.,
M.N.Achasov et.al., November 1997. e-Print Archive: hep-ex/9711023.

\bibitem{PDG}
{\it Particle Data Group.} Review of Particle Physics. Parts I and II.
Physical Review D, Particles and Fields v.54 (1996).

\bibitem{Landsberg}
L.G.Landsberg, Phys.Rep 128, N.6 (1985) pp.301--376.


\bibitem{NDETAEE} V.B.Golubev et al., Sov.journal of Nucl.Phys., 41, 
(1985) p.1183--1186.
 

\bibitem{CMD96}
 R.R.Akhmetshin et al., Recent results from the CMD-2
 detector at VEPP-2M collider. Proc. of the 28-th International
 Conference on High Energy Physics, Warsaw, Poland, 25--31
 July 1996, 522.

\bibitem{Ph.Rep}
S.I.Dolinsky et al., Phys. Rep. 202 (1991) 99.

\bibitem{Serednyakov/H-97}
S.I.Serednyakov, Plenary talk given at HADRON-97 Conference, Brookhaven,
August 25-30, 1997.

\bibitem{CMD-2}
R.R.Akhmetshin et al, Physics Letters, B415 (1997) p.445.

\bibitem{Q-model}
P.J.O'Donnell. Rev.Mod.Phys., v.53 (1981) p.673.

\bibitem{DES82} 
N.G.Deshpande and G.Eilam, Phys. Rev. D, v.25 (1982) p.270.

\bibitem{IVN1} V.P.Druzhinin et al., Z.Phys. C37 (1987) pp.1--10.

\bibitem{IVN2} N.N.Achasov, V.N.Ivanchenko, 
Preprint Budker 87-129, Novosibirsk, 1987, 
Nucl.Phys.  B315 (1989) p.465.

\bibitem{IVN3} J.Weinstein, N.Isgur, 
 Phys. Rev. D41 (1990) p.2236.

\bibitem{IVN4} S.Fajfer, R.J.Oakes, 
Phys. Rev. D42 (1990) p.2392.

\bibitem{IVN5}  A.Bramon, A.Grau, G.Panchieri,
Phys. Lett., D283 (1992) p.416.

\bibitem{IVN6} F.E.Close, N.Isgur, S.Kumano,
Nucl.Phys., B389 (1993) p.513.

\bibitem{IVN7} N.Broun, F.E.Close, 
 {\it The second DA$\Phi$NE Physics Handbook, Vol.2},
   Frascati: INFN Frascati, 1995, pp.649--662.

\bibitem{IVN8} A.Bramon, M.Greco, ibid, pp.663--670.

\bibitem{IVN9} J.L.Lucio, M.Napsuciale,  
Phys.Lett., B331 (1994) p.418.

\bibitem{IVN12}  N.A.Tornquist,
 Phys. Rev. Lett., 49 (1982) p.624.

\bibitem{IVN13}  R.L.Jaffe,
Phys. Rev., D15 (1997) pp.267, 281.

\bibitem{IVN11}  N.N.Achasov, V.V.Gubin,
Phys. Rev., D56 (1997) p.4084.


\bibitem{IVN22}  N.N.Achasov,
March 1998. e-Print Archive: hep-ex/9802327;
Pisma v JETP 67 (1998) p.445.


\bibitem{theor1}
V.A.Karnakov, Yad. Fiz., 42 (1985) p.1001.

\bibitem{theor2}
N.N.Achasov, A.A.Kozhevnikov, Int. J. Mod. Phys., A 7 (1992) p.4825. 

\bibitem{RadCor}
E.A.Kuraev,V.S.Fadin, Sov. J. Nucl. Phys., 41 (1985) p.466.

\bibitem{TDUM}
A.D.Bukin, Simulation of the elastic scattering process 
$ e^+ e^- \to e^+ e^- $ with radiative corrections. 
Preprint BINP 85-124, Novosibirsk, 1985.

\bibitem{TDNEEG}
M.R.Jane et al., Phys.Lett., V.59B (1975), p.103.
M.R.Jane et al., Phys.Lett., V.73B (1978), p.503, erratum.

\bibitem{TDEEGG}
E.A.Kuraev, A.N.Peryshkin, Sov.J.of Nucl.Phys., v.42 (1985) p.1195
(in russian).

\bibitem{3g1} Behrend H.-J. et al.,Phys. Lett., B202 (1988) p.154.

\bibitem{3g2} Fernandes et al., Phys.Rev., D35 (1987) p.1.

\bibitem{VSClegg1} A.B.Clegg, A.Donnachie, Z.Phys., C 62 (1994) p.455.

\bibitem{VSAchasov1} N.N. Achasov and A.A.Kozhevnikov, 
Phys.Rev., D55 (1997) p.2663.

\bibitem{VSDonnachie2} A.Donnachie and Yu.S.Kalashnikova, 
Z.Phys. C 59 (1993) p.621.

\bibitem{VSDonnachie3} A.Donnachie, Yu.S.Kalashnikova and Clegg, 
Z.Phys. C 60 (1993) p.187.

\bibitem{VSAston} D.Aston et al., Nucl.Phys. B, Proc. 
Supplement 21 105 (1991).

\bibitem{VSDonnachie1} A.Donnachie, A.B.Clegg, 
Phys.Lett. B 269 (1991) p.450.

\bibitem{VSOLYA_1} L.M.Kurdardze et al., JETP Lett. 47 (1988) p.512. 

\bibitem{VSOLYA_2} L.M.Kurdardze et al., JETP Lett. 43 (1986) p.643.

\bibitem{VSDM2_01} D.Bisello et al., preprint LAL 90-35 (1990). 

\bibitem{VSYY2_1}  C.Bacci et al., Nucl.Phys. B184 (1981) p.31.

\bibitem{3pi}
N.N.Achasov, N.M.Budnev et al., 
An electromagnetic  $\rho-\omega$ mixing as a tool for investigation
of the reaction $e^+e^-\to V\pi \to 3\pi$, 
Sov.J. of Nucl. Phys., v.23 (1976) p.610 (in russian).

\bibitem{bkMane}
F.~Man\'{e} et al.,
Phys. Lett., B 99 (1981) p.261.

\bibitem{olya} P.M.Ivanov et al.,
Pisma v JETP, v.36, issue.3 (1982) pp.91--94 (in russian).

\bibitem{cmd} G.V.Anikin et al. Preprint INP 83-85, Novosibirsk, 1983.

\bibitem{dm1_1} B.Delcourt et al. Phys. Lett., 113B (1982) p.93.

\bibitem{dm1_2} F.Mane et al. Phys. Lett., 112B (1982) p.178.

\bibitem{dm2} D.Bisello et al. Z. Phys. C52 (1991) p.227.

\bibitem{c1480_1} S.I.Bitukov et al. Phys. Lett., 188B (1987) p.383.

\bibitem{c1480_2} N.N.Achasov, A.A.Kozhevnikov. 
Phys. Lett., 207B (1988) p.199.

\bibitem{vslyaf13}
A.I.Vainshtein, I.B.Khriplovich, Journal of Nuclear Physics, 
13, 2(1971) p.620.

\bibitem{vslyaf21}
V.N.Novikov, S.I.Eidelman, Journal of Nuclear Physics,
21, 5(1975) p.1029.

\bibitem{vslyaf48}
Vorobyev P.V., Golubev V.B., Dolinsky S.I. et al., 
Journal of Nuclear Physics, 48, 2(8) (1987) p.436.

\bibitem{vslyaf40}
Belkov A.A., Kuraev E.V., Pervushin V.N., Journal of Nuclear Physics,
40, 6(12) (1984) p.1483.


\end {thebibliography}
\end{document}